\RequirePackage[hyphens]{url}
\documentclass{ati}
\usepackage[T2A,T1]{fontenc}
\usepackage[utf8]{inputenc}
\usepackage[russian,estonian,english]{babel}
\usepackage{amsmath,amssymb,gensymb,amsthm,graphicx,color}
\usepackage{breakcites}

\usepackage{sectsty}
\chapternumberfont{\huge} 
\chaptertitlefont{\huge}

\usepackage[style=authoryear,citestyle=authoryear,backend=biber,maxbibnames=99,backref=true]{biblatex}

\let\citep\parencite
\let\cite\textcite
\bibliography{references.bib}

\usepackage{enumitem}
\usepackage{makecell}

\usepackage{rotating}
\usepackage{chngcntr}
\counterwithout{figure}{chapter}

\begin{document}

\def\seriesname{Dissertationes Informaticae Universitatis Tartuensis}
\def\seriesnumber{19}
\pagestyle{empty}\begingroup\centering\sffamily
\vbox{\small\MakeUppercase{\seriesname}}
\vbox{\bfseries\small\seriesnumber}
\vfill\newpage\null\newpage
\vbox{\small\MakeUppercase{\seriesname}}
\vbox{\bfseries\small\seriesnumber}
\vskip52mm
\vbox{\bfseries\LARGE\MakeUppercase{Ilya Kuzovkin}}
\vskip12mm

\vbox{\LARGE{Understanding Information Processing in Human Brain by Interpreting Machine Learning Models}}

\vskip7mm
\vbox{\large{A Data-Driven Approach to Computational Neuroscience}}
\vfill
\vbox{\MakeUppercase{Tartu \the\year}}
\endgroup\newpage

\begingroup\setlength\parindent{0pt}\setlength\parskip\baselineskip
Institute of Computer Science, Faculty of Science and Technology, 
University of Tartu, Estonia.

Dissertation has been accepted for the commencement of the degree of
Doctor of Philosophy (PhD) in informatics on June 18, 2020 by the Council of
the Institute of Computer Science, University of Tartu.

\emph{Supervisor}

\begin{tabular}{p{20mm}l}
Prof. Dr. & Raul Vicente Zafra\\
& Computational Neuroscience Lab\\
& University of Tarty, Estonia
\end{tabular}

\emph{Opponents}

\begin{tabular}{p{20mm}l}
Prof. Dr.& Fabian Sinz\\
&IRG Neuronal Intelligence\\
&University of Tübingen, Germany\\
\\
Dr. & Tim C Kietzmann\\
&Donders Institute for Brain, Cognition and Behaviour\\
&Radboud University, Netherlands
\end{tabular}

The public defense will take place on September 22, 2020 at 14:15 in the University of Tartu Delta Centre, room 1021, Narva mnt 18, Tartu, Estonia.

The publication of this dissertation was financed by the Institute of
Computer Science, University of Tartu.

\vfill
Copyright \textcopyright\ \the\year\ by Ilya Kuzovkin

ISSN 1024-4212\\
ISBN 978-9949-03-398-0 (print)\\
ISBN 978-9949-03-399-7 (PDF)

University of Tartu Press\\
\url{http://www.tyk.ee/}
\endgroup 


\newpage\null\vskip35mm
\begin{flushright}
\itshape To the academic spirit of the city of Tartu
\end{flushright}

\newpage\pagestyle{plain}

%
%

\chapter*{Abstract}

\vspace{-2.0em}
\small
The thesis explores the role machine learning methods play in creating intuitive computational models of neural processing. We take the perspective that, combined with interpretability techniques, machine learning could replace human modeler and shift the focus of human effort from creating the models to extracting the knowledge from the already-made models and articulating that knowledge into intuitive representations. Automatic model-building methods can process larger volumes of data and explore more computationally complex relationships than a human modeler could. This perspective makes the case in favor of the larger role that exploratory and data-driven approach to computational neuroscience could play while coexisting alongside the traditional hypothesis-driven approach. We provide an example of how an intuitive model can be extracted from machine-learned knowledge, explore major machine learning algorithms in the context of the knowledge representation they employ, and propose a taxonomy of machine learning algorithms based on the knowledge representation that is driving their decision-making process.

We exemplify the illustrated approach in the context of the knowledge representation taxonomy with three research projects that employ interpretability techniques on top of machine learning methods at three different levels of neural organization. In each case we demonstrate the applicability of the approach and present the neuroscientific knowledge it allowed us to extract. The first study (Chapter \ref{ch:spectral-signatures-based}) explores feature importance analysis of a random forest decoder trained on intracerebral recordings from 100 human subjects to identify spectrotemporal signatures that characterize local neural activity during the task of visual categorization. The second study (Chapter \ref{ch:intracranial-dcnn-based}) employs representation similarity analysis to compare the neural responses of the areas along the ventral stream with the activations of the layers of a deep convolutional neural network. The analysis allowed us to make conclusions and observations about the hierarchical organization of the human visual cortex and the similarities between the biological and an artificial system of vision. The third study (Chapter \ref{ch:mental-space-visualization-based}) proposes a method that allows test subjects to visually explore the state representation of their neural signal in real time. This is achieved by using a topology-preserving dimensionality reduction technique that allows to transform the neural data from the multidimensional representation used by the  computer into a two-dimensional representation a human can grasp.

Taken together, the approach, the taxonomy, and the examples, present a strong case for the applicability of machine learning methods in conjunction with interpretability techniques to automatic knowledge discovery in neuroscience. Seen from this perspective, machine learning models cease to be mere statistical black boxes and, by capturing the underlying dynamics of real life processes, reintroduce themselves as candidate models of reality.
\normalsize

\tableofcontents


\addcontentsline{toc}{chapter}{Introduction}
\chapter*{Introduction}

It has been a very long time since humans began to use their reasoning machinery -- the brain, to reason, among other things, about that same reasoning machinery itself. Some claim that such self-referential understanding is impossible to attain in full, but others are still trying and call it Neuroscience. The approach we take is the very same we use to understand almost any other phenomenon -- observe, collect data, infer knowledge from the data and formalize the knowledge into elegant descriptions of reality. In neuroscience we came to refer to this later component as \emph{modeling}. Many aspects of the phenomenon in question were addressed and explained using this approach by neuroscientists over the years. Some aspects remain unexplained, some others even unaddressed. 

Entering the era of digital computing allowed us to observe and collect data at ever-growing rate. The amount of data gave rise to the need, and the increase in computational power provided the means, to develop automatic ways of inferring knowledge from data, and the field of Machine Learning was born. In its essence it is the very same process of knowledge discovery that we have been using for years: a phenomenon is observed, the data is collected, the knowledge is inferred and a formal model of that knowledge is created. The main difference being that now a large portion of this process is done automatically.

Neuroscience is traditionally a hypothesis-driven discipline, a hypothesis has to be put forward first, before collecting and analyzing the data that will support or invalidate the hypothesis. Given the amount of work that is required to complete a study, the reason for the process being set up in this way has a solid ground. In a setting where collecting data and extracting the knowledge takes a long time, exploratory analysis would indeed have a low yield in terms of solid and actionable knowledge as exploratory analysis can often result in finding nothing of value. However, with the new ways of automatic knowledge discovery the time that is required to complete the process has decreased and the balance between hypothesis-driven and exploratory, data-driven, approach is starting to change. In this work we put forward the argument that machine learning algorithms can act as automatic builders of insightful computational models of neurological processes. These methods can build models that rely on much larger arrays of data and explore much more complex relationships than a human modeler could. The tools that exist to estimate model's generalization ability can act as a test of model's elegance and applicability to the general case. The human effort can thus be shifted from manually inferring the knowledge from data to interpreting the models that were produced automatically and articulating their mechanisms into intuitive explanations of reality. 

In Chapter \ref{ch:ml-ns-synergy} we explore the history of the symbiosis between the fields of neuroscience and machine learning, evidencing the fact that those areas of scientific discovery have a lot in common and discoveries in one often lead to progress in another. Chapter \ref{ch:ml-for-modeling} explores more formally what would it take to create an intuitive description of a neurological process from a machine-learned model. We present the subfield called \emph{interpretable} machine learning, that provides the tools for in-depth analysis of machine learning models. When applied to neural data, it makes those models to be a source of insights about the inner workings of the brain. We propose a taxonomy of machine learning algorithms that is based on the internal knowledge representation a model relies on to make its predictions. In the following chapters \ref{ch:spectral-signatures-based}, \ref{ch:intracranial-dcnn-based} and \ref{ch:mental-space-visualization-based} we provide examples of scientific studies that gained knowledge about human brain by interpreting machine learning models trained on neurological data. The studies present the applicability of this approach on three different levels of organization: Chapter \ref{ch:spectral-signatures-based} shows how the analysis of a decoder trained on human intracerebral recordings leads to a better understanding of category-specific patterns of activity in human visual cortex. Chapter \ref{ch:intracranial-dcnn-based} compares the structure of human visual system with the structure of an artificial system of vision by quantifying the similarities between knowledge representations these two systems use. The final chapter makes a step into even higher level of abstraction and employs topology-preserving dimensionality reduction technique in conjunction with real-time visualization to explore relative distances between human subject's mental states.

With this work we aim to demonstrate that machine learning provides a set of readily available tools to facilitate automatic knowledge discovery in neuroscience, make a step forward in our ways of creating computational models, and highlight the importance and unique areas of applicability of exploratory data-driven approach to neuroscientific inquiry.

\chapter{Synergy between neuroscience and machine learning}
\label{ch:ml-ns-synergy}

Both, neuroscience and artificial intelligence, share, as one of their goals, the purpose of uncovering and understanding the mechanisms of intelligence\footnote{Here and throughout this work we adhere to using this loosely defined term to denote the collection of properties and behavior patterns that we attribute to systems that have analytic capabilities, can operate using abstract notions and carry out high level planning. The search for mechanisms of intelligence is congruent to the search for the precise definition of what the intelligence is, until that search is over, we need a term we can use, we use \emph{intelligence}.}. Neuroscience analyzes the existing examples of intelligent systems that animals and humans have, and tries to figure out how these systems work. Artificial intelligence approaches the task by searching through the space of possible solutions, implementing them one by one and using incremental improvements in performance as the guiding light. Sharing a common goal makes it inevitable that the paths of those two fields of scientific inquiry will cross.

\section{Neuroscience-inspired machine learning}
\label{sec:ns-to-ml}

Before exploring the ways machine learning can contribute to neuroscientific research, we first review the role neuroscience has played in establishing one of the most important machine learning methods of the present day. Since both fields contribute to the quest of solving intelligence, we find that it is important to explore the symbiosis between the fields, establish the benefit it had and highlight the importance of maintaining that symbiotic relationship going forward. This section provides the context for our work and helps to advocate in favor of interdisciplinary scientific inquiry, by which the results and methods of one field can greatly benefit the progress in another.

\subsection{Historical influence of neuroscience}

The first contribution from the field of neuroscience to the field of logical calculus, and thus to the early stages of AI research, can be traced to \cite{mcculloch1943logical}, where the authors describe the nervous system as \emph{``a net of neurons, each having a soma and an axon. Their adjunctions, or synapses, are always between the axon of one neuron and the soma of another. At any instant a neuron has some threshold, which excitation must exceed to initiate an impulse''}. They then show that \emph{``to each reaction of any neuron there is a corresponding assertion of a simple proposition''}, propose a mathematical model of an artificial neuron that is capable of the same behaviour as simplified biological neuron in the description, and postulate \emph{``Theorem II: Every temporal propositional expression is realizable by a net of order zero.''}, allowing to draw parallels between mathematical logic and inner workings of human brain.

Growing attention towards the \emph{``feasibility of constructing a device possessing human-like functions as perception, recognition, concept formation, and the ability to generalize from experience''}~\citep{rosenblatt1957perceptron} led to the first mechanism that was able to modify it's behavior by learning from examples -- perceptron~\citep{rosenblatt1958perceptron}, a physical system built from artificial neurons that were able to adjust their weights (artificial simplistic analog of synaptic connections).

According to \cite{schmidhuber2015deep} early works on animal visual cortex such as the ones by \cite{hubel1959receptive, hubel1962receptive} inspired layered architectures in artificial neural networks that became known as \emph{multilayer perceptrons} \citep{rosenblatt1961principles}, which, paired with the power of backpropagation algorithm \citep{werbos1974beyond, rumelhart1985learning}, are the backbone of modern deep learning~\citep{lecun2015deep}. The concept of the perceptive field from the same work has contributed to the notion and success of convolutional neural networks in computer vision \citep{fukushima1980neocognitron, lecun1998gradient, krizhevsky2012imagenet} by proposing a way of how visual information is being processed in animal brain.

A second pillar of contemporary AI \citep{hassabis2017neuroscience} is the field of reinforcement learning (RL). Dating back to the work on animals done by \cite{pavlov1903experimental} that later became known as \emph{classical conditioning} \citep{rescorla1972theory} the principles of reinforcement learning made their way into computer science and machine learning with the works of \cite{sutton1990time, sutton1998introduction}. Paired with deep learning, reinforcement learning was instrumental for achieving such results as computers learning to play computer games with no prior knowledge \citep{mnih2015human, alphastarblog, OpenAI_dota}, winning world champion at Go \citep{silver2017mastering}, and others. 

The historical lens that we have presented here allows us to appreciate the enormous impact neuroscience had on the development of the fields of machine learning and artificial intelligence.

\subsection{Examples of modern machine learning techniques\\inspired by neuroscientific insights}

There exists a certain difference in opinion when it comes to the question of how brain-like the \emph{modern} artificial learning systems are. In its most popular form the question is ill-posed and does not look into the matter deep enough to make that debate useful. We would like to attempt to rectify that by highlighting that it is important to keep the discussion separate for different levels of analysis \citep{marr1976understanding}: the level of implementation, the level of algorithm and representation, and the most abstract -- the computational level.

On the level of \emph{implementation} (following Marr's tri-level taxonomy), while there is a superficial similarity between biological neural networks and modern machine learning architectures, the specifics of engineering detail differ a lot. At this lowest level of analysis we would side with the claim that apart from the superficial similarity between a biological neuron and an artificial neuron, the systems are fundamentally different. However, as we move to a higher level of abstraction at the \emph{level of algorithm and representation}, the design principles, representations and strategies of information processing of biological systems sometimes start to resemble the architectural principles that the best artificial systems rely on. We will show several such examples later in this chapter. On the \emph{computational} level, that reflects the goal and purpose of the computation, biological and artificial system are often identical: object and speech recognition, speech synthesis, decision-making based on observations, spatial orientation -- these are some of the examples of computational goals that biological and artificial systems share.

In this section we will demonstrate several examples where from the similarity on the computational level (the goal of the computation) emerges the similarity on the level of algorithm and representation. In other words, when the goal of an artificial system coincides with the goal of the corresponding biological system, then the algorithmic mechanism of achieving that goal in an artificial system follows the mechanism we know to exist in its biological counterpart. These examples extend the discussion of the similarities between artificial and biological systems and demonstrate that there is more to this question than the simplistic comparison between neurons and units in an artificial neural network.
\ \\

 \textsc{Working memory.} The mechanism of \emph{working memory} is an important cognitive system that allows us to hold and use the information that is immediately relevant to the task at hand. It can contain context, recent occurrences, and bits of information that preceded the current moment. It also allows to hold pieces of information back while running another cognitive process and then recall the held-back information. This ability is crucial for reasoning and decision-making where the next logical step might depend on results of another, intermediate, process. The very similar challenge exists in artificial learning systems: an algorithm might need to remember some information to use it later, to match information across time and make decision based on temporally disjoint inputs. Recurrent Neural Networks (RNNs) \citep{hopfield1982neural} and later Long Short-Term Memory (LSTM) networks \citep{hochreiter1997long} were proposed to address that challenge. LSTM network consists of extended artificial neurons, that have a memory cell to hold a certain values and a set of gates, that regulate under which conditions the content of the memory cell can be modified or released back into the network. Since we do not know how biological working memory is working, we cannot claim the similarity on algorithmic level, but the similarity on computational level is clearly present.
\ \\

 \textsc{Associative memory.} It has been conjectured that there are multiple memory types in a human brain \citep{tulving1985many}. Other types of biological memory gave rise to various ideas in machine learning and reinforcement learning. \emph{Associative memory}, characterized by the ability to recall a certain piece of information by triggering a certain stimulus, found its reflection in an artificial memory model called \emph{Hopfield network}~\citep{hopfield1982neural} -- a neural network that can store different patterns and a given partial pattern return the whole. According to \cite{hassabis2017neuroscience}, \emph{experience replay}, a critical component of Deep Q-Network (DQN) \citep{mnih2013playing}, was \emph{``directly inspired by theories that seek to understand how the multiple memory systems in the mammalian brain might interact''} and draw the parallel between the role of hippocampus and experience replay buffer: \emph{``the replay buffer in DQN might thus be thought of as a very primitive hippocampus, permitting complementary learning in silico much as is proposed for biological brains''}. Persistent, \emph{long-term memory} is also a crucial part of a biological intelligent system, and, although the biological mechanisms of it did not yet find direct reflection in artificial intelligence systems, the conceptual necessity for this type of memory is widely acknowledged and was implemented in Neural Turing Machines~\citep{graves2014neural} and later in an architecture called Differentiable Neural Computer~\citep{graves2016hybrid}.
\ \\
\ \\

 \textsc{Predictive coding.} The theory of \emph{predictive coding} \citep{rao1999predictive} proposes that the brain learns a statistical model of the sensory input and uses that model to predict neural responses to sensory stimuli. Only in the case when the prediction does not match the actual response the brain propagates the mismatch to the next level of the processing hierarchy. By building and memorizing an internal model of the sensory input such mechanism would reduce the redundancy of fully processing each sensory input anew at all levels and thus greatly reduce the processing load on the sensory system. A recently proposed AI agent architecture called MERLIN \citep{wayne2018unsupervised} achieves a marked improvement on the tasks \emph{``involving long delays between relevant stimuli and later decisions: <...> navigation back to previously visited goals, rapid reward valuation, where an agent must understand the value of different objects after few exposures, and latent learning,
where an agent acquires unexpressed knowledge of the environment before being probed with
a specific task''} by introducing the similar principle into the architecture of the system. The authors point out that using reinforcement learning to learn the entire system at once, including the representations of the input, recurrent computation, rules for accessing the memory, and the action-making policy is indirect and inefficient. They propose to decouple the learning of the sensory data from learning the behavior policy that drives the decision-making by creating a subsystem that learns to compress sensory observations into efficient representation in an unsupervised manner. The decision-making policy is a recipient of already encoded information and thus does not have to learn the encoding through trial and error. The authors acknowledge the theory of predictive coding as one of the inspirations for the architecture.
\ \\

 \textsc{Successor representations.} The trade-off between \emph{model-based} and \emph{model-free} methods is a long-standing question in the field of RL. As the name suggests, the agents in the model-based methods have to learn (or have access to) the model of the environment, while model-free agents try to map observations directly onto actions or value estimates. While having a model would allow the agent to use it to plan ahead and be more \emph{sample efficient} during learning, it also poses significant challenges as learning a model of the environment, especially if the environment is complex, is a very hard task. Many successful results were achieved with model-free methods as those are easier to implement and learning the mapping between the observations and the actions is in most cases sufficient and is easier than properly learning the model of the environment. The idea of \emph{successor representations} \citep{dayan1993improving} lies in-between those two approaches. During the learning the agent counts how often the transition between a state $s_a$ and state $s_b$ has occurred. After interacting with the environment for some time the agent forms what is called an \emph{occupancy matrix} $M$, which holds empirical evidence of transitioning between the states. This matrix is much easier to obtain than a full model of the environment and at the same time it provides some of the benefits of the model-based approach by allowing to model which transition is likely to occur next. The hypothesis that the brain is using successor representations proposes that brain stores in some form the occupancy probabilities of future states and is supported by behavioral \citep{tolman1948cognitive, russek2017predictive, momennejad2017successor} and neural evidence \citep{alvernhe2011local, stachenfeld2017hippocampus}. Using this statistics the brain can estimate which states are likely to occur next, serving as a computationally efficient approximation of a full-fledged environment model. The revival of the original concept in the context of RL~\citep{momennejad2017successor} proposes a way to introduce some of the benefits of model-based methods without sacrificing the efficiency and ease of implementation of model-free methods. 
\ \\

 \textsc{Grid cells.} In 2014 the Nobel Prize in Physiology or Medicine was awarded for the discovery of cells that constitute the positioning system in the brain \citep{o1976place, sargolini2006conjunctive}. In the recent work by \cite{banino2018vector} it was demonstrated that an artificial agent trained with reinforcement learning to navigate a maze starts to form periodic space representation similar to that provided by grid cells. This representation \emph{``provided an effective basis for an agent to locate goals in challenging, unfamiliar, and changeable environments''}.
\ \\

 \textsc{Attention.} After providing the initial motivation for convolutional neural networks (CNN) via the ideas of hierarchical organization and the concept of a receptive field, neuroscience served a source of ideas for further improvement though the concept of \emph{attention} \citep{desimone1995neural, posner1990attention, olshausen1993neurobiological}. Adding the similar functionality to CNNs \citep{mnih2014recurrent, ba2014multiple} helped to further improve the performance of visual object recognition. The same concept was found to be useful in artificial neural networks designed for natural language processing tasks \citep{bahdanau2014neural, vaswani2017attention} and as a component of the memory module of differentiable neural computers \citep{graves2016hybrid}.
\ \\

\textsc{Memory consolidation.} The standard model of \emph{systems memory consolidation} \citep{squire1995retrograde} suggests that a novel memory is first retained in hippocampus, and then, with each new recollection of that memory, its engram is strengthened in neocortex, making the memory permanent \citep{dudai2004neurobiology}. On one hand this mechanism ensures that the important memories, that are being recalled often, become permanent, but it also keeps neocortex free of clutter and thus makes it more stable. If every new memory would be immediately consolidated we would remember too much ``noise''. A similar principle is used in Double DQN architecture \citep{van2016deep} to make deep reinforcement learning process more stable: two networks instead of one are maintained at the same time, the \emph{online} network is used to pick actions and its weights are updated immediately, while the second, \emph{target} network, is used to evaluate the selected actions and is updated periodically. Periodic update of the network, in contrast to updating the network immediately, provides more stable evaluation of actions -- within the period between the updates the actions are evaluated by the same network, allowing those evaluations to have a common reference point and thus serving as a better relative measure of the quality of an action.
\ \\

The examples we discussed in this section demonstrate that on algorithmic level, the biological and artificial systems sometimes share curious similarities. This observation holds a very promising message in the context of our work: since the systems share some of the properties, it can be informative to analyze one in order to gain knowledge about the other. In our case -- to analyze artificial learning systems, explore their mechanisms and hypothesize the similarities between those mechanisms and the cognitive processes of biological systems.

\section{The role of machine learning in neuroscience}
\label{sec:ml-to-ns}

Approximately 20 years after \cite{hodgkin1952quantitative} published their fundamental single neuron model that inspired multiple works in mathematical modeling of neural dynamics, the field has accumulated enough methodology and data to start looking into the models of neuronal populations~\citep{wilson1972excitatory, wilson1973mathematical, nunez1974brain}. Due to the volume of that data and the complexity of the systems being modeled, the community turned to statistical methods to provide approximations of aggregate behavior~\citep{lopes1974model, buice2009statistical, pillow2005neural, rolls2010noisy}. See \cite{van2013modeling} for more examples. The adoption of statistical modeling, which is a precursor of modern machine learning, has established a link between neuroscience and statistical learning. The advancement of computational power and the growing amount of digital data fueled the development of data processing tools, pattern recognition algorithms and data analysis methods. These tools found multiple applications in various fields, including, of course, the field of neuroscience~\citep{vu2018shared, hassabis2017neuroscience, paninski2017neural, hinton2011machine, glaser2019roles}. According to Semantic Scholar\footnote{https://www.semanticscholar.org}, the percentage of neuroscience papers that mention machine learning has risen from 1.3\% to 7.6\% over the last two decades. In this section we give an overview of the major roles machine learning methods play in neuroscientific inquiry. We then suggest that there is a methodological component that is readily available, would benefit the study of neural systems, and would extend the role of machine learning in neuroscience even further, but, in our observation, this methodology is lacking mass adoption.
\ \\

\emph{Neural decoding} represents the most direct application of machine learning methods to neurological data. A dataset of neural responses to predetermined stimuli is collected, and a machine learning method is tasked with building a model that can reverse the mapping -- given a neural signal it has to learn to identify which stimulus caused that neural response. It does so by inferring a set of rules or statistical associations that map neural responses to the corresponding stimuli in the dataset. One of the earliest applications of data analysis to characterize stimulus specific cortical activity can be traced back to \cite{mountcastle1969cortical} and displays a case of manual data analysis. With the rise of machine learning techniques the process of looking for patterns in vast arrays of data became automated, and now it is safe to say that machine learning is the default approach to neural decoding.  While the studies that employ the approach are too numerous to list here, we would like to mention a few. The algorithm proposed by \cite{bialek1991reading} is one of the first direct attempts to read the neural code to identify movement detection in blowfly's visual system. Already in the work by \cite{seung1993simple} statistical modeling based on maximum likelihood estimation was applied to a decode direction from the activity of sensory neurons. \cite{zhang1998interpreting} had successfully applied the decoding methods to identify animal's location based on the activity of the place cells. \cite{haxby2001distributed} have demonstrated that it is possible to decode fMRI recordings of responses to 8 different visual categories with average accuracy of $96$\%. Decoding of prefrontal activity of rats learning an alternating goal task allowed to predict rat's decision, effectively reading rat's future intention directly from brain activity \citep{baeg2003dynamics}. In the works of \cite{nishimoto2011reconstructing, shen2019deep} it was demonstrated that it is possible to train a decoder that can, albeit with a limited quality, reconstruct visual information such as images or movies directly from fMRI recordings from occipitotemporal visual cortex of human subjects who watched natural movies or images. In their extensive fMRI study, \cite{huth2012continuous} mapped 1705 object and action categories to the changes they evoke in human test subjects watching natural movies, allowing them to map the semantic space of those categories onto cortical maps of human brain. Applying decoding toolbox to the responses of cells to facial stimuli allowed \cite{chang2017code} to identify the code for facial identity in the primate brain. The uncovered code allowed the authors to both predict the neural responses that a particular facial stimulus will elicit and also to decode facial identity from the neural activity. Using recurrent neural networks to decode articulatory movements from cortical activity allowed \cite{anumanchipalli2019speech} to decode the intended utterances and synthesize audible speech.

Neural decoding of the activity of the motor cortex into the intended movement of a subject has branched into its own field, called \emph{brain-computer interfaces}~\citep{wolpaw2002brain}. \cite{fetz1969operant} first demonstrated that a monkey can learn to operate a robotic hand that was controlled by the activity of single cells in the motor cortex, effectively learning to operate an artificial limb. With more advanced multi-site neural ensemble recoding capabilities \cite{wessberg2000real} were able to make accurate real-time predictions of the trajectories of arm movements of a non-human primate and successfully use those predictions for the control of a robotic arm. Similar work by \cite{serruya2002brain} demonstrated even wider applicability of the method by showing that the same approach allows a monkey to move a computer cursor to any location on the computer screen. Finally in the work by \cite{hochberg2006neuronal} the technology was successfully applied to a human subject, allowing to operate a robotic limb with nothing else other than the mental intention to do so.

Performance of a decoding model can be used as a way to quantify the lower bound on the \emph{amount of information or selectivity} of a certain brain region~\citep{hung2005fast, raposo2014category, rich2016decoding}. Providing a learning algorithm with the data from the region of interest and tasking it with decoding forces the algorithm to uncover the information that is pertinent to the process of decoding. The level of performance of the final model informs the investigator on the existence and the quality of relevant information in that region.

Difference in performance of the decoders trained under different experimental conditions provides a way to \emph{quantify that difference and allow for quantitive comparison}. For example, \cite{hernandez2010decoding} recorded the neuronal activity of diverse cortical areas while monkeys performed a certain task. The level of performance of the decoding models trained on the activity from different cortical areas was used as an indicator of the involvement of each particular area in that particular task. Similar approach is used by \cite{van2010triple} to analyze the contribution of hippocampus, ventral striatum, and dorsal striatum into the information processing during a spatial decision task. By comparing the results of decoding the activity of posterior parietal cortex (PPC) under two different tasks, \cite{quiroga2006movement} were able to establish that activity in PPC predicts the location of targets significantly worse than it predicts the intended movements, providing insight into the functional role of that area.

In section \ref{sec:ns-to-ml} we have described multiple models that the field of artificial intelligence has produced in attempts to solve various perceptual and behavioral tasks. Most of the problems we challenge the artificial intelligence systems with are the ones that we, humans, already are capable of solving. This fact naturally leads to the idea that it could be interesting to compare the biological mechanisms of solving these problems with the mechanisms that are employed and learned by artificial systems. The modeling branch of computational neurosciences approaches this question by proposing models of biological systems and comparing the behavior of the proposed models with biological data. We find, and this is one of the main arguments we would like to put forward in this thesis (see Chapter \ref{ch:ml-for-modeling}), that the rise of the fields of artificial intelligence and machine learning awarded us with an alternative way to investigate that question. For example, to quantify the similarity between the hierarchies of a convolutional neural network (CNN) and human ventral stream system \citep{yamins2013hierarchical} employed representational similarity analysis (RSA) \citep{kriegeskorte2008representational}) to find that the representations of that are formed in a CNN were similar to the representations in the ventral stream. Similar and more detailed findings were reported by \cite{cadieu2014deep, yamins2014performance, gucclu2015deep, seeliger2017cnn, kuzovkin2018activations} confirming the evidence in favor of the similarities in hierarchical organization of both biological and artificial systems of vision. \cite{khaligh2014deep} compared representational dissimilarity matrices (RDM) of 37 computational models of vision reaching the same conclusion, that deep convolutional neural networks explain activations of inferior temporal cortex during visual object recognition task. Similar to visual perception there are comparisons between the hierarchical structure of human auditory cortex and hierarchy of artificial neural networks trained to process auditory data \citep{kell2018task, huang2018connecting}.

A new potential role of machine learning in neuroscience was alluded to in the works on Neurochip~\citep{jackson2006long, zanos2011neurochip, nishimura2013spike, zanos2018phase}. Being an example of a bidirectional brain-computer interface, Neurochip both reads the inputs from biological neurons, and, after running on-chip computations on those inputs, stimulates the cortex with its output connections. Seeing the similarities between some computational mechanisms of biological and artificial systems we are very curious to see the development of that idea and creation of a computational system that is a hybrid of biological and artificial circuits. 

\emph{Biologically plausible deep learning} is a direction of research that develops artificial learning architectures under the restrictions the biological systems are prone to. \cite{hinton2007backpropagation} outlined a list of reasons why the processes employed by modern deep learning methods cannot be running in the brain. The question was further explored and reviewed by~\cite{bengio2015towards}. This sparked multiple works~\citep{urbanczik2014learning, lillicrap2016random, liao2016important, scellier2017equilibrium} where those limitations were addressed to demonstrate that it is still possible to achieve learning in an artificial system while respecting some of the biological constraints. This line of research creates yet another way for machine learning to play a role in creating plausible computational models of neuronal processing thus advancing our understanding of the brain.
\ \\

This overview of the major paths of how machine learning benefits the advancement of neuroscience highlights the fact that, for various reasons, numerous machine learning models are being trained on neurological data. While all of those models serve their purpose in the above-mentioned studies, many of them are being treated as ``black box'' tools, where the input is provided and the output is tested and accepted to be used for further analysis, interpretation and confirmation of the experimental findings. In the next chapter we will argue that some of the models that were created in the above-mentioned and other scientific studies have inadvertently captured some of the key computational mechanisms of the phenomena the models were being trained on. The analysis of how exactly these models achieve their results and reach their predictions could lead to unearthing those captured computational mechanisms. We find, that while many research groups are working in this direction, more rigorous and widespread adoption of the tools that facilitate interpretation of machine learning models would require little effort but could lead to new and unexpected byproducts of the main investigation.

\chapter{Machine learning as automatic builder of computational models}
\label{ch:ml-for-modeling}

\begin{flushright}
\emph{``All models are wrong,\\but some are useful.''}\\
-- George E. P. Box
\end{flushright}
\ \\

Building a model of a complex phenomenon is an ancient way for humans to gain knowledge and understanding of that phenomenon. Models of planetary motion~\citep{kepler1621epitome}, gravity~\citep{newton1687philosophiae, einstein1915feldgleichungen}, standard model of particle physics~\citep{wilczek1975weak} are prominent examples of this approach. By comparing the predictions made by a model to observations in the real world, we theorize that the mechanism driving the model could be the same as the one driving the phenomenon. By building more and more accurate models, we approach the true mechanism closer and closer, hoping to get to the point of being able to artificially replicate the phenomenon in full.

This line of scientific inquiry is being widely applied to brain studies as well. The method of mathematical modeling spans across the whole field of computational neuroscience and includes single neuron models~\citep{lapicque1907recherches, hodgkin1952quantitative, koch2004biophysics, herz2006modeling}, network models~\citep{white1986structure, hagmann2008mapping, bullmore2009complex, sporns2010networks, bassett2017network, bassett2018nature}, models of memory~\citep{durstewitz2000neurocomputational, frank2001interactions, chaudhuri2016computational}, cognition~\citep{smith2004psychology, oaksford2009precis, tenenbaum2011grow, palmeri2017model, kriegeskorte2018cognitive} and learning~\citep{hebb1949organization, raisman1969neuronal, zilles1992neuronal, fuchs2014adult}, sensory precessing~\citep{barlow1959sensory, barlow1967neural, ernst2002humans, weiss2002motion, olshausen2004sparse, kording2004bayesian, kriegeskorte2018cognitive} and other neural phenomena. Both computational and structural modeling lead to numerous discoveries and important contributions to our understanding of the nervous system.

The most prized property of a model is our ability to understand its mechanism and thus understand the phenomenon that is being modeled. Coming up with a theory of how a particular phenomenon works and proposing a model that describes it has always required careful and extensive observation, good intuitive understanding of the process and almost artful ability to consolidate the intuition with the observation into a formal description that generalizes well across all instances of the phenomenon. A good sign of a model being successful is its ability to make predictions about future observations and results of interactions, making predictability the first litmus test of any model or theory. Models and theories that do not pass that test are usually discarded from the pool of scientific knowledge.

A typical machine learning pipeline involves such consecutive steps as data acquisition, data preprocessing, training a model that employs statistical tools to describe the data, and testing of the resulting model on a hold-out subset of the data~\citep{murphy2012machine}. This latter step is of particular interest to us in the context of the argument we put forward in this chapter. Statistical learning theory~\citep{vapnik2013nature} addresses the problem of selecting such a model that minimizes its error on the data, while keeping bias and variance of the model as low as possible. Further set of techniques, such as \emph{training/test split}, \emph{cross-validation} and others are then applied to estimate model's performance and its generalization ability. All this theoretical machinery serves one purpose -- the resulting model should accurately describe the data at hand and make correct predictions on previously unseen data samples. A model that does not sufficiently satisfy this requirement is discarded the same way as non-predictive models and theories we discussed in the previous paragraph.

The consequence of the machine learning approach being set up in this way is that all of the successful models that were ever built on neural data, including the ones we have discussed in Section~\ref{sec:ml-to-ns}, do, by design, satisfy the primary requirement of a good model and pass the litmus test of generalizability. In this section we put forward the argument that in addition to solving the primary issue those models were created to address (being it neural decoding, comparison of experimental conditions, quantification of information, or else), they also are models (or reflect the dynamics) of the computational mechanisms that gave rise to that neural data. Our predecessors had to analyze such data manually and use their insight to come up with a good model they understood in great detail. In the era of big data and high performance computing we are facing the opposite -- the analysis of the data and building of a model that satisfies that data is done automatically, but what we sacrifice is the understanding of the resulting model. Thankfully, modern machine learning toolbox does include various methods to achieve model interpretability, which, combined with the abundance of data and computing power leaves us with the best of two worlds -- we can build models on a grand scale and at a fast pace and interpret those models to read out the formalisms they develop, informing us on the underlying neural mechanisms.

\section{Gaining an intuitive understanding of computation carried out by machine-learned models}
The definition of a \emph{mathematical model} is a broad one and includes statistical models, differential equations, computational models, dynamical systems and more. The precise nature of a model produced by a machine learning approach depends on the particular machine learning algorithm that generated the model. In this section we will describe general mechanics of machine learning process, provide an example based on the decision tree algorithm that demonstrates how a computational model is born from local statistical decisions, and describe major families of machine learning methods to understand what kind of model is being created by each of those families when applied to a set of data.

To illustrate the necessity and motivation for the following material let us introduce a hypothetical situation. Let us assume that during a study a group of researchers have obtained vast volumes of data, preprocessed it and successfully trained a machine learning model that accurately differentiates between the experimental conditions and generalizes well to previously unseen data. Now we are in a peculiar situation, where the group of researchers, given the same data, will not be able to decode it, but in their hands they have a model, which does ``know'' how to do it. The content of this section explores the feasibility of transferring that knowledge from within the model and into the researchers.

\subsection{General mechanics of the machine learning approach}
\label{sec:general-ml}
The process starts with a \emph{dataset} of observations, where each particular observation is called a \emph{sample} and is described by a set of values called \emph{features}. A sample can also have a \emph{label} associated with it, that, depending on type of the learning problem, can represent the category of the sample (\emph{supervised} learning, \emph{classification} problem), numerical outcome (supervised learning, \emph{regression} problem), reward from the environment (\emph{reinforcement} learning), or not be present at all (\emph{unsupervised} learning). An example of a neural dataset could be a set of observations in the frequency domain, where the features are particular frequency bands, a sample is described by the powers of those frequencies and has a label indicating whether test subject's eyes were open or closed when that sample was recorded. A straightforward application of machine learning on such data would be to train a decoder (a model), that can identify whether test subject's eyes are open or closed based on the neurological data alone.

Once the initial set of observations is collected, the next steps are \emph{feature engineering} and \emph{feature selection}. During feature engineering one has to come up with the best way to represent the data from the perspective of a machine learning algorithm. In the example above we took powers of the frequency bands as our features, but that was not the only choice available and we did it only because we know that the information about whether the test subject's eyes are closed or open is readily available in the alpha frequencies. We made the decision to represent our data in this particular form because we know that this representation will make it easy for the learning algorithm to identify the pattern that separates closed eyes recordings from open eyes recordings. Feature engineering is often a creative process, that requires both domain knowledge and understanding of the machine learning method that will be subsequently applied. One of the reasons for the popularity of deep learning methods is the ability of deep artificial neural networks to automate feature engineering and learn good features directly from data. This methodology has revolutionized the fields of computer vision~\citep{krizhevsky2012imagenet} and speech recognition~\citep{hinton2012deep}, and proved to be applicable in other areas as well~\citep{lecun2015deep}. 

The subsequent (or alternative) step of \emph{feature selection} is a related, but conceptually different process, where we seek to identify the most representative features and remove the rest to make the learning problem easier. This can be done manually by employing human domain knowledge, or with the help of statistical techniques~\citep{blum1997selection, hall1999correlation}.

The next, central, step is running a machine learning algorithm on preprocessed data. The choice of the algorithm will depend on the type of the learning problem (supervised, unsupervised or reinforcement, classification or regression) and on the types of the features we use to describe the data (numerical or categorial, continuous or discrete, etc). The exact learning mechanism can be quite different depending on the chosen algorithm, but the underlying framework of \emph{mathematical optimization}~\citep{snyman2005practical} is common to all of them. Every machine learning algorithm has two essential components: an \emph{objective function} (also called the \emph{loss function}) that the algorithm has to optimize and a set of \emph{parameters} that it can change in order to optimize the objective. Depending on the algorithm, the parameters can be numerical weight coefficients (examples: linear and logistic regression as presented in~\cite{murphy2012machine}; neural networks), categorical variables and numerical thresholds (decision trees by~\cite{breiman1984classification}; random forest by~\cite{breiman2001random}), points in the feature space (K-means clustering by~\cite{hartigan1979algorithm}; support vector machines by~\cite{cortes1995support}; linear discriminant analysis by~\cite{fisher1936use}) or have one of the multiple other possible representations. The final configuration of parameters in conjunction with the computational process that the algorithm runs is the final model that the algorithm will output. Changing the parameters affects the value of the objective function, so all the algorithm has to do is to find the parameters that work best. To give an example of how this can be achieved we consider the case when the objective function is differentiable and the parameters are continuous, which is the case for such algorithms as artificial neural networks, linear and logistic regression and many others. In such a case gradient-based optimization methods can be applied to iteratively approach better and better model parameters. Each configuration of parameters is a point in the parameter space where the objective function is defined. Since the function is differentiable we can compute the gradient (derivative) of the function at every possible configuration point. That gradient is a vector in the parameter space, that tells us which way we should move the point in order to increase the value of the objective function. Depending on whether we want to maximize or minimize the objective function we respectively move in the direction of the gradient or in the direction opposite to it. This optimization technique is called \emph{gradient descent} (or \emph{gradient ascent}). For a more detailed and formal description of this and other optimization methods see~\cite{vanderplaats2001numerical, snyman2005practical}. Once the optimization process has approached the global or a local optimum within a predefined tolerance threshold, or is unable to improve the result any further, the learning algorithm stops and outputs the configuration of parameters that has achieved the best result so far.

The final step of the process is the evaluation of model's performance and generalization ability. When a human is designing a model, he or she takes particular care to make their model general, so that it would not only describe the data at hand, but also work correctly on future data samples. A machine learning algorithm has no such natural inclination and, if it has sufficient expressive power, tends to memorize the whole dataset, as such representation will be, in most cases, the most accurate one from the optimization perspective. This phenomenon is called \emph{overfitting} and has to be avoided if we want the resulting model to capture the underlying dynamics or patterns in the data. The ability of a model to do so is called \emph{generalization ability} and is as important as accuracy of the representation of the training data. A common approach to estimate generalization ability of a model is to reserve a portion of the data, a \emph{test set}, run the learning procedure on the remaining \emph{training set}, and use the performance of the final model on the test set as the estimate of generalization ability. In most cases the very first algorithm we try will not be successful at finding a good model and we will try many different ones before the one that works is found. In the process of doing so we can overfit to the test set as well. To avoid that the training set is further split in two parts: a smaller training set and a \emph{validation} set. The complete training procedure looks as follows: learning algorithms are trained only on the smaller version of the training set, then their performance is estimated on the validation set and, if desired, the process is repeated until a good model is found. And only then the test set is used once to gauge model's true performance. There are variants to this procedure such as \emph{cross-validation}, \emph{leave-one-out} and a few others, all of which were developed to ensure that the model that was built by an artificial learning system is able to generalize and make accurate predictions on previously unseen data. This process is set in place to emulate human modeler's natural strive towards general and elegant models.

The process we have outlined above is being applied across multiple branches of neuroscientific research. Often, in the context of a particular scientific study that employs machine learning approach, the question of \emph{how} the resulting model achieves its result is not in the spotlight, because the focus is on the result itself. However, behind each successful model, lies, encoded in the values of those parameters, the computational principle that allowed the model to succeed. Often trivial, sometimes revelational -- we will only know once we have interpreted the parameters and unearthed the principle.

\subsection{An example of intuitive understanding emerging from a machine-built decision tree}
The decisions a machine learning algorithm makes during the process of fitting the model to the data are driven by local statistical, information-theoretic, probabilistic or combinatorial rules. The question of whether a combination of such decisions can amount to a comprehensive mathematical model is a valid one to ask. In this section we argue in favor of the positive answer to that question and illustrate our reasoning using one particular learning algorithm -- a decision tree~\citep{breiman1984classification}.

Consider the task of decoding a neural recording to determine whether a test subject's eyes are open or closed, that we introduced above. Assume that the data for that task was recorded using an EEG device, the raw signal was cleaned and transformed to frequency domain, and the power spectral densities of 30 frequencies (1 Hz to 30 Hz) constitute the feature space. Building of a decision tree using ID3~\citep{quinlan1986induction} algorithm would proceed as follows:
\ \\

\begin{enumerate}[label=(\alph*)]
	\item Given the dataset $S$, for each feature $f$ compute, using entropy $\text{H}$, the \emph{information gain} $\text{IG}(S, f) = \text{H}(S) - \text{H}(S|f)$. That number shows the amount of additional information that will be obtained if the dataset $S$ is split into two disjoint subsets $S_\text{right}$ and $S_\text{left}$ using the value $v_f$ of the feature $f$ as the splitting criterion. Maximal information gain will be achieved by splitting at optimal (in terms of information gain) value $v^\ast$. The data samples that have $v_f \geq v^\ast$ are assigned to $S_\text{right}$ and the rest to $S_\text{left}$.
	\item If all of the samples in $S_\text{right}$ belong to the same class (eyes closed, for example) the branching process stops and this subset becomes a \emph{leaf} that contains samples of the ``eyes closed'' category. The same is done with $S_\text{left}$.
	\item If a subset contains samples from both classes, the algorithm goes recursively into this subset and repeats the procedure starting from step (a).
\end{enumerate}

Assume that we have completed the training process, tested the resulting model on a test set and found that the model is very accurate and can reliably identify if test subject's eyes are open or closed. If the purpose of our study was to prove that such decoding is possible, or it was an engineering project for clinical purposes (for example to automatically detect whether a patient is asleep), then we have successfully achieved the goal of our study. Many real-world studies do stop at this stage.

\begin{figure}[h!]
    \centering
    \includegraphics[width=0.9\linewidth]{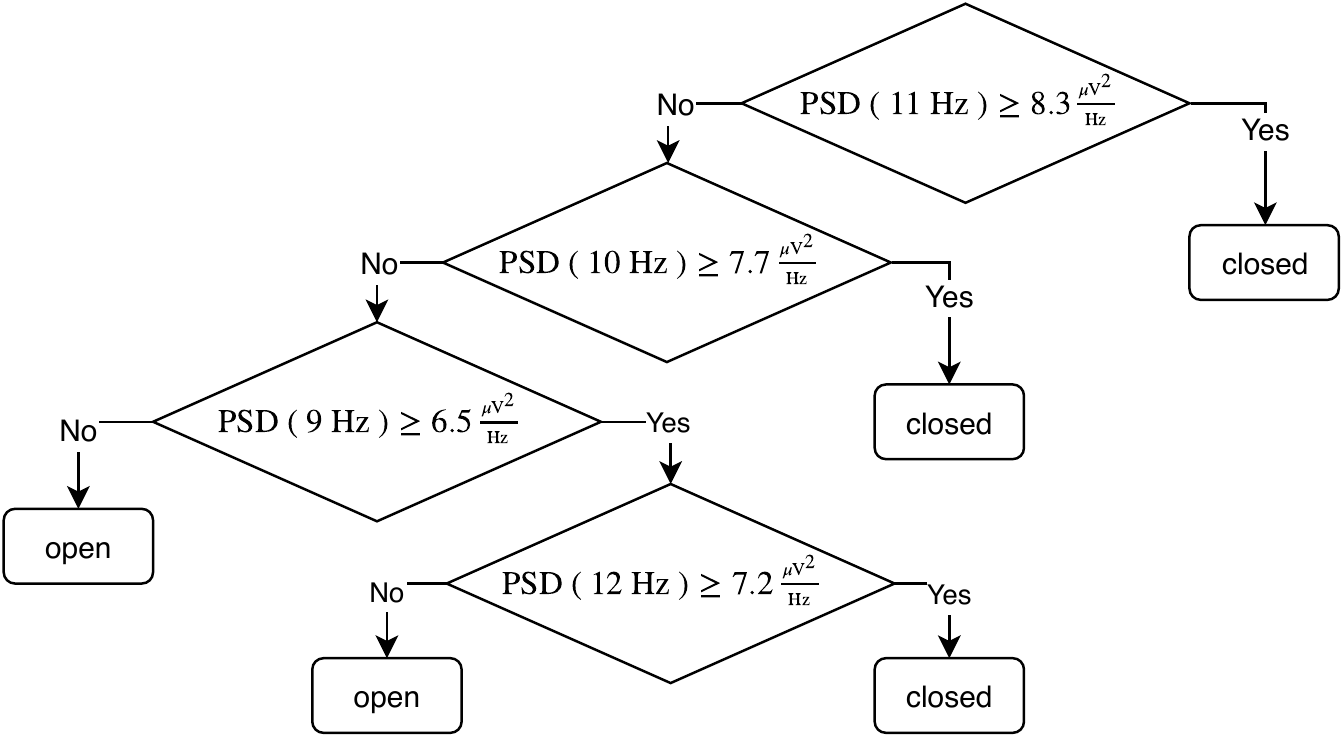}
    \caption{A made-up example of a decision tree built on a dataset of recordings of power spectral density features under two experimental conditions: eyes open and eyes closed. Traversal of this tree provides us with a rule-based computational model and supplies knowledge about neuronal dynamics -- it indicates which frequencies are relevant to the task and which thresholds are the best discriminators of the two experimental conditions.}
    \label{fig:decision-tree}
\end{figure}

We would like to note, that at this point we do have a model that works, but we do not know \emph{why} or \emph{how} it works. An additional step of interpreting the model should be taken in order to answer those questions. In the case of a decision tree the analysis is very simple -- we can visualize the tree that is the final model. Figure \ref{fig:decision-tree} illustrates a made-up example of how such a tree might look like.
This analysis will reveal to us, that the model has 8 parameters -- the four features that are put in the branching points and the four threshold values of these features for making the branching decisions. Over the whole set of frequencies from 1 Hz to 30 Hz the model deemed important only the 11 Hz, 10 Hz, 9 Hz and 12 Hz. This informs us that these are the frequencies which are indicative of the ``eyes closed'' experimental condition. Furthermore we learn that the power spectral density values those frequencies need to reach in order to indicate the ``eyes closed'' condition are, respectively, 8.3, 7.7, 6.5 and 7.2 $\tfrac{\mu\text{V}^2}{\text{Hz}}$. We also find out that the 11 Hz feature provides the highest information gain (since it was selected first and was placed at the root of the tree), and is followed by 10 Hz, and then by 9 Hz and 12 Hz. We can also see, that in the case of 9 Hz reaching the threshold of 6.5 $\tfrac{\mu\text{V}^2}{\text{Hz}}$ there is still a chance that this could happen even under ``eyes open'' condition and thus the further check of whether 12 Hz is higher than 7.2 $\tfrac{\mu\text{V}^2}{\text{Hz}}$ is required, thus indicating that only in conjunction those two features can reliably indicate the ``eyes closed'' condition. All these observations carry information about the neurological correlates of our experimental conditions and all those details would be missed if we would not pursue the analysis and have stopped as soon as the primary goal of the project has been achieved. Pursuing the analysis, however, allowed to postulate an intuitive rule-based computational model of the neural conditions characteristic of the ``eyes closed'' state.

Although this example is trivial, its simplicity allows us to describe the process in full detail. In Chapter \ref{ch:spectral-signatures-based} we provide the details and the findings of a study, that employed similar approach to analyze the contributions of spectral components into the process of visual categorization based on a dataset of 11000 local field potential (LFP) recordings from intracerebral electrodes across 100 human subjects.
\ \\ 

\subsection{Understanding the models built by different machine learning algorithms}
\label{sec:representation-taxonomy}
The example in the previous section has demonstrated that the way to understand a particular machine learning model and the way to interpret it will depend a lot on the algorithm and the architecture that generated the model. The architecture of a decision tree enabled us to readily convert the output of the algorithm into a set of intuitive rules that provide neurological information to a domain expert. Applying other machine learning methods would result in very different representations of the computation that is required to solve the task. The core challenge in gaining an intuitive understanding from observing model parameters lies in the requirement to know the details of the inner mechanism in order to see what is \emph{it} that the model has learned that allows it to make the decisions.
\ \\

\emph{Interpretability} becomes more and more important topic in the machine learning community, across scientific communities that employ machine learning methods, and even in the global community as machine learning models become embedded in our everyday lives~\citep{doshi2017towards}. \emph{``Interpret means to explain or to present in understandable terms. In the context of ML systems, we define interpretability as the ability to explain or to present in understandable terms to a human''} (ibid). Multiple general-purpose methodologies on how interpretability could be achieved have been suggested over the years~\citep{vellido2012making, ribeiro2016model, ribeiro2016should} along with numerous domain specific approaches. Since the notion of an understandable explanation in ambiguous, it is hard to come up with a rigorous method to quantify and measure interpretability of a machine learning model. As a result of this ambiguity, multiple review articles~\citep{lipton2016mythos, bibal2016interpretability, doshi2017towards, guidotti2018survey, gilpin2018explaining, murdoch2019interpretable} proposed different taxonomies to help systematize the way we think about interpretability. Surveys like the one by \cite{narayanan2018humans} are being conducted to empirically estimate interpretability via user-studies. \cite{bibal2016interpretability} systematically explore various terms that are used in machine learning literature to denote interpretability and make suggestions how to bring the terminology in order. The same motivation drives~\cite{lipton2016mythos} and leads to suggesting desiderata for interpretability: trust, causality, transferability, informativeness and ethics, followed by a taxonomy of the properties of interpretable models. Another study by~\cite{doshi2017towards} argues for the need of a rigorous approach and introduces the notion of \emph{incompleteness} of problem formalization. Incompleteness encompasses unquantifiable gaps in knowledge that a model might have and has to be addressed in order to reach the desiderata of a comprehensive model. The outstanding survey by~\cite{guidotti2018survey} proposes a classification of approaches to model interpretability based on the type of the problem, type of the \emph{explanator} adopted, type of the model and type of data. The most recent review~\citep{gilpin2018explaining} provides a good summary of the taxonomies proposed in the previous studies and puts forward a distinction between interpretability and \emph{explainability} -- ability of a model to summarize the reasons for the behavior of the model.

Exploring the question of interpretability in the context of neuroscience allows us to narrow down the scope of applicable desiderata, properties and methods and focus on the ways to uncover knowledge from the models that are the products of automatic scientific discovery. In our case the question we want to answer is \emph{``when translated back from model's representation into neuroscientific domain, what is it that allows the model to make accurate predictions?''}. In such form the question of interpretability is perhaps best covered by \emph{multivariate pattern analysis}~\citep{ritchie2017decoding} on fMRI data~\citep{haxby2012multivariate}, where simple linear methods allowed researchers to decode mental states and analyze their representations~\citep{haynes2006neuroimaging, norman2006beyond, o2007theoretical, kriegeskorte2013representational, haxby2014decoding}. Applying other machine learning methods with the direct goal of extracting neuroscientific knowledge was also attempted by interpreting SVM~\citep{grosenick2008interpretable, hardoon2010decomposing, haufe2014interpretation}, decision trees and random forests~\citep{richiardi2010brain, oh2003estimating}, artificial neural networks~\citep{sturm2016interpretable, samek2017explainable}, probabilistic models~\citep{ma2006bayesian, doya2007bayesian, wolpert2007probabilistic, griffiths2010probabilistic}, dimensionality reduction techniques~\citep{freeman2014mapping, cunningham2014dimensionality}, graphical models~\citep{bullmore2011brain}, and other methods.

The choice of the algorithm for building an interpretable model on neural data is guided by the nature of knowledge representation the authors of the above-mentioned studies were aiming to extract. Such reasoning for the choice of the algorithm leads to yet another basis for a taxonomy of interpretable machine learning models. Given the abundance of different methods and the freedom to choose any of them for a particular neurophysiological study, the obvious choice would be in favor of the method that will uncover the representation that is most interpretable in the context of this particular study. If an investigator is interested in which neural features are the most informative for a given task -- they should choose a method that is naturally suited for feature importance analysis (e.g. Random Forest). If the aim of the investigation is to identify the data samples that are crucial for correct performance -- a method that identifies such samples during the learning process (e.g. SVM). Here we propose a preliminary taxonomy (Table \ref{tab:representation-taxonomy}) of machine learning methods that forgoes classical distinctions such as supervised or unsupervised, predictive or generative and instead organizes the methods into the groups based on the representation of the core knowledge that the model learns in order to make its decisions.
\ \\

\begin{table}[]
	\centering
	\begin{tabular}{|p{7.2em}|p{23.1em}|}
	
		\hline
		\thead{\textbf{Knowledge}\\\textbf{representation}} & \thead{\textbf{Learning algorithms}  } \\
		\hline
		
		\thead{Linear coefficients} &
		\thead{Linear regression\\
		           Logistic regression} \\
		\hline
		
		\thead{Points in feature\\space} &
		\thead{Linear discriminant analysis~\citep{fisher1936use}\\
		           Support vector machines~\citep{cortes1995support}\\
		           K-means~\citep{macqueen1967some}\\
		           Self-organizing maps~\citep{kohonen1990self}\\
		           Density-based spatial clustering of applications with noise\\\hspace*{\fill}(DBSCAN,~\cite{ester1996density})} \\
		\hline
		
		\thead{Distance between\\samples} &
		\thead{Hierarchial clustering~\citep{ward1963hierarchical}\\
		            K-nearest neighbors (kNN, \cite{cover1967nearest})\\
		            Representational similarity analysis\\\hspace*{\fill}(RSA, \cite{kriegeskorte2008representational})} \\
		\hline
		
		\thead{Distribution in\\feature space} &
		\thead{Bayesian learning~\citep{murphy2012machine}, e.g. Na\"ive Bayes classifier\\
		           Gaussian and non-gaussian mixture models\\\hspace*{\fill}\citep{mclachlan2019finite}} \\
		\hline
		
		\thead{States and\\transitions} &
		\thead{Probabilistic graphical models~\citep{koller2009probabilistic},\\\hspace*{\fill}e.g. HMM~\citep{rabiner1989tutorial}\\
		           Reinforcement learning~\citep{sutton2018reinforcement}} \\
		\hline
		
		\thead{Tree structure,\\important features,\\thresholds} &
		\thead{Decision trees~\citep{breiman1984classification}\\
		           Random forest~\citep{breiman2001random}} \\
		\hline
		
		\thead{Distributed\\representations\\over inputs or\\latent variables} &
		\thead{Deep learning (\cite{lecun2015deep}):\\
			\phantom{----}Feed-forward neural network (ANN)\\
		           \phantom{----}Convolutional neural networks (CNN, \cite{fukushima1980neocognitron})\\
		           \phantom{----}Recurrent neural networks (RNN, \cite{hopfield1982neural})\\
		           \phantom{----}Long short-term memory\\\hspace*{\fill}(LSTM, \cite{hochreiter1997long})} \\
		\hline
		
		\thead{Compressed\\feature space} &
		\thead{Autoencoder~\citep{vincent2008extracting}\\
		           Restricted Boltzmann Machines\\\hspace*{\fill}(RBM, \cite{hinton2006reducing})\\
		           Multidimensional scaling (MDS, \cite{mead1992review})\\
		           Principal component analysis (PCA, \cite{pearson1901liii})\\
		           Independent component analysis (ICA, \cite{comon1994independent})} \\
		\hline
		
		\thead{Embeddings} &
		\thead{Deep learning methods, such as:\\
			\phantom{----}word2vec~\citep{mikolov2013efficient}\\
			\phantom{----}Convolutional neural networks\\
		           \phantom{----}Graph convolutional networks~\citep{kipf2016semi}\\
		           \phantom{----}node2vec~\citep{grover2016node2vec}}\\
		\hline
		
		\thead{Functions} &
		\thead{Gaussian processes~\citep{rasmussen2003gaussian}} \\
		\hline
		
	\end{tabular}
	\caption{Taxonomy of machine learning algorithms based on the way they represent the knowledge they have gathered during inference (with some examples).}
	\label{tab:representation-taxonomy}
\end{table}

 \textsc{Linear coefficients.} One of the most straightforwardly interpretable, but also the least expressive in terms of encoded knowledge, are the algorithms like logistic and linear regression, that encode the learned inferences in \emph{linear coefficients} of features. The learning process directly optimizes the coefficients to minimize (in the most common formulation) the \emph{cross-entropy} or, in the case of linear regression, the \emph{mean squared} error. The final output of, for example, a logistic classifier can be represented as a separating hyperplane in the feature space. Please note that while the final decision rule of many classification algorithms can be represented by a separating hyperplane, the underlying principles and knowledge on which the separating plane is built are very different across different learning algorithms. From the interpretability perspective linear coefficients indicate each feature's contribution into the final decision, and, especially if features were normalized before training, can allow for comparison between feature importances. Whether a coefficient is positive or negative provides an additional dimension for interpretation.
\ \\

\textsc{Points in the feature space.} Many popular classification and clustering algorithms encode their findings in particular \emph{points in the feature space}. Support Vector Machines find the samples in the training dataset that are next to the decision boundary, thus indicating which samples either have particular significance, or are the fringe members of class categories. Linear Discriminant Analysis (also known as Fisher's discriminant) locates \emph{centroids} of the samples in each category and then devises the separating boundary that is perpendicular to the straight line connecting the centroids. The centroids are thus characteristic of the groups of samples they represent. Similar knowledge representation can be found in one of the most popular clustering algorithms -- K-Means. The algorithm finds the predetermined number of cluster centers and places them at the locations in the feature space that optimize the objective function. Similar, but extended, concept is employed by self-organizing maps, where each unit of a map is assigned to a centroid in a feature space that represents center of a cluster. In the context of interpretability centroids can obtain special meaning when interpreted by a domain expert that has intuitive understanding of the feature space.
\ \\

\textsc{Distance between samples.} Multiple methods make conclusions about similarity of data samples based on pairwise distances between them. The family of hierarchical clustering
algorithms and distance-based classification algorithms such as K-Nearest Neighbors are good examples of this knowledge representation. In neuroscientific domain the method of Representation Similarity Analysis facilitated scientists to compare representational geometry of samples that have different representations. The interpretability of the findings that those methods make is straightforward as the intuitive meaning of a distance between the samples is directly applicable in almost any domain of knowledge.
\ \\

\textsc{Distributions in the feature space.} Extending the idea of important points in the feature space some methods store distributions in the same space and each distribution is assigned to a category, for example to an experimental condition. Mixture models describe data centroids as distributions, providing much more information that a point-based centroid would. Parameters of a distribution will capture statistical properties of each group of samples, modeling the values of the features that describe the samples, their orientation and extent in the feature space. Bayesian learning methods, such as Na\"ive Bayes, learn, before applying the Bayes' rule, the distribution of observations conditioned on the category these observations belong to. The interpretative value of this knowledge representation is similar to that provided by the points in the feature space, but carries considerably more information.
\ \\

\textsc{States and transitions}. Graph-like representation of possible states of a system and transitions between them is a very flexible way to capture the rules inferred by a learning model. Probabilistic graphical models, such as Bayesian networks or Hidden Markov Models, condense the dynamics they observe in the data into probabilistic state machines and are interpretable by human investigator as a set of rules that governs the underlying data generation process.
\ \\

\textsc{Tree structure.} Decision trees and their ensembles such as Random Forest, represent the inferred decision making in a form of a hierarchically organized sequences of threshold-based rules, where at each step of the process a certain rule is applied to the value of a feature and depending on the outcome the sample is assigned to a certain category. Decision trees are considered to be one of the most interpretable algorithms, they store the whole process of decision in an intuitive form, are directly translatable to deterministic decision rules, rank the features by their importance (the more important a feature is, the closer it will be to the root of a tree) and find the threshold values of those features that are meaningful in the context of the decision-making process. All this information can be easily accessed by the investigator and provide domain-specific insight.
\ \\

\textsc{Distributed representations.} The enormous capacity~\citep{vapnik2015uniform} of modern deep learning models has led, on one hand, to their success in the last decade, and to obscuring the reasons for model's decisions on the other. Both the intermediate, and the final knowledge is stored in the model as a set of weights, often measured in millions. However, the final decision of a deep learning model can often be decomposed into (a) hierarchical organization and (b) set of local decision within each layer of the hierarchy. This allows to portray the knowledge stored by the model as a collection of distributed representations, each of which describes one of the decision rules that the model makes during computation. Direct interpretation of this knowledge, is, however, impossible and an investigator has to use additional tools to extract the knowledge. We will review these tools in Section \ref{sec:interpretability-tools}.
\ \\

\textsc{Compressed feature space.} This group of algorithms employs various methods to reach the same end-goal -- find a feature space of a smaller dimensionality than the original, but preserve as much of the information as possible when the data samples are transformed from the original space to the new one. These methods are often referred to as \emph{dimensionality reduction techniques}. An \emph{autoencoder} is an artificial neural network that receives a sample as an input, performs a series of transformations to encode that sample using smaller number of artificial neurons (the \emph{bottleneck} layer), and then decodes it back from the bottleneck layer to the original representation. The difference between the original sample and its reconstruction serves as the objective function that the algorithm has to minimize. \emph{Principle component analysis} finds an orthogonal transformation of the feature space, such that the axis of the new space correspond to those data dimensions with largest variance. The new axis are called the \emph{principle components}. The first component explains largest amount of variance, the second represents the dimension with second-largest variance, etc. After the transformation the user can estimate how many components are to be kept in order to preserve the predetermined percentage of variance (usually 90\%, 95\% or 99\%) and discards the rest. \emph{Multidimensional scaling} takes another approach where instead of the variance, the new feature space, while being reduced in size, preserves the pairwise distances between the samples as well as possible. There are other example algorithms, but the important common property that allows us to group those algorithms together is that in order to preserve the information given limited expressive power, these algorithms are forced to detect an underlying pattern, or a principle, following which the data can be best reconstructed. Capturing that principle allows to reduce the size of the feature space, but also, importantly for the interpretability context, distills the underlying patterns from the less important ones and from the noise.
\ \\

\textsc{Learned embeddings.} While being similar to the previous group in form, \emph{embeddings} are numerical features vectors that are built, differently from dimensionality reduction techniques, by following a rule that captures a specific property, of even semantics, of the original data. Modern word embeddings~\citep{mikolov2013efficient}, for example, are built by teaching a neural network to predict the word that appears in the context of other words in a language corpus. Words that appear in similar context will have similar internal representation in the embedding (feature) space. Since appearing in a similar context carries semantic meaning in natural language, this representation even allows for arithmetic operations on words, such as the one demonstrated by the famous example $\text{E}(king) - \text{E}(man) + \text{E}(woman) = \text{E}(queen)$, where $\text{E}$ is the function that maps the word to its numerical vector in the embedding space. Subtracting $man$ from $king$ leaves us with the embedding that represents notion of ``kingness'', adding $woman$ to that leads to an embedding that combines notion of ``femenineness'' with ``kingness'', leading to the word $queen$. The ability of embedding to capture semantic similarity has been also demonstrated in visual domain~\citep{deselaers2011visual, frome2013devise}. When applied to graphs, an embedding can reflect topological properties of graph nodes~\citep{grover2016node2vec}, or combine topological data with node attributes~\citep{kipf2016semi}. The key property of embeddings for the interpretability efforts is their ability to capture semantic similarity between the data samples.
\ \\

\textsc{Functions.} \emph{Gaussian processes} is an example of a learning method that models a distribution over the possible functions, that are consistent with the observed data. After applying initial constraints to reduce the set of possible functions, the learning process narrows the distribution by eliminating the functions that are not consistent with the data. The final output is a set of possible functions that can capture the learnt knowledge.
\ \\

This brief overview of the basis of the proposed taxonomy of machine learning algorithms provides a useful guide for selecting the method that is appropriate for a learning problem at hand when a certain interpretation of the learnt knowledge is desired. While some of the representations above are straightforward to interpret, others require additional tools, such as visualization, rule extraction or simplification, in order to extract intuitive understanding of the knowledge represented by the model. In the next section we provide an overview of such tools.

%
%
\subsection{Techniques to analyze machine learning models and extract knowledge from representations}
\label{sec:interpretability-tools}

At this point we have completed three out of four steps along the path towards insightful automatically built computational models of neural computation: (1) selected the learning algorithm with accordance to the learning problem and desired end-knowledge representation (Section \ref{sec:representation-taxonomy}), (2) trained the model to fit the data, and (3) evaluated model's performance and generalization ability (Section \ref{sec:general-ml}). The last step is gaining intuitive understanding of the model's knowledge. Depending on the algorithm we used to train the model this knowledge can be already in a human-tractable form, or it might require some additional steps.

Linear models, single decision trees and rules are recognized by the machine learning community as most interpretable \citep{guidotti2018survey} and empirical experiments were conducted to estimate comprehensibility of these models by humans~\citep{huysmans2011empirical}. Following the nomenclature presented in Table \ref{tab:representation-taxonomy} we extend the list of straightforwardly interpretable knowledge representations to include linear coefficients, points in the feature space, distances between samples, and distributions in the feature space. For a domain expert, who has the understanding of the data the model was trained on, these representations have direct meaning and no further steps are required to spot an intuitive meaning if the model has uncovered one.

\emph{Feature importance analysis} is applicable to any model that has a quantifiable way to estimate each feature's contribution into the final decision. Linear models, provided that the data was normalized across the dataset prior to training the model, provide this information in the values of linear coefficients. Decision trees and random forests make branching decision based on the information gain, which acts as a measure of how big of a role a feature plays in the decision-making process. In Chapter \ref{ch:spectral-signatures-based} we use this method to identify time-frequency patterns of neural activity that are important for perceptual categorization in human brain. If a model does not provide quantitative information about each feature's contribution, \emph{sensitivity analysis}~\citep{saltelli2002sensitivity} gives us the means to obtain the same information by measuring how the output of a model is affected by altering the input. If altering the input values of a certain feature (or a set of features) does not change model's behavior, one can conclude that this feature (or features) is not important for the model. If features of the dataset represent specific domain knowledge, their importance is directly interpretable by a domain expert.

\emph{Visualization} of model parameters and internal data representation is one of the main tools for achieving interpretability. Simplest methods include plotting, if dimensionality permits, the data points, decision boundary and the knowledge representation (support vectors, centroids, etc). In most cases, however, dimension of the feature space is too high to be visualized directly and investigators resort to visualizing aggregated statistics, such as histograms of value distributions. Dimensionality reduction techniques, that are, on one hand, learning algorithms in their own regard as we noted in the previous section, can, on the other hand, be essential for visualization efforts. Reducing the dimensions down to 2 or 3 allows to plot the data, decision boundaries and internal data representation in a human-readable form. Different dimensionality reduction techniques focus on preserving different properties of the object that undergoes the reduction, allowing the investigator to choose the right one depending on what property should be preserved. Topology preserving algorithms such as t-distributed stochastic neighbor embedding (t-SNE, \cite{maaten2008visualizing}), self-organizing maps~\citep{kohonen1990self}, and multi-dimensional scaling keep the objects that were similar in the original space also close in the new space. Principle component analysis identifies the data dimensions that have the highest variance and keep the transformation that allows to see the contribution of the dimensions of the original space into the dimensions of the new space. Please refer to \cite{guidotti2018survey} for a comprehensive review of visualization methods in the context of interpretability.

\emph{Automatic rule extraction} is another way to achieve interpretability by converting complex model into a set of rules. In additional to trees and tree ensembles that can be naturally converted to this representation, the approach has been proposed for SVMs~\citep{nunez2002rule} and neural networks~\citep{andrews1995survey, tsukimoto2000extracting, zilke2015extracting}.

The enormous number of trainable parameters and high capacity of deep learning models is one of the reasons for the success of these methods in the last decade. It is also the reason why the decisions made by these methods are even less transparent than the ones made by other machine learning methods. This predicament has led to an explosion in the number of studies proposing, in addition to model-agnostic, also neural network-specific interpretability methods. \emph{Activation maximization} methods identify input patterns that maximize the activation values of a particular neuron, later visual inspection of those patterns can be very insightful, especially in visual input domain~\citep{zeiler2014visualizing}. \emph{Attention} mechanism~\citep{mnih2014recurrent} allows the investigators to analyze the areas of the input that the model deemed worth of its attention and understand which part of the feature space, or what data content, was most relevant for the model. \emph{Proxy models} are trained to approximate the behavior of a neural network on a full, or partial, set of data and build a simpler representation of network's behavior. Due to ease of interpretation, linear~\citep{ribeiro2016should} or decision tree~\citep{craven1996extracting, schmitz1999ann, zilke2016deepred} models are most commonly used as proxy models. \emph{Readout} technique employs linear models to take activations of a subset of neurons as input and train to predict the final outcome (for example the category of a sample), the method draws inspiration from \emph{reservoir computing}~\citep{schrauwen2007overview}. Ability or inability of a readout model to perform well is an indicator of the involvement of the chosen subset of neurons in the decision-making process of the model. \cite{koul2018learning}  demonstrate how to extract \emph{finite state representation} of a recurrent neural network. Please see \cite{gilpin2018explaining} for an extensive survey of interpretability techniques developed specifically for deep learning methods.

\section{Interpretation of machine-learned models for neuroscientific inquiry at different levels of organization}

In the next three chapters of this thesis we present detailed description of three studies that demonstrate how the approach we have described in this chapter can be realized in the context of neuroscientific inquiry to analyze, interpret and understand neural processes at three different levels of organization. Each study shows how the in-depth analysis of the machine learning models can lead to insights or confirmations of conjectures about inner workings of the human brain.

Chapter \ref{ch:spectral-signatures-based} demonstrates the analysis on a level of local field potentials. We first train a decoder to predict visual category based on spectro-temporal activity of single iEEG probes using a random forest classifier. We then perform feature importance analysis to understand which locations are relevant for the task of visual decoding and which, while being active and responsive compared to the baseline, do not carry relevant information. Further analysis of the important parts of TF spectrum shows difference in roles of different neuronal locations and uncovers category-specific patterns we call \emph{spectral signatures} of visual perceptual categorization.

Chapter \ref{ch:intracranial-dcnn-based} shows how the comparison of activations of an artificial model of vision (convolutional neural network) with the activations in human visual cortex allows to draw the analogy between the hierarchical structures of those two systems. This study serves as an example of how machine learning methods can be used to analyze the brain on the level of functional organization.

The third example, demonstrated in Chapter \ref{ch:mental-space-visualization-based}, shows how using dimensionality reduction for clustering and visualization of high-dimensional EEG feature space helps to gain a high-level understanding of relative properties of mental concepts encoded in that space. By visualizing mental state space we can browse the signals generated by the human brain under different conditions and visually assess which ones are close to each other and which ones are further apart. The work shows an application of this concept to the field of brain-computer interfaces and serves as an example of applicability of the interpretable machine approach on the level of mental concepts.

\chapter{Feature importances of random forest models inform on localized cortical activity}
\label{ch:spectral-signatures-based}

Human brain has developed mechanisms to efficiently decode sensory information according to perceptual categories of high prevalence in the environment, such as faces, symbols, objects. Neural activity produced within localized brain networks has been associated with the process that integrates both sensory bottom-up and cognitive top-down information processing. Yet, how specifically the different types and components of neural responses reflect the local networks’ selectivity for categorical information processing is still unknown. In this work we train Random Forest classification models to decode eight perceptual categories from broad spectrum of human intracranial signals ($4-150$ Hz, 100 subjects) obtained during a visual perception task. We then analyze which of the spectral features the algorithm deemed relevant to the perceptual decoding and gain the insights into which parts of the recorded activity are actually characteristic of visual categorization process in human brain. We show that network selectivity for a single or multiple categories in sensory and non-sensory cortices is related to specific patterns of power increases and decreases in both low ($4-50$ Hz) and high ($50-150$ Hz) frequency bands. By focusing on task-relevant neural activity and separating it into dissociated anatomical and spectrotemporal groups we uncover spectral signatures describing neural mechanisms of visual category perception in human brain that have not yet been reported in the literature.

Previous works have shown where and when perceptual category information can be decoded from the human brain, our study adds to that line of research by allowing to identify spectrotemporal patterns that contribute to category decoding without the need to formulate a priori hypothesis on which spectral components and at which times are worth investigating. Application of this method to an extensive dataset of human intracerebral recordings delineates the locations that are predictive of several perceptual categories from the locations that have narrow specialization, identifies spectral signatures characteristic of each of 8 perceptual categories and allows to observe global and category-specific patterns of neural activity pertinent to visual perception and cognition.

%
%
\section{Spectral and temporal signatures\\of human brain activity}
\label{sec:signatures-intro}
Our capacity to categorize sensory information allows us to quickly process and recognize complex elements in our environment. Early studies revealed strong relations between the brain activity within certain localized networks and the neural representations of certain stimulus categories, as for example faces, bodies, houses, cars, objects and words \citep{kanwisher1997fusiform, epstein1999parahippocampal, peelen2009neural, malach1995object, haxby2001distributed, ishai1999distributed, cohen2000visual}. These early assessments also revealed brain networks' capability to rapidly extract categorical information from short exposure to natural scenes \citep{potter1975time, thorpe1996speed, li2002rapid} based on models of parallel processing across neural networks \citep{rousselet2002parallel, peelen2009neural}. In both animal and human studies, visual cortices and particularly inferior temporal cortex (ITC) appear as a key region to integrate information at the object-level \citep{grill2014functional, tanaka1996inferotemporal, dicarlo2012does}. In humans, a great deal of observations of cortical response selectivity have been achieved using fMRI, but measuring direct neuronal activity \citep{quiroga2005invariant, kreiman2000category} also revealed similar patterns. To further understand how stimulus features and perceptual experience is processed in neural networks, brain activity, especially in sensory cortices, has been decoded using a variety of methods and signals \citep{haynes2006neuroimaging, kriegeskorte2006information, kamitani2006decoding}. This decoding often relies on machine learning to avoid a priori selection of partial aspects of the data by the human observer, and unless additional analysis is performed on the model itself it does not emphasize the mechanisms of neuronal communication within and between neural networks involved in this processing.

A pervasive feature of electrophysiological neural activity are its spectral fingerprints. Neural oscillations have been proposed to reflect functional communication processes between neural networks \citep{fries2009neuronal, buzsaki2006rhythms, siegel2012spectral, michalareas2016alpha}. Certain frequency bands are selectively associated with the operating of different cognitive processes in the human and animal brain \citep{vidal2006visual, wyart2008neural, jensen2010shaping, vanrullen2016perceptual, engel2010beta, dalal2011spanning}, and lately, direct recordings from the human cortex have revealed the remarkable representation selectivity of broadband high-gamma activity ($50-150$ Hz) \citep{lachaux2012high, parvizi2018promises, fox2018intracranial}. Human intracranial recordings have previously shown evidence of functional processing of neural networks related to perceptual category representation \citep{mccarthy1997face} and lately the prominence of broadband high-gamma activity in selective category responses in visual areas \citep{vidal2010category, davidesco2013spatial, hamame2014functional, privman2007enhanced, fisch2009neural}. Yet, very little is known about the specific relation between the different components of the full power-spectrum, including high-gamma activity, and their level of selectivity in processing perceptual categories. Previous works have shown where and when perceptual category information can be decoded from the human brain, the approach introduced in this work adds to that line of research by allowing to identify spectrotemporal patterns that contribute to category decoding without the need to formulate a priori hypothesis on which spectrotemporal regions of interest are worth investigating.
 
In this work we capitalize on an extensive dataset of deep intracranial electrical recordings on 100 human subjects to decode neural activity produced by 8 different stimulus categories. We analyzed the decoding models built by a random forest classifier to disentangle the most informative components of the time-frequency spectrum related to the simultaneous classification of 8 different perceptual categories. Via \emph{feature importance} analysis we quantified the contribution of each TF component into the decoding decision, which allowed us to identify the activity patterns that were either characteristic of the processing of a specific visual category or were shared by several categories. In addition to feature importance we analyzed the predictive power of each activity pattern and identified how informative was their spectral signature for the classification of visual categories. We tested the predictive power of broadband high-gamma activity in comparison to lower frequency activity as they reflect different communication mechanisms elicited by networks seemingly involved in distinct temporal windows of functional neuronal processing. Through the analysis of feature importance we show the specific neuronal spectral fingerprints from highly distributed human cortical networks that were elicited during automatic perceptual categorization. The uncovered spectral signatures provide insight into neural mechanisms of visual category perception in human brain.

%
%
\section{Large-scale intracortical recordings during\\visual object recognition task}
\label{sec:raw-intracranial-data}

One of the important steps leading up to being able to interpret the results and representations of a machine learning model is the correct choice of representation of the input data. In this section we explain the origin of our dataset and preprocessing choices that were made in order to present the data in a form that is both informative for the algorithm and directly interpretable by human domain expert.

\subsection{Patients and recordings}
100 patients of either gender with drug-resistant partial epilepsy and candidates for surgery were considered in this study and recruited from Neurological Hospitals in Grenoble and Lyon (France). All patients were stereotactically implanted with multi-lead EEG depth electrodes (DIXI Medical, Besan\c con, France). All participants provided written informed consent, and the experimental procedures were approved by local ethical committee of Grenoble hospital (CPP Sud-Est V 09-CHU-12). Recording sites were selected solely according to clinical indications, with no reference to the current experiment. All patients had normal or corrected to normal vision.

\subsubsection{Electrode implantation}
11 to 15 semi-rigid electrodes were implanted per patient. Each electrode had a diameter of 0.8 mm and was comprised of 10 or 15 contacts of 2 mm length, depending on the target region, 1.5 mm apart. The coordinates of each electrode contact with their stereotactic scheme were used to anatomically localize the contacts using the proportional atlas of Talairach and Tournoux \citep{talairach1993referentially}, after a linear scale adjustment to correct size differences between the patient's brain and the Talairach model. These locations were further confirmed by overlaying a post-implantation MRI scan (showing contact sites) with a pre-implantation structural MRI with VOXIM$^\text{\textregistered}$ (IVS Solutions, Chemnitz, Germany), allowing direct visualization of contact sites relative to brain anatomy.

All patients voluntarily participated in a series of short experiments to identify local functional responses at the recorded sites \citep{vidal2010category}. The results presented here were obtained from a test exploring visual recognition. All data were recorded using approximately $120$ implanted depth electrode contacts per patient using SD LTM Express, Micromed system for signal acquisition with a sampling rate of $512$ Hz, high-pass filter 0.15 Hz, low-pass filter 500 Hz. Data were obtained from a total of $11321$ recording sites.

\subsubsection{Stimuli and task}
\label{sec:stimuli-and-task}
The visual recognition task lasted for about $15$ minutes. Patients were instructed to press a button each time a picture of a fruit appeared on screen (visual oddball paradigm). Non-target stimuli consisted of pictures of objects of eight possible categories: houses, faces, animals, scenes, tools, pseudo words, consonant strings, and scrambled images. All the included stimuli had the same average luminance. All categories were presented within an oval aperture of $2\degree \times 3\degree$ in visual angle (illustrated on Figure~\ref{fig:methods-pipeline}a) at a distance of $70-90$ cm using NeuroBehavioral Systems (NBS) Presentation$^\text{\textregistered}$ software. Stimuli were presented for a duration of $200$ ms every $1000-1200$ ms in series of 5 pictures interleaved by 3 second pause periods during which patients could freely blink. Patients reported the detection of a target through a right-hand button press and were given feedback of their performance after each report. A 2 second delay was placed after each button press before presenting the follow-up stimulus in order to avoid mixing signals related to motor action with signals from stimulus presentation. Altogether, responses to 400 unique natural images were measured per subject, 50 from each category.

\subsection{Processing of neural data}
\label{sec:signatures-ns-processing}

The analyzed dataset consisted of $4528400$ local field potential (LFP) recordings -- responses from $11321$ recording sites to $400$ stimuli. To remove the artifacts the signals were linearly detrended and the recordings that contained values $\ge10\sigma_{images}$, where $\sigma_{images}$ is the standard deviation of voltage values (in the time window from $-500$ ms to $1000$ ms) of that particular probe over all stimuli, were excluded from data. All electrodes were re-referenced to a bipolar reference and the reference electrodes were excluded from the analysis. The signal was segmented in the range from $-500$ ms to $1000$ ms, where $0$ marks the moment when the stimulus was shown. The $-500$ to $-100$ ms time window served as a baseline.

To quantify the power modulation of the neural signals across time and frequency we used standard time-frequency (TF) wavelet decomposition \citep{daubechies1990wavelet}. The signal $s(t)$ was convoluted with a complex Morlet wavelet $w(t, f_0)$, which has Gaussian shape in time $(\sigma_t)$ and frequency $(\sigma_f)$ around a central frequency $f_0$ and defined by $\sigma_f = 1/2 \pi \sigma_t$ and a normalization factor. To achieve good time and frequency resolution over all frequencies we slowly increased the number of wavelet cycles with frequency, $\frac{f_0}{\sigma_f}$ was set to: 6 for high ($61 - 150$ Hz) and low ($31 - 60$  Hz) gamma, 5 for beta ($15 - 30$ Hz), 4 for alpha ($9 - 14$ Hz) and 3 for theta ($4 - 8$ Hz) frequency ranges. This method allowed to obtain better frequency resolution than applying a constant cycle length \citep{delorme2004eeglab}. The square norm of the convolution results in a time-varying representation of spectral power, given by: $P(t, f_0) = |w(t, f_0) \cdot s(t)|^2$. Baseline normalization was performed by dividing the average power after stimulus onset ($0$ to $1000$ ms) in each frequency by the average power of that frequency in the baseline window ($-500$ to $-100$ ms). Each LFP recording was transformed from 768 data points (1.5 seconds of voltage readings at 512 Hz sampling rate) into a matrix of size $146 \times 48$ where each row represents a $1$ Hz frequency band from $4$ Hz to $150$ Hz and columns represent $31.25$ ms time bins. Value in each cell of that matrix is the power of that specific frequency averaged over $16$ time points.

Further analysis was done only on the electrodes that were responsive to the visual task. In each frequency band we compared each electrode's average post-stimulus band power to the average baseline power with a Wilcoxon signed-rank test for matched-pairs. Only the probes that showed a post-stimulus response that is statistically significantly (p-value $\leq 0.005$, corrected for multiple comparisons with the false discovery rate (FDR) procedure \citep{genovese2002thresholding}) different from the baseline response in at least two frequencies were preserved for future analysis. Please note that eliciting a significant response in at least 2 out of 146 frequencies is a relaxed requirement. The use of such a relaxed criterion allowed us to include into analysis not only the areas that had a strong response in the visual areas, but also the responses from other brain areas that might reflect downstream processes related to automatic perceptual categorization. This was possible due to the fact that the proposed method, given sufficiently large dataset, will not be hindered by the additional volume of irrelevant data and is able to detect narrow phenomena even in the large corpus of data.

To anatomically localize the source of each signal in subject's brain each electrode's MNI coordinates were mapped to a corresponding Brodmann brain area~\citep{brodmann1909vergleichende} using Brodmann area atlas from MRICron~\citep{rorden2007mricron} software. 

To confirm that probe's predictiveness of a certain category implies that the probe belongs to the network selective of that category we ran a set of experiments on three well-known functional areas: Fusiform Face Area (FFA) \citep{kanwisher1997fusiform}, Visual Word Form Area (VWFA) \citep{cohen2000visual} and Parahippocampal Place Area (PPA). Following Montreal Neurological Institute (MNI) coordinates of FFA reported in \citep{harris2012morphing} and \citep{axelrod2015successful} we defined FFA bounding box as $x \in [-44,-38]$, $y \in [-61, -50]$, $z \in [-24, -15]$ in the left hemisphere and $x \in [36, 43]$, $y \in [-55, -49]$, $z \in [-25, -13]$ in the right hemisphere. Based on the Table 1 from \citep{price2003myth} we defined VWFA area as MNI bounding box $x \in [-50,-38]$, $y \in [-61, -50]$, $z \in [-30, -16]$ in the left hemisphere. From MNI coordinates reported in \citep{bastin2013temporal} and \citep{park2009different, hamame2013dejerine} we defined PPA bounding box to be $x \in [-31,-22]$, $y \in [-55, -49]$, $z \in [-12, -6]$ in the left hemisphere and $x \in [24, 32]$, $y \in [-54, -45]$, $z \in [-12, -6]$ in the right hemisphere.

%
%
\section{Feature importances of a decoder are indicative of task-relevant brain activity}

In the taxonomy of knowledge representation (Section \ref{sec:representation-taxonomy}), decision trees were brought forward as the representation best suited for feature importance analysis (see Section \ref{sec:interpretability-tools}). In this work we use spectral power readings in time-frequency domain as our input and look to identify which time-frequency features contribute most to the task at hand. Since feature importance is our desired interpretation we have chosen Random Forest learning algorithm to build the decoding model. In this section we explain the inner workings of this algorithm and present our feature analysis approach in details.

\subsection{Random Forest as a decoding model}
\label{sec:signatures-rf-decoder}

A Random Forest~\citep{breiman2001random} is a collection of decision trees, where each tree gets to operate on a subset of features. Each tree is assigned a random set of features and it has to find the decision boundaries on those features that lead to best classification performance. At each branching point the algorithm must decide which feature will be most efficient in terms of reducing the entropy of class assignations to the data points in the current branch of the decision tree. To achieve that, the feature that is most useful is selected first and will be responsible for largest information gain. For example, if the activity of a probe at 52 Hz at 340 ms is high when a subject is presented with a face and low for all other categories, decision tree will use that fact and rely on the ``$52$ Hz at $340$ ms'' feature, thus assigning it some importance. How high the importance of a feature is depends on how well does this feature distinguish faces from all other categories. As Random Forest is a collection of trees and the same feature will end up being included into several different trees, being important in many trees contributes to the overall importance of a feature (for the exact computation see the section on feature importance below).

\begin{figure}
    \centering
    \includegraphics[width=1.0\linewidth]{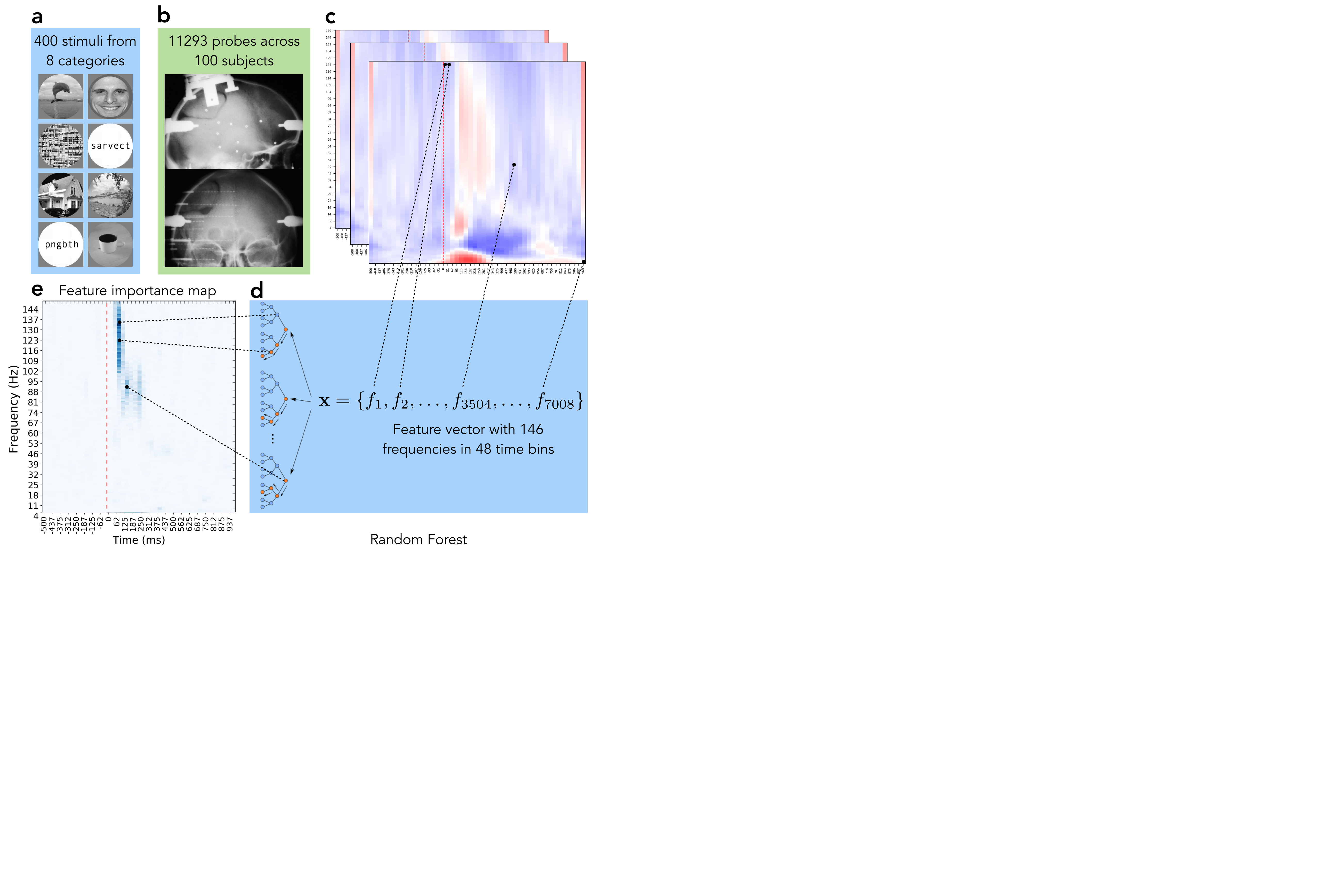}
    \caption{Major steps of the data processing pipeline. \textbf{a}: Image stimuli from 8 categories were presented to test subjects. \textbf{b}: Human brain responses to images were recorded with deep intracranial electrodes. \textbf{c}: LFP signals were preprocessed and transformed into time-frequency domain. \textbf{d}: Random Forest models were trained to decode image category from each electrode's activity. \textbf{e}: Feature importances of each model were calculated to identify the region on each electrode's activity map that was relevant to visual object recognition. Notice how the final results on panel \textbf{e} tell us that high gamma activity in $90-120$ ms window and the subsequent activity in the low gamma range in $120-250$ ms window are the only bands and time windows in that particular electrode's activity that are relevant for the classification task, while the spectrogram on panel \textbf{c} also shows that there was activity in early theta, beta and low gamma bands. Our analysis revealed that not all activity was relevant (or useful) for the classification of an object and showed which parts of the activity are actually playing the role in the process.}
    \label{fig:methods-pipeline}
\end{figure}

We treated each electrode's responses as a separate dataset consisting of 400 data points (one per stimulus image), and 7008 features -- time-frequency transformation of LFP response into 146 frequencies and 48 time bins. For each electrode we trained a Random Forest with 3000 trees and used 5-fold cross-validation to measure the predictive power of the neural activity recorded by each of the electrodes. Per-class F\textsubscript{1} score, a harmonic mean of precision and recall of a statistical model, provides us with a metric of success of the classification. The parameters were selected by performing informal parameter search. Random Forest was the algorithm of choice for our analysis due to interpretability of the resulting models, that allowed us to track the process that led each particular model to a decoding decision and due to its previous application to spectrotemporal features \citep{westner2018across}. We used \texttt{scikit-learn}~\citep{scikit-learn} implementation of the above-mentioned methods with default parameters unless indicated otherwise. 

As the first step of the decoding analysis we estimated which of $11321$ electrodes have predictive power. For that we split each electrode's 400-sample dataset into 320 samples for training and 80 for prediction estimation. Repeating this procedure 5 times provided us with 400 predictions that we could compare to the true categories. By running a permutation test 100000 times on electrodes with randomly permuted class labels we estimated that 99.999th percentile (equivalent to significance threshold of $p \leq 0.00001$) of \emph{F\textsubscript{1} score} is $0.390278$. F\textsubscript{1} score is an aggregated metric of the performance of a classifier that combines both the \emph{precision} (the ratio of the data samples that truly belong to a category among the ones that were assigned to that category by the model) and \emph{recall} (the ratio of data samples that were correctly identified to belong to a category to the total number of samples of that category in the dataset) into one number: $\text{F}_1 = 2\cdot\frac{\text{precision}\cdot\text{recall}}{\text{precision}+\text{recall}}$.  In total $787$ electrodes had a predictive power of F\textsubscript{1} $> 0.390278$ in at least one of the categories. For each of those electrodes a Random Forest model was retrained once more on whole data (400 samples instead of 320) and that model was used for calculating feature importances and, ultimately, for understanding which parts of the recorded activity were relevant for visual object recognition in human brain.

%
%
\subsection{Feature importance for the analysis of task-relevant neural activity}
During the process of constructing the decision trees, Random Forest relies on some features more than on the others. We chose \emph{Gini impurity} \citep{breiman2017classification} as a measure of which features should be used to make the branching decisions in the nodes of a tree. This score, along with the number of times each particular feature was used across trees, informed us on the relative importance of each particular feature with respect to other features. Gini impurity $G$ is calculated as
\begin{equation}
    G = \displaystyle\sum^{i=n_c}_{i=1}p_i(1 - p_i),
\end{equation}
where $n_c$ is the number of categories and $p_i$ is the proportion of class $i$ in a node. To pick a feature for a parent node, Gini impurity of both child nodes of that parent are calculated and used to estimate the \emph{reduction in impurity} that would be achieved by picking that particular feature as the branching factor for the node. The feature that decreases impurity the most is selected to be the branching factor of that parent node. The reduction in impurity is calculated as
\begin{equation}
    I = G_{\text{parent}} - G_{\text{left child}} - G_{\text{right child}}
\end{equation}
and is called \emph{node importance}. \emph{Feature importance} of a feature $f$ is estimated by calculating the sum of Gini impurity reductions over all samples in the dataset that were achieved with the use of a particular feature $f$ and normalizing it by the total number of samples. Figure~\ref{fig:methods-pipeline}e is a visual representation of relative feature importance, color intensity shows the importance of each of 7008 (146 frequencies $\times 48$ time bins) spectrotemporal features from one probe. In total our analysis has produced $787 \times 8$ such images -- one for each probe-class pair.

\begin{figure}
    \centering
    \includegraphics[width=1.0\linewidth]{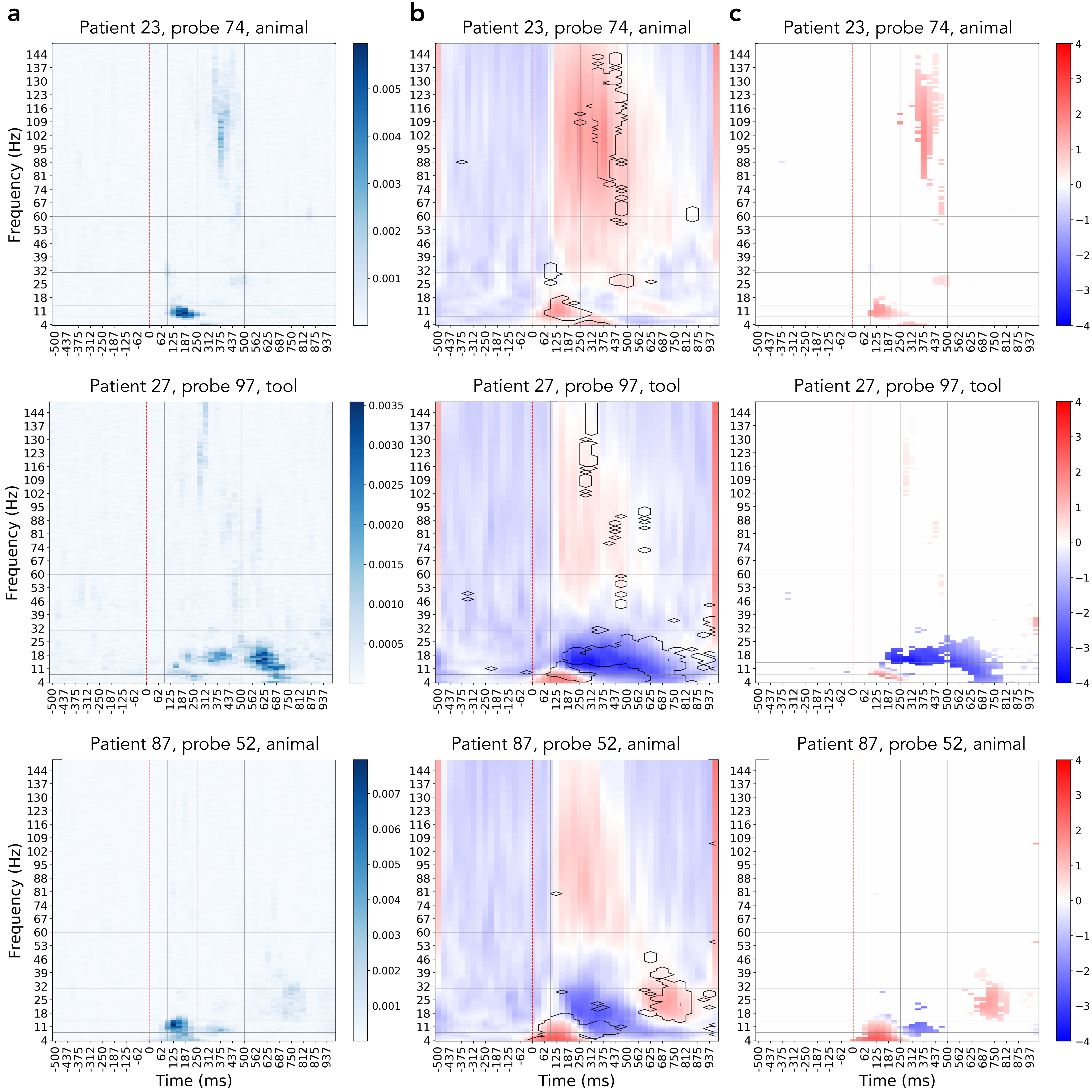}
    \caption{Using the importance map to filter out irrelevant activity. The three rows show three different examples of how filtering of the activity by importance is beneficial: in patient 23, probe 74 we see that only later portion of the broadband gamma activity increase was useful for identifying this activity as a response to the \texttt{animal} stimulus; patient 27, probe 97 shows that although there is an increase in broadband activity, the actually useful information contained in decrease in the lower frequency bands; patient 87, probe 52 demonstrates that for decoding this particular probe's activity one must focus on the activity in lower frequencies at specific time and, despite prominent presence, ignore the increase in broadband gamma. \textbf{a.} Probe's importance map, color codes the relative importance of each spectrotemporal feature within the map. \textbf{b.} Full spectrotemporal activity of the probe, features with importances one standard deviation higher than the average (in contour) mark the regions of activity that were useful for the decoding model. \textbf{c.} Activity of the probes filtered by the importance mask, only the relevant activity is preserved.}
    \label{fig:filter-by-importance}
\end{figure}

The importance map computed as depicted on Figure~\ref{fig:methods-pipeline} is an example of a global map for all 8 categories. The regions that are highlighted on the map are important for distinguishing between all 8 categories. There is, however, a way to look at category-specific importances as well. The final set of nodes of a decision tree, called \emph{leaves}, are the end-points of the classification process and each leaf is associated with a certain category. If we take one TF activity map (TF components are the features) and start traversing a decision tree following the rules set by the nodes of the tree, we will end up in a certain leaf. That leaf will be associated with a certain category, for example, with faces. The fact that we followed the rules and ended up in that leaf indicates that the TF map we used as the input to the tree probably comes from a trial where a \texttt{face} stimulus was shown to the subject. In order to get category-specific feature importance map we took all the leaves associated with a category, traversed the tree backwards and tracked all the features that were used on the path from the leaf to the root of the tree. This way we got a list of features (TF components) that were used to identify a neural response as belonging to a certain category. Random Forest feature importance allowed us to identify which sub-regions of neural activity (TF maps) are relevant for decoding. It showed that only a small portion of activity is actually crucial for identifying the categories (see Figure \ref{fig:filter-by-importance}).

To compare importance maps between each other we fit a normal distribution on the difference between two maps and considered statistically significant the differences that are bigger than $\mu + 4\sigma$. One spectrotemporal importance map consists of 7008 values. To filter out false positives we stipulated that only 1 false positive out of 7008 pixels can be tolerated and tuned the threshold accordingly. That requirement resulted in the p-value of $0.0001427$ and confidence level of $99.99\%$, corresponding to $3.89\sigma$, which we rounded up to $\sigma=4.0$.

%
%
\subsection{Hierarchical clustering to reveal types of activity patterns}
\label{sec:signatures-clustering}
To further analyze the spectrotemporal signatures elicited by different visual categories in different parts of human brain we clustered filtered activity patterns and identified the most prominent groups. The result of this analysis is shown in the second column of Figure~\ref{fig:importances-clusters-mnis}. For each category, the four most populated (in terms of the number of probes) clusters of activity patterns elicited by this category are shown.

To do the clustering we first took each probe's category-specific activity separately by averaging probe's responses to 50 images of each particular category in time-frequency domain. We then masked the activity with the category importance map (as shown on Figure~\ref{fig:filter-by-importance}), leaving only those features out of $146 \times 48$ that have importance score larger that $\mu + \sigma$, where $\mu$ is the average importance score for that category and $\sigma$ is one standard deviation of the score distribution.

Masked activity patterns were hierarchically clustered using Eq~\ref{eq:clustering-distance} to calculate the distance between a pair of clusters $U$ and $V$ as the maximal cosine distance between all of the clusters' member observations (complete linkage clustering):
\begin{equation}
    \label{eq:clustering-distance}
    d(U, V) = max\Big(\displaystyle\frac{\mathbf{u} \cdot \mathbf{v}}{\|\mathbf{u}\| \|\mathbf{v}\|}\Big)\ \forall \mathbf{u} \in U,\ \forall \mathbf{v} \in V
\end{equation}

\texttt{SciPy}~\citep{scipy} implementation of the hierarchical clustering methods was used in this work. Resulting clustering assignments were visually inspected and corrected.

%
%
\section{The role and diversity of time-frequency patterns of individual locations and area networks in perceptual categorization}

By choosing the machine learning algorithms with subsequent need for interpretability in mind (random forest, hierarchical clustering) and application of interpretability techniques (feature importance analysis, visualization) we were able to extract the knowledge that the model had obtained, present it in a way that is understandable to a neuroscientist and articulate the neurological insights that the model has found. The three main observations we made as a result of this analysis were: (a) the difference between responsiveness of a neural location and its ability to predict an experimental condition (visual category), (b) existence of monopredictive and polypredictive neural locations, where the former are specialized and only are relevant for processing specific visual categories, while the latter carry information that is relevant to decoding of several categories, and (c) to extensively map and describe time-frequency patterns that are characteristic of cognitive processing of each particular visual category. In this section we present these findings in full detail.

\subsection{Feature importance allows to separate out the neural signals that are predictive of perceptual categorization from the mixture of stimulus-induced responses}

To identify spectrotemporal features that are characteristic of automatic perceptual categorization of a particular category we relied on time-frequency (TF) maps of the neural responses of intracranially implanted electrodes. Out of the total set of 11321 probes 11094 (98\%) were responsive (see the section \ref{sec:signatures-ns-processing} on processing of neural data for details) to the stimuli from at least one of the categories. On one hand this provides us with abundance of data, on the other raises the question whether all of that activity was relevant to the processes that encode and process visual input.

Training a decoding model (see the section \ref{sec:signatures-rf-decoder} on Random Forest as decoding model) for each of the probes allowed us to dissociate the \emph{predictive probes} that exhibited activity that was useful for decoding from the rest of the \emph{responsive probes} that did not carry such activity.

\begin{figure}[h!]
    \centering
    \includegraphics[width=1.0\linewidth]{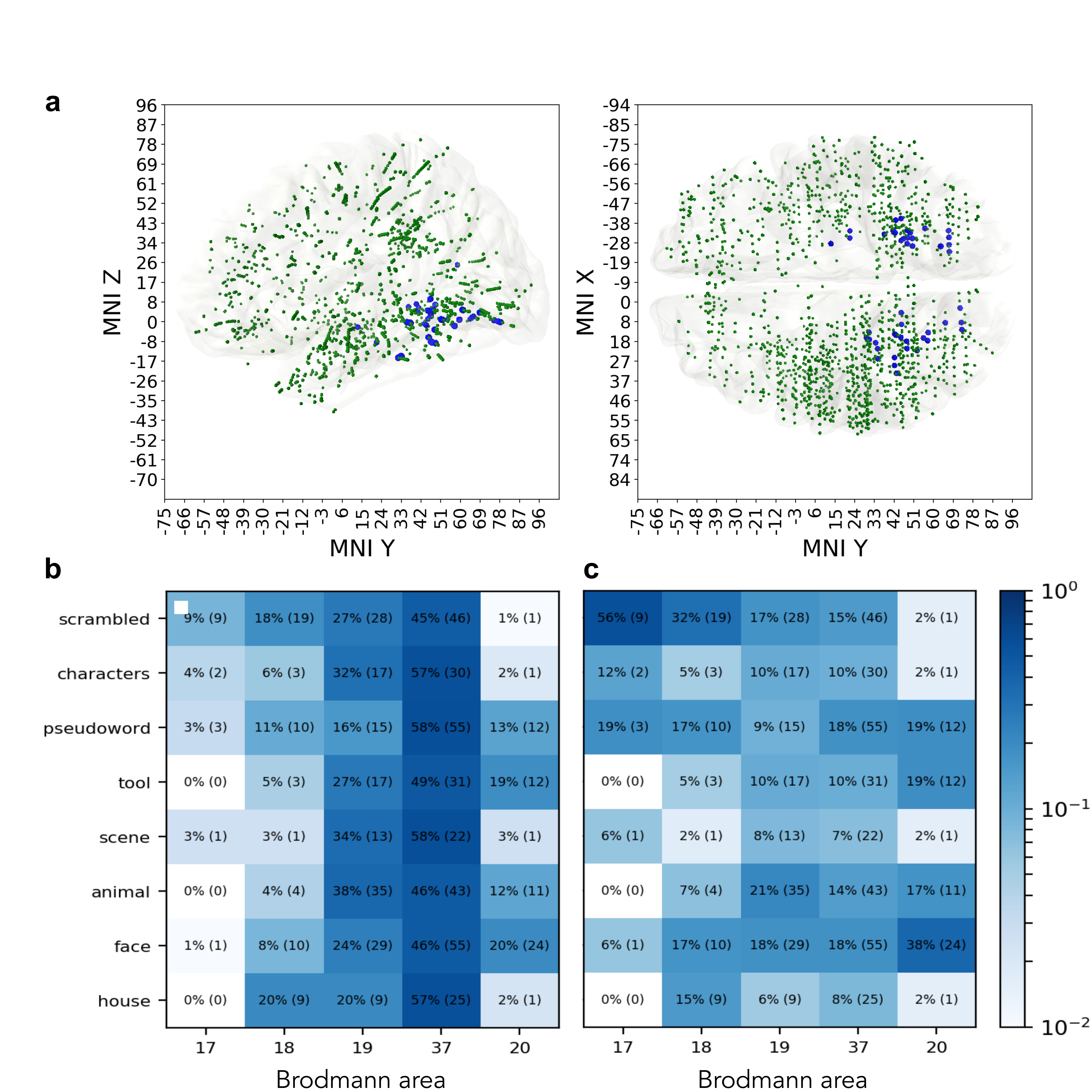}
    \caption{Distribution of predictive probes. \textbf{a}. Green markers indicate all of the probes that were responsive to stimuli from the \texttt{house} category. Blue markers indicate only the predictive probes that carry information that is relevant to decoding the neural response as reaction to \texttt{house} stimulus. \textbf{b}. Distribution of predictive probes over areas within each category (each row sums up to 100\%). Color shows the percentage across categories. \textbf{c}. Distribution of predictive probes over a category within each area (each column sums up to 100\%). Color shows the percentage across areas.}
    \label{fig:responsive-vs-predictive}
\end{figure}

Green markers on Figure~\ref{fig:responsive-vs-predictive}a show the set of probes that are responsive to the \texttt{house} category, while the blue markers are the probes that are predictive of that category (4.8\%, 535 probes). Decoding models built on the neural responses of the predictive probes were successful at classifying at least one perceptual category ($\text{F}_1 > 0.39$ for one or more classes), focusing on them in our further analysis allowed to work only with the locations that carry information relevant to the task of perceptual categorization.

Predictive probes had a heterogeneous distribution in the brain, yet remained mostly concentrated in visual cortices and inferior temporal regions (76\%), from BA17 to BA20, including early visual areas (BA 18, 19), fusiform gyrus (BA 37) and inferior temporal cortex (BA 20). A majority of the predictive probes were in fusiform cortex (average of 52\% over all categories, Figure~\ref{fig:responsive-vs-predictive}b), followed by BA 19 (27\%), across all category networks. 

Within the primary visual cortex, BA 17 and 18, the \texttt{scrambled} was the stimulus that elicited most predictive probes (28) amongst all stimulus categories (Figure~\ref{fig:responsive-vs-predictive}c), followed by \texttt{pseudowords} (13). Probes predictive of \texttt{faces} were mostly concentrated in BA19, BA37 and BA20 (72\%, 108 out of 150). The low number of predictive probes in area 17 is explained by the fact that less than 1\% of the implantation sites in the original dataset were located in primary visual cortex. 

Previous studies have shown that perceptual category-selective networks are located in occipito-temporal cortex \citep{grill2014functional, ishai1999distributed, malach1995object}. To test whether predictive power of the Random Forest model trained to decode activity of probes is coherent with known functional processing by cortical networks we evaluated the selectivity of the predictive power in three known functional networks: Fusiform Face Area (FFA) \citep{kanwisher1997fusiform}, Visual Word Form Area (VWFA) \citep{cohen2000visual} and Parahippocampal Place Area (PPA) \citep{epstein1998cortical}. We checked whether the probes located in each of these areas and the Random Forest model trained on these probe's activity to discriminate between 8 categories produces the highest predictive power for the category for which this area is known to be selective. Probes in FFA are associated with facial recognition and encoding facial information \citep{parvizi2012electrical, ghuman2014dynamic, kadipasaoglu2016category, jonas2016face, jonas2015beyond} and thus we expect their activity to be predictive of the \texttt{face} category, probes in VWFA should be predictive of \texttt{characters} and \texttt{pseudowords} categories \citep{kadipasaoglu2016category, lochy2016left, hirshorn2016decoding} and probes in PPA should be responsive to \texttt{scenes} and \texttt{houses} \citep{aguirre1998area, megevand2014seeing, epstein1998cortical, bastin2013timing}.

There were 12 probes in the FFA that were significantly (permutation test $p < 1e-4$) predictive (classification score $\text{F}_1 > 0.39$) of a category: 5 were predictive of \texttt{faces}, 4 of \texttt{animals} (which mostly have faces on the image), 2 of \texttt{pseudowords} and 1 of \texttt{scrambled} images. Most probes that were in FFA and were predictive, carried information of the categories containing facial features.

There were 8 probes in the VWFA that were predictive of a category: 5 were predictive of \texttt{pseudowords}, 2 of \texttt{characters} and 1 of \texttt{faces}. This points to the fact that the predictive probes in VWFA are predictive of the stimuli with written characters on them. These results confirm that predictive power of a Random Forest model trained on probes activity in VWFA reflects the functional role known to be carried by this area.

For probes in the PPA results were less selective. There were 23 probes inside that area that were predictive of a category: 5 were predictive of \texttt{houses}, 4 of \texttt{scenes}, 5 of \texttt{characters}, 5 of \texttt{scrambled} images, 2 of \texttt{tools} and 2 of \texttt{pseudowords}. The probes from PPA predicted not only \texttt{houses} and \texttt{scenes}, but also other categories. However, \texttt{houses} and \texttt{scenes} were among the categories that the probes from PPA were able to identify successfully in highest proportion as compared to the other categories.

These confirmatory findings give credibility to the methodology by which the probes that are identified as predictive of a certain category are involved in the processing of the stimuli that belong to that category.

Training per-probe decoding models not only allowed us to identify the predictive locations, but also to apply feature importance analysis to decoding models trained on local activity. Computing the feature importance across the time-frequency map ($4 - 150$ Hz and $-500$ to $1000$ ms) allowed us to see which parts of neural activity are crucial for the decoding. Overlaying the importance over time-frequency map showed at which frequencies and at what times the activity that was important for the algorithm has occurred. This can be applied both on aggregated level, where the importance map is averaged over probes, and on individual probe level. Figure~\ref{fig:filter-by-importance} illustrates the application of probe importance map to filter irrelevant activity and obtain spectrotemporal signature of a particular category on a particular probe. Now we can use the feature importance map as a mask and perform the analysis of the activity itself, focusing only on the relevant parts of it. When applicable, this methodology helps to filter out irrelevant activity and allows to focus on the activity that is important to the scientific question under investigation. 

\begin{figure}[htb]
    \centering
    \includegraphics[width=1.0\linewidth]{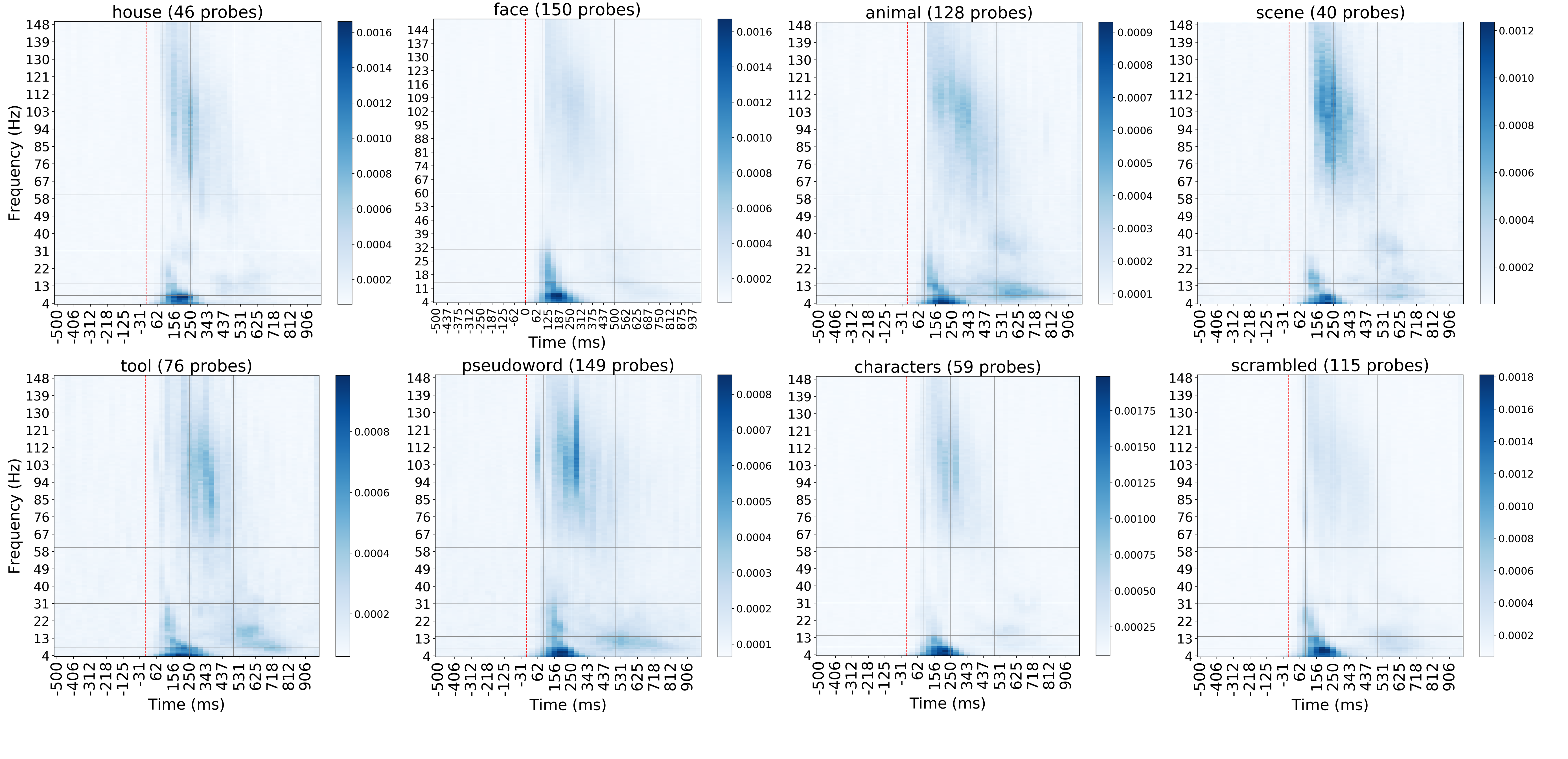}
    \caption{Average importance map of each of eight categories over probes predictive of that category. The color shows the relative importance of each spectrotemporal feature, indicating how informative that particular feature was for the task of decoding.}
    \label{fig:eight-importance-maps}
\end{figure}

We took an average over importance maps of all individual probes within each category to obtain the global picture of where the category-specific activity lies in time and frequency space. Figure \ref{fig:eight-importance-maps} summarizes such analysis and singles out the spectrotemporal signatures that are unique to specific categories and those that are less selective. From these importance maps we notice that certain TF components are distinctly present per category, as for example high (significantly higher than 81 out of 112 regions of interest, Mann-Whitney U $p < 8.9\mathrm{e}{-7}$, corrected) importance of the transient theta activity in all categories, or the almost absence of importance of broadband gamma (significantly lower than 10 out of 12 other regions of interest, Mann-Whitney U $p < 8.3\mathrm{e}{-5}$, corrected) in the control scrambled condition.

In the following sections we expand our analysis to the comparison of the feature maps and analyzing the activity in the regions that we have identified as important.

%
%
\subsection{Polypredictive and monopredictive probes}

\begin{figure}[htb]
    \centering
    \includegraphics[width=1.0\linewidth]{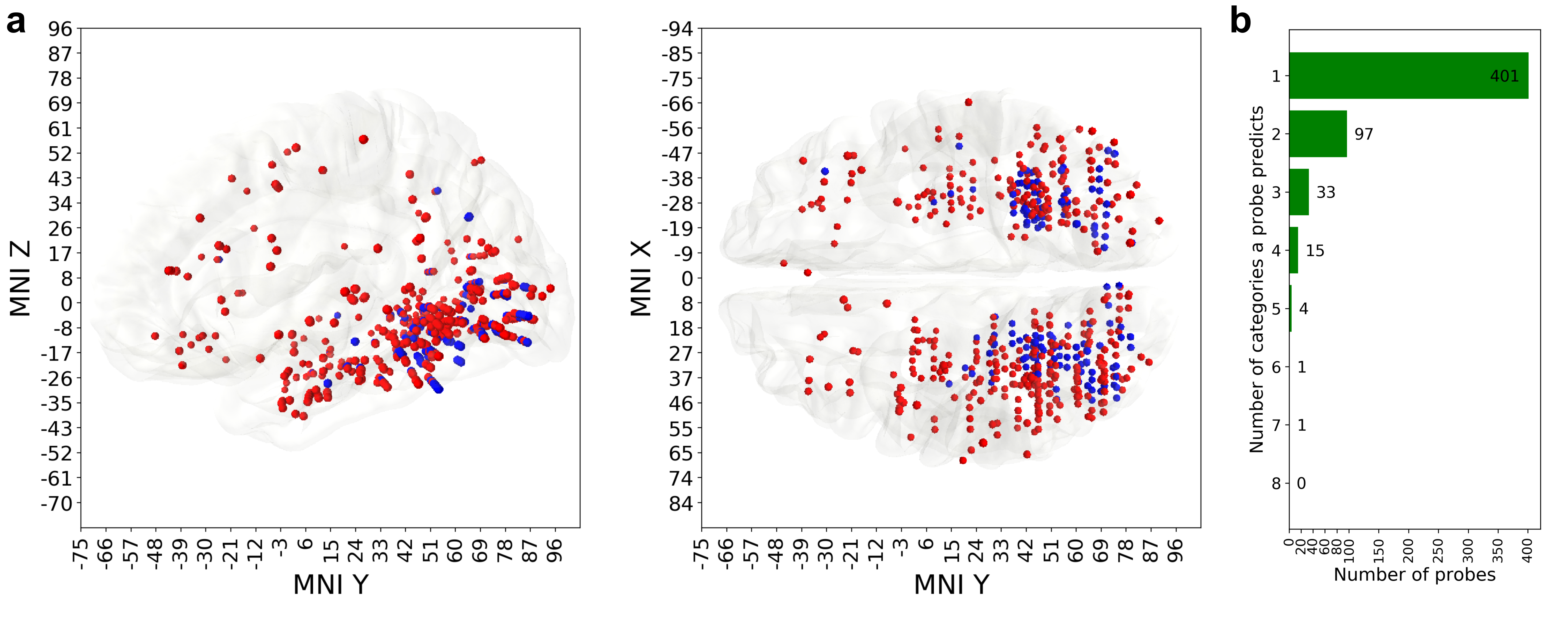}
    \caption{Anatomical distribution of mono- and polypredictive locations. \textbf{a:} Red markers are the locations of monopredictive probes, blue markers are the locations of polypredictive ones. Polypredictive probes (145 unique locations) are mostly confined to visual areas and temporal lobe (both parts of the ventral stream), while monopredictive (specialized, 401 unique locations) probes are, in addition to visual areas, also found in frontal and parietal cortical structures. \textbf{b:} The histogram shows how many categories are predictable by how many probes.}
    \label{fig:poly-mono-location}
\end{figure}

The analysis revealed two types of neural locations: \emph{polypredictive} probes are predictive of multiple visual categories, while \emph{monopredictive} are useful for decoding only one out of 8 different types of stimuli revealing a high degree of specialization (Figure~\ref{fig:poly-mono-location}b). We considered a probe to be predictive of a category if cross-validation F\textsubscript{1} score for that category was higher than $0.39$ (see the section \ref{sec:signatures-rf-decoder} for the details on the threshold selection), which is a stricter condition than above-chance criterion ($\text{F}_1 > 0.125$). Figure~\ref{fig:poly-mono-location}a shows that polypredictive probes reside mainly (94\%, 136 out of 145) in posterior occipital and posterior temporal, while the monopredictive probes extend, in addition to occupying similar posterior occipital and temporal locations, to frontal cortex (92\%, 45 out of 49 probes in this area are monopredictive) and anterior temporal cortex (88\%, 51 out of 58 probes). Both mono- and polypredictive probes are also observed in parietal cortex. Monopredictive probes that extend beyond ventral stream and temporal cortex pertain to the following perceptual categories: \texttt{faces} (orbitofrontal cortex), \texttt{animals} and \texttt{pseudowords} (dorsofrontal cortex, inferior frontolateral cortex, premotor cortex), and, to a smaller extent, \texttt{scrambled} images (prefrontal cortex).  

\begin{figure}[h!]
    \centering
    \includegraphics[width=1.0\linewidth]{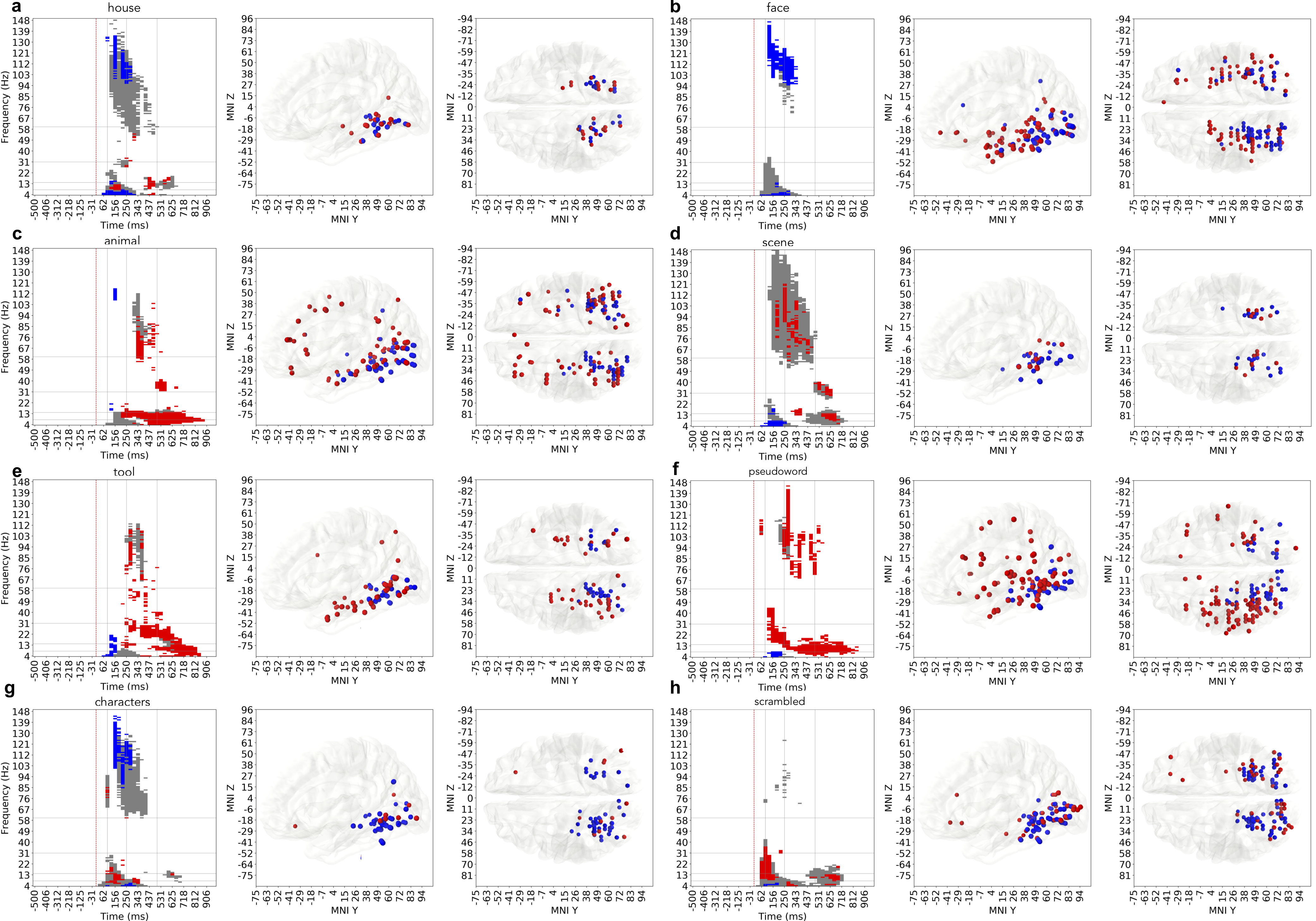}
    \caption{Statistically significant differences between the importance of monopredictive and polypredictive probes' activity. Gray regions indicate the areas of TF spectrum where both monopredictive and polypredictive probes exhibit important ($2\sigma$ from the mean) activity. On top of it, if one of the groups is statistically significantly ($4\sigma$ difference) more important than another, the region is colored with blue (polypredictive) or red (monopredictive) to show which of two kinds of neural specialization dominates this TF region in terms of importance. For example decoding of \texttt{scenes} (d) involves early theta activity of polypredictive (blue) probes, followed broadband gamma activity that is significantly important (gray), and is slightly dominated by monopredictive (red) probes, then followed by late alpha activity produced predominantly by monopredictive neural locations. \textbf{a:} \texttt{house}, \textbf{b:} \texttt{face}, \textbf{c:} \texttt{animal}, \textbf{d:} \texttt{scene}, \textbf{e:} \texttt{tool}, \textbf{f:} \texttt{pseudoword}, \textbf{g:} \texttt{characters}, \textbf{f:} \texttt{scrambled}. }
    \label{fig:poly-mono-fitfs}
\end{figure}

The unique association of specific TF feature importance components with either polypredictive and monopredictive probes was category specific, as shown in figures \ref{fig:poly-mono-fitfs}a to \ref{fig:poly-mono-fitfs}h.
While all of the data presented on these figures shows statistically significant differences between monopredictive and polypredictive neural locations, we will focus only on a few that were supported by the strongest signal in the data. For \texttt{face} stimuli, most of the feature importance in the early broadband gamma response was significantly ($4\sigma$) higher in polypredictive probes as compared to monopredictive probes, indicating that the most useful information for distinguishing \texttt{faces} from other visual categories is coded in that region of time-frequency space and is carried by polypredictive probes (Figure~\ref{fig:poly-mono-fitfs}b). Decoding of \texttt{animals} and \texttt{tools} relied on the activity patterns produced by monopredictive neural locations in late broadband gamma range ($> 300$ ms) and in even later ($350 - 600$ ms) alpha/beta range, with very little involvement of the activity of polypredictive probes. \texttt{Scenes} and \texttt{houses} also show strong feature importance in late alpha and beta band responses of monopredictive probes ($4\sigma$ higher). Interestingly, for \texttt{characters} (Figure~\ref{fig:poly-mono-fitfs}g), feature importance in the early broadband gamma range was dominant for polypredictive probes ($4\sigma$ higher than monopredictive), while the opposite was true for the \texttt{pseudowords} (Figure~\ref{fig:poly-mono-fitfs}f) -- the late broadband gamma revealed to be dominant for monopredictive probes, also note the difference in the anatomical locations that were the most useful for the decoding of the \texttt{pseudowords} compared to the locations that were useful for decoding \texttt{characters}. \texttt{Pseudowords} also elicited a significantly stronger TF feature importance in monopredictive probes in late ($350 - 750$ ms) low-frequency ($4 - 12$ Hz) range, similar to \texttt{animal} and \texttt{tool} stimulus categories. Finally, an interesting observation was that \texttt{animals} and \texttt{faces} share most of their polypredictive probes (51\%) indicating a large overlap of categorization networks of these two categories.

%
%
\subsection{Further decomposition of important activity reveals clusters of distinct time-frequency patterns}

\begin{figure}[h!]
    \centering
    \includegraphics[width=1.0\linewidth]{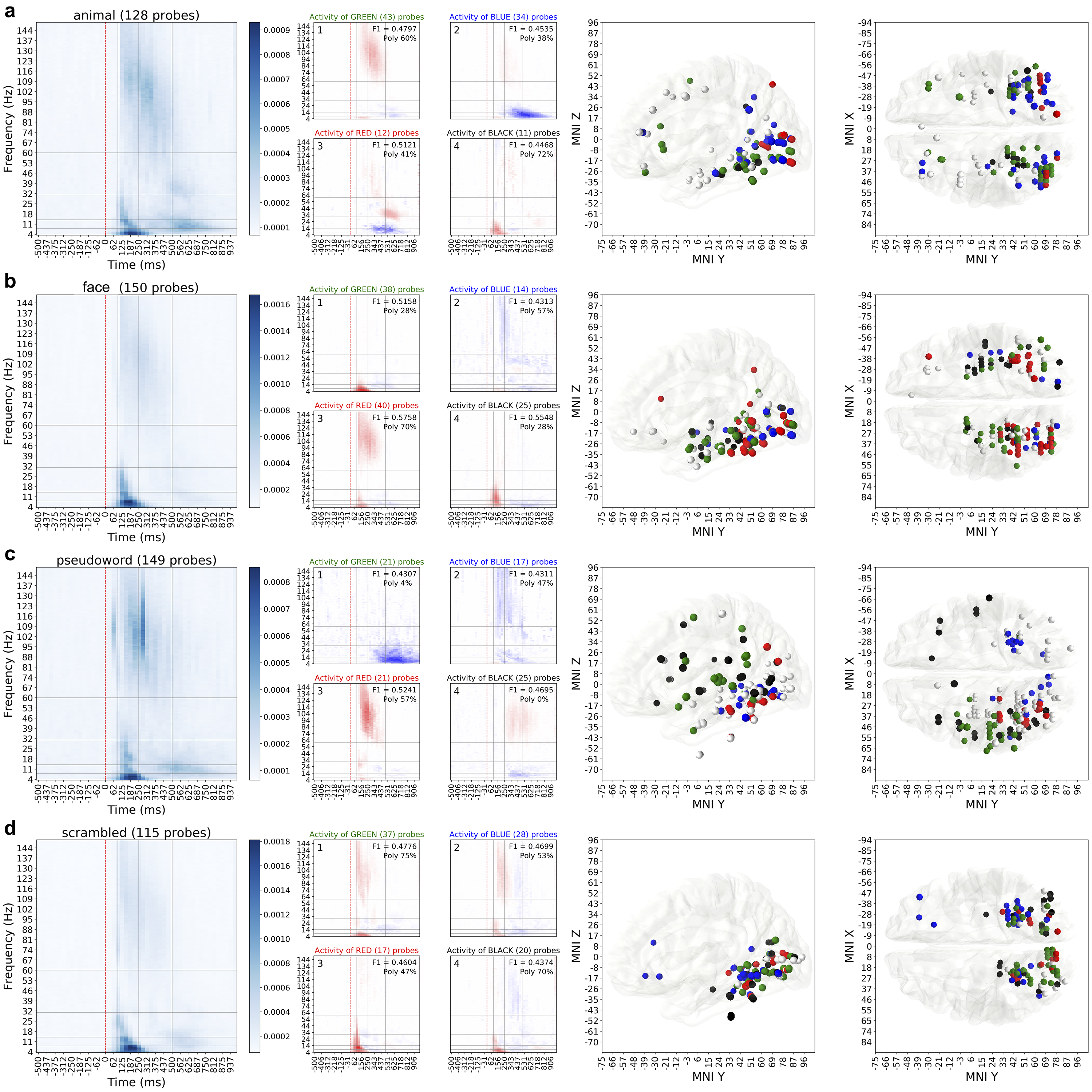}
    \caption{Detailed analysis of spectral activity of \textbf{(a)} \texttt{animals}, \textbf{(b)} \texttt{faces}, \textbf{(c)} \texttt{pseudowords} and \textbf{(d)} \texttt{scrambled} images. Leftmost column contains the importance maps extracted from Random Forest models and shows where in time and frequency the important activity is. Second column visualizes the four largest (by the number of recording sites) clusters of activity patterns inside those spectrotemporal regions that are deemed important. The numbers in the top right corner of each cluster's activity pattern show the average predictive power (F\textsubscript{1} score) of the probes in that cluster and proportion of polypredictive locations that exhibited this particular pattern of activity. Note how every cluster has a designated color: green, blue, red or black. This color of the cluster matches the color of MNI location markers in the last two columns, that show sagittal and dorsal views of the brain. White markers show the probes that have predictive power, but their activity pattern does not belong to any of the four major clusters.}
    \label{fig:importances-clusters-mnis}
\end{figure}

We ran clustering analysis of the probes predictive of a category based on their activity to see which probes in the category-network behave in a similar way. Left column of Figure~\ref{fig:importances-clusters-mnis} shows an averaged feature importance map for a given category. We look into the regions of the time-frequency map that are indicated as important by the feature importance map, extract baseline-normalized activity in those regions and cluster the probes according to that activity using hierarchical complete linkage clustering with cosine distance (see the section \ref{sec:signatures-clustering} on hierarchical clustering for details). The second column of Figure~\ref{fig:importances-clusters-mnis} shows the activity of four most populated clusters for each category. Each cluster represents the activity pattern exhibited by the probes in that cluster. Only the probes whose activity had predictive power ($\text{F}_1 > 0.39$) are included in this analysis. As the final step we identified the anatomical locations of the probes from each cluster to see whether difference in the activity patterns could be attributed to the functional regions of the brain. The visualization of this step in the last two columns of Figure~\ref{fig:importances-clusters-mnis}.

This analysis allowed us make a number of \emph{global} and \emph{category-specific} observations. The set of visual categories presented in our data is diverse enough to consider category-specific findings to be general and emerge under any comparable set of visual stimuli.

The first global observation was that it is not only broadband gamma activity that is useful for the decoder's (Random Forest) performance, but low-frequency activity also contributed significantly (41\% of predictive probes exhibited only low-frequency activity in the regions of importance), sometimes overshadowing the activity of higher frequency bands altogether (for \texttt{face} and \texttt{scrambled} stimuli low frequency activity was significantly more important than broadband gamma activity, Mann-Whitney U test $p < 1\mathrm{e}{-7}$, corrected). Most clusters were composed of a combination of low and high-frequency components (Figure~\ref{fig:importances-clusters-mnis}, second column) and were mostly (87\%) located in occipito-temporal cortices, though some electrodes in parietal and frontal cortex (7\%) also appeared to contribute with predictive responses in the decoding process (Figure~\ref{fig:importances-clusters-mnis}, two right columns), especially for such stimulus categories as \texttt{animal} and \texttt{pseudoword}. 

The second observation spanning across all categories was that the classifier used not only the increases in power to perform the classification, but also relied on power decreases in different brain networks (7 out of 32 dominant activity clusters consisted solely from the activity patterns characterized by power decrease). The most prominent examples are the clusters \texttt{faces-2} (Figure~\ref{fig:importances-clusters-mnis}b), \texttt{animals-2} (Figure~\ref{fig:importances-clusters-mnis}a), \texttt{tools-2}, \texttt{pseudowords-1}, \texttt{pseudowords-2} (Figure~\ref{fig:importances-clusters-mnis}c), \texttt{scrambled-1} and \texttt{scrambled-2} (Figure~\ref{fig:importances-clusters-mnis}d). For example, to decode \texttt{face} or \texttt{pseudowords} from the activity of the blue cluster network, the RF classifier used broadband gamma power decreases located in posterior inferior temporal cortex and inferior occipital gyrus. None of the probes for which the decrease in activity was identified as important for decoding were located in classically defined Default Mode Network \citep{buckner2008brain, raichle2015brain}. 

Across all categories, the earliest component that often appeared in clusters was the brief power increase (mean non-zero power increase was $2.8$ times the baseline in the region of interest) in the low-frequency interval (4-25 Hz), which for one group of probes can be associated to an almost instantaneous broadband gamma power increase (\ref{fig:importances-clusters-mnis}b, cluster 3, mean broadband gamma increase of 1.9 times the baseline), but remains the only source of important activity for another group of probes (\ref{fig:importances-clusters-mnis}b, cluster 1).

Studying the anatomical locations of the probes belonging to different clusters of activity revealed interesting observations. Figure~\ref{fig:importances-clusters-mnis}c, \texttt{pseudowords}, clusters 1 and 3 show a clear example how clustering by activity patterns leads to assigning the probes into functionally different anatomical areas. The gamma-band increase signature captured by cluster 3 occurs only in the left hemisphere (red markers on Figure~\ref{fig:importances-clusters-mnis}c), the late theta-alpha power decrease captured by cluster 1 also occurs only in the left hemisphere (green markers) and is spatially clearly distinct from probes in cluster 3. Because it is known that \texttt{pseudoword} stimuli elicit top-down language-related (orthographic, phonological and semantic) analysis, which elicits highly left-lateralized networks identifiable in iEEG recordings \citep{juphard2011direct, mainy2008cortical}, we know that this observation reflects a functional brain process. This dissociation in both the spectrotemporal and anatomical domains provides us with valuable data on the locations and associated activity patterns emerging during automatic perceptual categorization and highlights the benefit of disentangling the activity into functionally and anatomically disassociated clusters.

\begin{figure}[h!]
    \centering
    \includegraphics[width=1.0\linewidth]{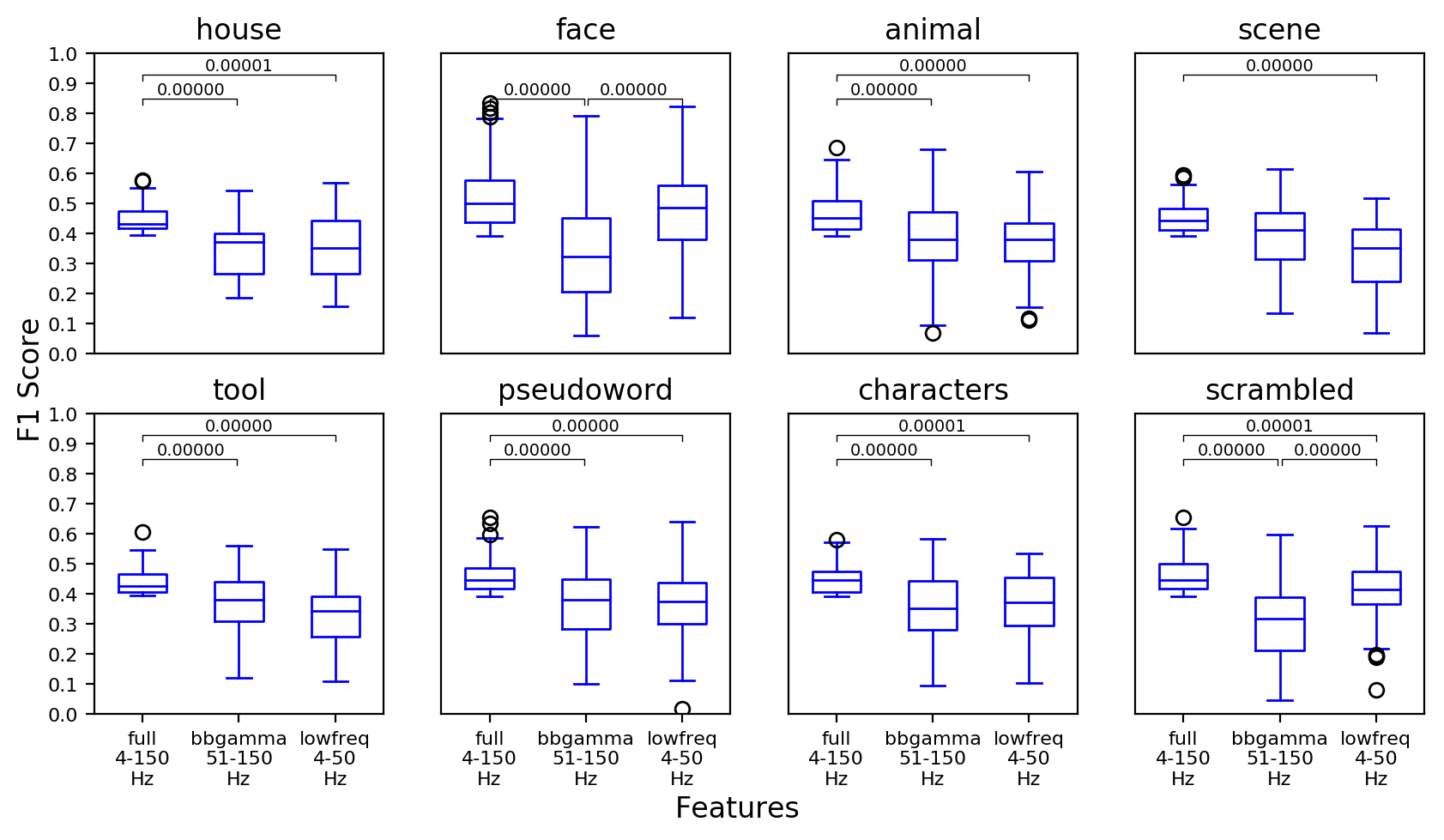}
    \caption{Comparison of predictive power of the electrodes from three different sets of features: full spectrum ($4 - 150$ Hz), broadband gamma alone ($50 - 150$ Hz) and lower frequencies alone ($4 - 50$ Hz) across categories. The bracket with the p-value indicates a significant difference according to Mann--Whitney U test.}
    \label{fig:predictiveness-bbgamma-vs-low}
\end{figure}

Finally, the relevance of the different components in the TF domain for the Random Forest classification process was assessed. Specifically, we tested whether the activity in the broadband gamma range, commonly present on most clusters across categories, is in general the most valuable neural signature for category networks as compared to the low-frequency parts of the spectrum. To test whether broadband gamma was solely the most informative frequency interval we statistically compared predictive power of three intervals: broadband gamma ($50 - 150$ Hz), low-frequency ($4 - 50$ Hz) and full spectrum ($4 - 150$ Hz). Overall, across 7 perceptual categories out of 8 (except for \texttt{scenes}), using the full spectrum was more informative than using the broadband gamma interval or the low-frequency interval alone (Mann--Whitney U test, $p < 0.001563$, corrected to the number of clusters compared, see Figure~\ref{fig:predictiveness-bbgamma-vs-low}), which is in line with the results reported by \cite{miller2016spontaneous}. Importantly, for \texttt{scrambled} images and \texttt{faces} the broadband gamma carried \emph{less} (Mann-Whitney U $p < 1\mathrm{e}{-7}$, corrected) decoding-relevant information than the lower frequencies.

%
%
\section{Significance of bottom-up approach to the\\analysis of human intracerebral neural activity}

In this chapter we explored the bottom-up approach to the analysis of human intracerebral neural activity. Due to a rich dataset and powerful methodology we were able to uncover facts about neural processing of automatic visual categorization that we would not necessarily address in a hypothesis-driven study. Previous works have shown where and when perceptual category information can be decoded from the human brain, our study adds to that line of research by identifying spectrotemporal patterns that contribute to category decoding without the need to formulate a priori hypothesis on which spectral components and at which times are worth investigating.

The classifier model first allowed us to globally identify two types of neural responses: those that were predictive of a certain category and those that did not predict any category despite eliciting strong amplitude modulation across multiple frequency bands. Surprisingly, when comparing the level of predictability of probe responses we found that only 4.8\% of the responsive probes were predictive of a category. This very low percentage highlights an important fact regarding the level of ``selectivity'' of a neural responses. In this decoding approach, the level of single-probe neural response selectivity depends on the diversity and overall quantity of the comparison/reference group to which it is compared to. Stimulus-induced neural signal selectivity is thus a graded quality that can be assessed through multiple comparisons with a broad variety of stimulation conditions.  This result also implies that although any stimulus can elicit a local neural response throughout the cerebral cortex, in the light of our results, there is a high probability of it being non-predictive of any of the categories or being polypredictive of several categories at once.  

In line with a vast literature on the localization of category related networks \citep{kanwisher1997fusiform, epstein1999parahippocampal, malach1995object, haxby2001distributed, ishai1999distributed, cohen2000visual, peelen2009neural, grill2014functional, tanaka1996inferotemporal, dicarlo2012does} predictive probes concentrated mostly in the inferior temporal cortex, namely the fusiform gyrus (BA 37), yet surprisingly for some categories, probes in primary visual cortex were also predictive of these categories. This effect is probably related to the specifics of the physical content of certain images that uniquely characterize certain categories amongst all others, as for example the content in high-contrast edge information in \texttt{scrambled} and written text stimuli.  

Predictive probes were subsequently classified according to their level of selectivity towards a single or multiple visual categories. Polypredictive probes (36\%) clustered in visual cortices and inferior temporal cortex and were associated with early spectral components ($< 300$ ms) such as broadband gamma power increases and a transient theta burst shortly after stimulus presentation. Monopredictive probes (64\%) were abundant in these same regions, but extending uniquely in frontal, parietal, superior temporal and anterior limbic cortex. Their activity was strongly associated with the later ($> 300$ ms) time and with power suppression of spectral importance features, versus baseline, in the theta ($4-7$ Hz), alpha ($8-15$ Hz) and beta bands ($16-40$ Hz). In a subgroup of probes the associated power suppression of the feature importances extended into the broad gamma band ($50-150$ Hz).

Importantly, the capacity to ascribe category selectivity to predictive probes (mono vs polypredictive probes) arises from the fact that the decoding model was trained to discriminate between all 8 categories simultaneously. The separation between mono and polypredictive probes revealed specific effects in terms of network localization and time-frequency components. The high concentration of polypredictive probes (and local networks) in early visual cortices, from primary visual cortex up to inferior temporal cortex is coherent with the idea that networks in the ventral visual stream progressively integrate more complex features into object representations, thus becoming progressively more selective, and converge within median temporal lobe to more stimulus-invariant representations \citep{quiroga2005invariant}. This progressive information integration by spectral features of neuronal responses across the visual hierarchy has been recently connected with the computations carried out by deep convolutional neural networks trained to solve the task of visual recognition \citep{kuzovkin2018activations}. 

Globally, the random forest data classification provided results that are coherent with current knowledge on 1) the implication of networks located in visual cortex and inferior temporal cortex in processing visual categories, 2) the timing of object categorization in the human brain and 3) the role of broadband gamma responses in processing category-selective information within these networks. Previous studies have shown that certain stimulus categories elicit clustered cortical responses of highly localized networks in the occipito-temporal ventral stream such as the fusiform-face-area (FFA) and the visual-word-form area (VWFA) \citep{kanwisher1997fusiform, cohen2000visual}. Yet, other studies have broadened this scope by showing that certain categories, as for example faces, rely on the involvement of a larger brain-wide distributed network \citep{ishai2005face, vidal2010category}.  Our classification analysis shows that the spatial extent of this network distribution is category specific, certain stimuli eliciting larger network responses, such as for \texttt{faces}, \texttt{animals} and \texttt{pseudowords}, as compared to \texttt{scenes}, \texttt{houses} and \texttt{scrambled} images which concentrate in the fusiform cortex, the parahippocampal cortex and primary visual cortex respectively. 

Our results largely agree with previous works trying to decode visual object categories over time with magnetoencephalography (MEG) \citep{carlson2013representational, cichy2014resolving} or intracranial recordings \citep{liu2009timing}. All these studies converge on the result that perceptual categories can be decoded from human brain signals as early as 100 ms. Our current work goes a step beyond these previous investigations by demonstrating which spectral components underlie this fast decoding. Previous intracranial studies have also shown that broadband gamma is modulated by information about object categories \citep{vidal2010category, privman2007enhanced, fisch2009neural}. Moreover, broadband gamma has been suggested as a proxy to population spiking output activity \citep{manning2009broadband, ray2011different, lachaux2012high, ray2008neural}. It has since then been considered as a hallmark of local population processing \citep{parvizi2018promises}. Our classification results however show that broadband gamma is not the sole selectivity marker of functional neural processing, and that higher decoding accuracy can be achieved by including low-frequency components of the spectrum. For certain stimulus categories, as scrambled images, the broadband gamma range is even outperformed by the predictive power of the low-frequency range.   

To understand which spectral components play a specific role in stimulus categorization we analyzed the decision process that drives the decoding model and identified the combined spectrotemporal regions that are informative for the output of the random forest classification procedure. This allowed us 1) to identify the category-selective spectral components of high importance for the automatic visual categorization process, and 2) identify the correlates functional involvement of positive as well as negative power modulations (increases and decreases versus baseline) in early and late time windows of neural processing involved in visual categorization.  

While the distinctive activity of polypredictive neural locations is mostly reflected by early TF components (i.e. broadband gamma and theta burst in \texttt{faces}), the sustained decrease in power in the alpha/beta band was extended in space and time. This process is probably dependent on the degree of difficulty for the networks in reaching a perceptual decision and which appeals to the involvement of top-down processing required to resolve perceptual ambiguity elicited by the different stimulus categories. For example, \texttt{animal} and \texttt{tool} stimuli are highly diverse in their physical image structure, as compared to \texttt{face} stimuli. This affects the efficiency of bottom-up process in extracting category information, often associated with increase in gamma activity, and probably in parallel triggers top-down processes through selective activity modulation in low-frequency channels \citep{bastos2015visual}. In our data, this latter phenomenon could be mirrored by a decrease of predictive power in the low-frequency range. Studies have shown that power modulations reflect changes in network connectivity \citep{tewarie2018relationships} and that top-down processes, eliciting a decrease in power in the alpha-beta band, are accompanied by an increase in distant network connectivity \citep{gaillard2009converging}.
   
Finally, we also show that certain probes elicit decreased broadband gamma responses (versus baseline) while representing a significant feature importance for the classification model. It has been shown that neural activity in the Default Mode Network can be negatively modulated by attending sensory stimulation \citep{buckner2008brain}, and intracranial studies have found that this was reflected by decreases (versus baseline) in the broad gamma range \citep{ossandon2011transient, jerbi2010exploring, dastjerdi2011differential}. Here we found no evidence of such power decreases in probes located in the DMN \citep{buckner2008brain}. However, the random forest classifier singled-out broad spectral patterns of power decreases at probes located in visual regions and beyond for categories \texttt{faces}, \texttt{pseudowords} and \texttt{characters}. This is the first time, to our knowledge, that power decreases in the broadband gamma range outside the DMN have been associated with highly functional neural signal classification of perceptual categories. Their functional significance should be studied in the future as they could reflect an important phenomenon of communication regulation between networks during perceptual decision making of visual categories.  

Expanding on this work and established methodology by including more subject data in the future might allow us to make a transition from the observations of local activity and the analysis of its role to being able to detect signatures of global decision-making processes. It is possible that these signatures would be reflected in specific spectral fingerprints as many classic theories would suggest \citep{rodriguez1999perception, varela2001brainweb, engel2001dynamic, siegel2012spectral}. The methodology proposed in this study can facilitate the search of those fingerprints without the need to formulate a priori hypothesis about which spectrotemporal components are worth investigating.

\chapter{Representational similarities between biological visual processing and artificial neural networks inform on structural organization of human visual cortex}
\label{ch:intracranial-dcnn-based}

In the previous chapter we have demonstrated a way to interpret an automatically built model to gain knowledge about neurological mechanisms on the level of local field potentials. In this chapter we move to a higher level of abstraction and employ machine learning and interpretability to peek into functional organization of human visual cortex. Using a metric based on representational similarity analysis we compare the activations of biological neurons in the layers of human visual cortex with the activations of artificial neurons in the layers of a deep convolutional neural network. This comparison allows us to find an alignment between those two hierarchical structures and to investigate which spectrotemporal regions of human brain activity are aligned the best, confirming the role of high gamma activity in visual processing. Using \emph{deconvolution} technique to interpret the behavior of the DCNN we were able to visualize visual inputs that the artificial neurons are tuned to and observe the similarities with the reported tuning of biological neurons when processing visual inputs.

%
%
\section{The search for the model of\\human visual system}

Biological visual object recognition is mediated by a hierarchy of increasingly complex feature representations along the ventral visual stream \citep{dicarlo2012does}. Intriguingly, these transformations are matched by the hierarchy of transformations learned by deep convolutional neural networks (DCNN) trained on natural images~\citep{gucclu2015deep}. It has been shown that DCNN provides the best model out of a wide range of neuroscientific and computer vision models for the neural representation of visual images in high-level visual cortex of monkeys \citep{yamins2014performance} and humans \citep{khaligh2014deep}. Other studies with functional magnetic resonance imaging (fMRI) data have demonstrated a direct correspondence between the hierarchy of the human visual areas and layers of the DCNN \citep{gucclu2015deep, eickenberg2016seeing, seibert2016performance, cichy2016comparison}. In sum, the increasing feature complexity of the DCNN corresponds to the increasing feature complexity occurring in visual object recognition in the primate brain \citep{kriegeskorte2015deep, yamins2016using}. 

However, fMRI based studies only allow one to localize object recognition in space, but neural processes also unfold in time and have characteristic spectral fingerprints (i.e. frequencies). With time-resolved magnetoencephalographic recordings it has been demonstrated that the correspondence between the DCNN and neural signals peaks in the first 200 ms \citep{cichy2016comparison, seeliger2017cnn}. Here we test the remaining dimension: that biological visual object recognition is also specific to certain frequencies. In particular, there is a long-standing hypothesis that especially gamma band ($30 - 150$ Hz) signals are crucial for object recognition \citep{singer1995visual, singer1999neuronal, fisch2009neural, tallon1997oscillatory, tallon1999oscillatory, lachaux1999measuring, wyart2008neural, lachaux2005many, vidal2006visual, herrmann2004cognitive, srinivasan1999increased, levy2015selective}. More modern views on gamma activity emphasize the role of the gamma rhythm in establishing a communication channel between areas \citep{fries2005mechanism, fries2015rhythms}. Further research has demonstrated that especially feedforward communication from lower to higher visual areas is carried by the gamma frequencies \citep{van2014alpha, bastos2015visual, michalareas2016alpha}. As the DCNN is a feedforward network one could expect that the DCNN will correspond best with the gamma band activity. In this work we used the DCNN as a computational model to assess whether signals in the gamma frequency are more relevant for object recognition than other frequencies.

To empirically evaluate whether gamma frequency has a specific role in visual object recognition we assessed the alignment between the responses of layers of a commonly used DCNN and the neural signals in five distinct frequency bands and three time windows along the areas constituting the ventral visual pathway. Based on the previous findings we expected that: mainly gamma frequencies should be aligned with the layers of the DCNN; the correspondence between the DCNN and gamma should be confined to early time windows; the correspondence between gamma and the DCNN layers should be restricted to visual areas. In order to test these predictions we capitalized on direct intracranial depth recordings from $100$ patients with epilepsy and a total of $11293$ electrodes implanted throughout the cerebral cortex. 

We observe that activity in the gamma range along the ventral pathway is statistically significantly aligned with the activity along the layers of DCNN: gamma ($31 - 150$ Hz) activity in the early visual areas correlates with the activity of early layers of DCNN, while the gamma activity of higher visual areas is better captured by the higher layers of the DCNN. We also find that while the neural activity in the theta range ($5 - 8$ Hz) is not aligned with the DCNN hierarchy, the representational geometry of theta activity is correlated with the representational geometry of higher layers of DCNN.

%
%
\section{Simultaneous recordings of human intracortical responses and of responses of an artificial neural network to the same visual stimuli}

The dataset that was created for this study consists of two components: recordings of local field potentials in human visual cortex and activations of artificial neurons of a deep convolutional neural network trained on a visual recognition task. The raw neurological data was the same as the one used in Chapter \ref{ch:spectral-signatures-based}, please refer to section \ref{sec:raw-intracranial-data} for the technical details on the subjects and data acquisition parameters. The preprocessing pipeline was mostly similar to the one performed in the previous study, however there were a few differences, please see the section below. Further in this section we present the protocol we used to obtain activations of artificial neurons once the artificial neural network was presented with the same visual stimuli as the the human subjects.

%
%
\subsection{Processing of neural data}

The final dataset consists of $2823250$ local field potential (LFP) recordings -- $11293$ electrode responses to $250$ stimuli. To remove the artifacts the signals were linearly detrended and the recordings that contained values $\ge10\sigma_{\text{images}}$, where $\sigma_{\text{images}}$ is the standard deviation of responses (in the time window from $-500$ ms to $1000$ ms) of that particular probe over all stimuli, were excluded from data. All electrodes were re-referenced to a bipolar reference. For every electrode the reference was the next electrode on the same rod following the inward direction. The electrode on the deepest end of each rod was excluded from the analysis. The signal was segmented in the range from $-500$ ms to $1000$ ms, where $0$ marks the moment when the stimulus was shown. The $-500$ to $-100$ ms time window served as the baseline. There were three time windows in which the responses were measured: $50 - 250$ ms, $150 - 350$ ms and $250 - 450$ ms.

We analyzed five distinct frequency bands: $\theta$ ($5 - 8$ Hz), $\alpha$ ($9 - 14$ Hz), $\beta$ ($15 - 30$ Hz), $\gamma$ ($31 - 70$ Hz) and $\Gamma$ ($71 - 150$ Hz). To quantify signal power modulations across time and frequency we used standard time-frequency (TF) wavelet decomposition \citep{daubechies1990wavelet}. The signal $s(t)$ is convoluted with a complex Morlet wavelet $w(t, f_0)$, which has Gaussian shape in time $(\sigma_t)$ and frequency $(\sigma_f)$ around a central frequency $f_0$ and defined by $\sigma_f = 1/2 \pi \sigma_t$ and a normalization factor. In order to achieve good time and frequency resolution over all frequencies we
slowly increased the number of wavelet cycles with frequency ($\frac{f_0}{\sigma_f}$ was set to 6 for high and low gamma, 5 for beta, 4 for alpha and 3 for theta). This method allows obtaining better frequency resolution than by applying a constant cycle length \citep{delorme2004eeglab}. The square norm of the convolution results in a time-varying representation of spectral power, given by: $P(t, f_0) = |w(t, f_0) \cdot s(t)|^2$. 

Further analysis was done on the electrodes that were responsive to the visual task. We assessed neural responsiveness of an electrode separately for each region of interest -- for each frequency band and time window we compared the average post-stimulus band power to the average baseline power with a Wilcoxon signed-rank test for matched-pairs. All p-values from this test were corrected for multiple comparisons across all electrodes with the false discovery rate procedure \citep{genovese2002thresholding}. In the current study we deliberately kept only positively responsive electrodes, leaving the electrodes where the post-stimulus band power was lower than the average baseline power for future work. Table \ref{tab:responsive-counts} contains the numbers of electrodes that were used in the final analysis in each of $15$ regions of interest across the time and frequency domains.

\begin{table}[h!]
    \centering
    \begin{tabular}{r|ccccc}
        & $\theta$ & $\alpha$ & $\beta$ & $\gamma$ & $\Gamma$ \\ \hline
         $50 - 250$ ms & 1299     & 709      & 269     & 348      & 504 \\
        $150 - 350$ ms & 1689     & 783      & 260     & 515      & 745 \\
        $250 - 450$ ms & 1687     & 802      & 304     & 555      & 775
    \end{tabular}
    \caption{Number of positively responsive electrodes in each of the $15$ regions of interest in a time-resolved spectrogram.}
    \label{tab:responsive-counts}
\end{table}

Each electrode's Montreal Neurological Institute coordinate system (MNI) coordinates were mapped to a corresponding Brodmann brain area~\citep{brodmann1909vergleichende} using Brodmann area atlas contained in MRICron~\citep{rorden2007mricron} software. 

To summarize, once the neural signal processing pipeline is complete, each electrode's response to each of the stimuli is represented by one number -- the average band power in a given time window normalized by the baseline. The process is repeated independently for each time-frequency region of interest.

%
%
\subsection{Processing of DCNN data}
We feed the same images that were shown to the test subjects to a deep convolutional neural network (DCNN) and obtain activations of artificial neurons (nodes) of that network. We use \texttt{Caffe} \citep{jia2014caffe} implementation of \texttt{AlexNet} \citep{krizhevsky2012imagenet} architecture (see Figure \ref{fig:layer_specificity_and_volume}) trained on \texttt{ImageNet} \citep{ILSVRC15} dataset to categorize images into 1000 classes. Although the image categories used in our experiment are not exactly the same as the ones in the \texttt{ImageNet} dataset, they are a close match and DCNN is successful in labelling them.

The architecture of the \texttt{AlexNet} artificial network can be seen on Figure \ref{fig:layer_specificity_and_volume}. It consists of 9 layers. The first is the input layer, where one neuron corresponds to one pixel of an image and activation of that neuron on a scale from 0 to 1 reflects the color of that pixel: if a pixel is black, the corresponding node in the network is not activated at all (value is 0), while a white pixel causes the node to be maximally activated (value 1). After the input layer the network has 5 \emph{convolutional layers} referred to as \texttt{conv1-5}. A convolutional layer is a collection of filters that are applied to an image. Each filter is a 2D arrangement of weights that represent a particular visual pattern. A filter is convolved with the input from the previous layer to produce the activations that form the next layer. For an example of a visual pattern that a filter of each layer is responsive to, please see Figure \ref{fig:layer_specificity_and_volume}b. Each layer consists of multiple filters and we visualize only one per layer for illustrative purposes. A filter is applied to every possible position on an input image and if the underlying patch of an image coincides with the pattern that the filter represents, the filter becomes activated and translates this activation to the artificial neuron in the next layer. That way, nodes of \texttt{conv1} tell us where on the input image each particular visual pattern occurred. Figure \ref{fig:layer_specificity_and_volume}b shows an example output feature map produced by a filter being applied to the input image. Hierarchical structure of convolutional layers gives rise to the phenomenon we are investigating in this work -- increase of complexity of visual representations in each subsequent layer of the visual hierarchy: in both the biological and artificial systems. Convolutional layers are followed by 3 \emph{fully-connected} layers (\texttt{fc6-8}). Each node in a fully-connected layer is, as the name suggests, connected to every node of the previous layer allowing the network to decide which of those connections are to be preserved and which are to be ignored. For both convolutional and fully-connected layers we can apply \emph{deconvolution} \citep{zeiler2014visualizing} technique to map activations of neurons in those layers back to the input space. This visualization gives better understanding of inner workings of a neural network. Examples of deconvolution reconstruction for each layer are given in Figure \ref{fig:layer_specificity_and_volume}b.

For each of the images we store the activations of all nodes of DCNN. As the network has 9 layers we obtain 9 representations of each image: the image itself (referred to as layer 0) in the pixel space and the activation values of each of the layers of DCNN. See the step 2 of the analysis pipeline on Figure \ref{fig:methods_pipeline} for the cardinalities of those feature spaces.

%
%
\section{The mapping between the Brodmann areas and layers of a Deep Convolutional Neural Network}

As a result of the preprocessing steps we were left with two sets of responses to the same set of stimuli: one from a biological system, one from an artificial one. Our ultimate goal was to compare those responses, but since the representations were very different a direct comparison was not possible. To overcome this we used representational similarity analysis -- a technique that relies on the distance measure between the data samples (see taxonomy in Table \ref{tab:representation-taxonomy} of Section \ref{sec:representation-taxonomy}) to provide a way to compare behaviors of two systems under the same set of stimuli while having different data representations.

%
%
\subsection{Mapping neural activity to the layers of DCNN}
\label{sec:mapping-brain-to-dcnn}
Once we extracted the features from both neural and DCNN responses our next goal was to compare the two and use a similarity score to map the brain area where a probe was located to a layer of DCNN. By doing that for every probe in the dataset we obtained cross-subject alignment between visual areas of human brain and layers of DCNN. There are multiple deep neural network architectures trained to classify natural images. Our choice of \texttt{AlexNet} does not imply that this particular architecture corresponds best to the hierarchy of visual layers of human brain. It does, however, provide a comparison for hierarchical structure of human visual system and was selected among other architectures due to its relatively small size and thus easier interpretability.

Recent studies comparing the responses of visual cortex with the activity of DCNN have used two types of mapping methods. The first type is based on linear regression models that predict neural responses from DCNN activations~\citep{yamins2014performance, gucclu2015deep}. The second is based on representational similarity analysis (RSA)~\citep{kriegeskorte2008representational}. We used RSA to compare distances between stimuli in the neural response space and in the DCNN activation space~\citep{cichy2016deep}.

\begin{figure}[h!]
    \centering
    \includegraphics[width=1.0\linewidth]{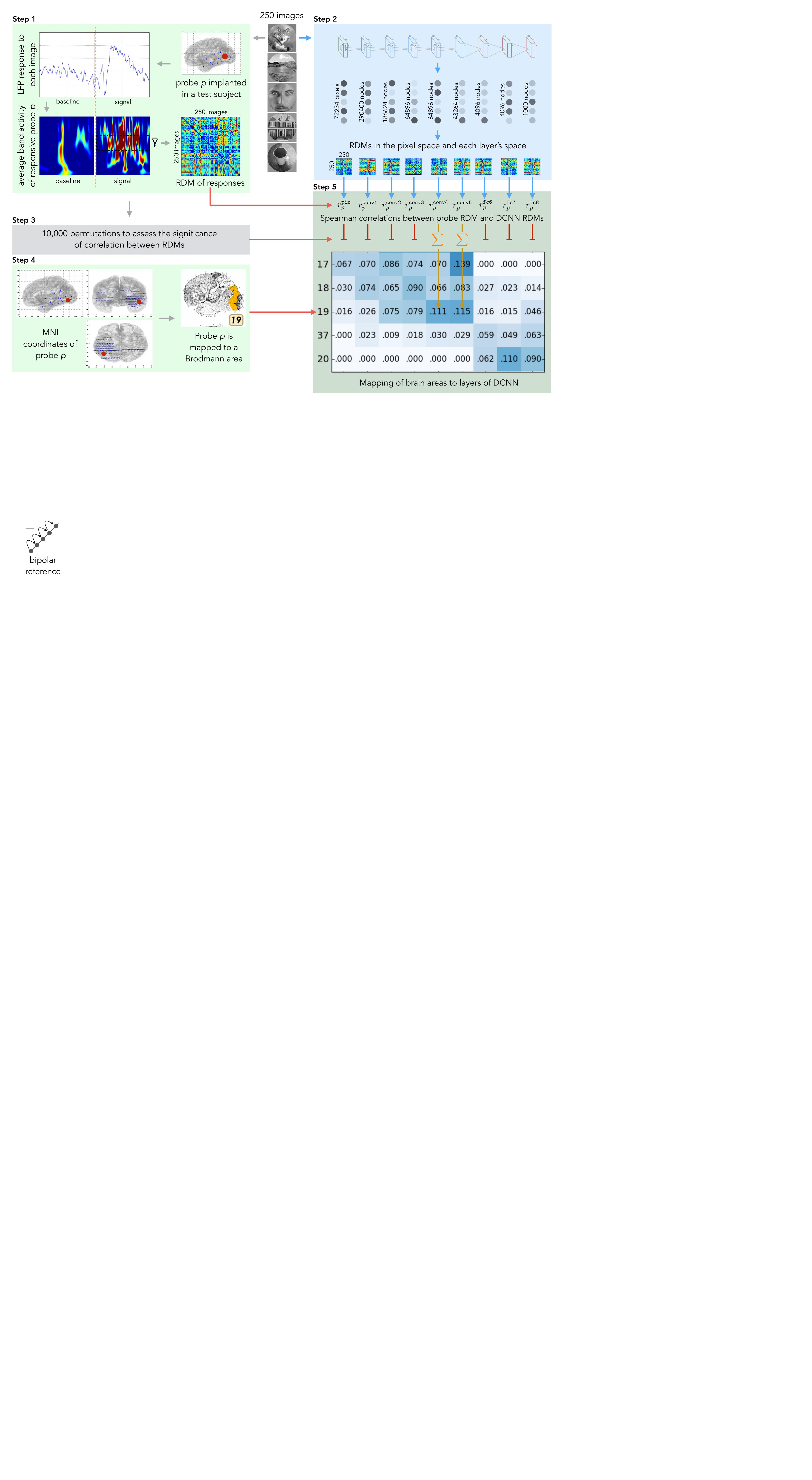}
    \caption{Overview of the analysis pipeline where $250$ natural images are presented to human subjects and to an artificial vision system. The activities elicited in these two systems are compared to map regions of human visual cortex to layers of deep convolutional neural network. \textbf{Step 1:} LFP response of each of $11293$ electrodes to each of the images is converted into the frequency domain. Activity evoked by each image is compared to the activity evoked by every other image and results of this comparison are presented as a representational dissimilarity matrix (RDM). \textbf{Step 2:} Each of the images is shown to a pre-trained DCNN and activations of each of the layers are extracted. Each layer's activations form a representation space, in which stimuli (images) can be compared to each other. Results of this comparison are summarized as a RDM for each DCNN layer. \textbf{Step 3:} Subject's intracranial responses to stimuli are randomly reshuffled and the analysis performed in step 1 is repeated $10000$ times to obtain $10000$ random RDMs for each electrode. \textbf{Step 4:} Each electrode's MNI coordinates are used to map the electrode to a Brodmann area. The figure also gives an example of electrode implantation locations in one of the subjects (blue circles are the electrodes). \textbf{Step 5:} Spearman's rank correlation is computed between the true (non-permuted) RDM of neural responses and RDMs of each layer of DCNN. Also $10000$ scores are computed with the random RDM for each electrode-layer pair to assess the significance of the true correlation score. If the score obtained with the true RDM is significant (the value of $p < 0.001$ is estimated by selecting a threshold such that none of the probes would pass it on the permuted data), then the score is added to the mapping matrix.The procedure is repeated for each electrode and the correlation scores are summed and normalized by the number of electrodes per Brodmann area.  The resulting mapping matrix shows the alignment between the consecutive areas of the ventral stream and layers of DCNN.}
    \label{fig:methods_pipeline}
\end{figure}

%
%
We built a representation dissimilarity matrix (RDM) of size \emph{number of stimuli} $\times$ \emph{number of stimuli} (in our case $250 \times 250$) for each of the probes and each of the layers of DCNN. Note that this is a non-standard approach: usually the RDM is computed over a population (of voxels, for example), while we do it for each probe separately. We use the non-standard approach because often we only had 1 electrode per patient per brain area. Given a matrix $\text{RDM}^\text{feature space}$ a value $\text{RDM}_{ij}^\text{feature space}$ in the $i$th row and $j$th column of the matrix shows the Euclidean distance between the vectors $\mathbf{v}_i$ and $\mathbf{v}_j$ that represent images $i$ and $j$ respectively in that particular feature space. Note that the preprocessed neural response to an image in a given frequency band and time window is a scalar, and hence correlation distance is not applicable. Also, given that DCNNs are not invariant to the scaling of the activations or weights in any of its layers, we preferred to use closeness in Euclidean distance as a more strict measure of similarity. In our case there are 10 different feature spaces in which an image can be represented: the original pixel space, 8 feature spaces for each of the layers of the DCNN and one space where an image is represented by the preprocessed neural response of probe $p$. For example, to analyze region of interest of high gamma in $50 - 250$ ms time window we computed $504$ RDM matrices on the neural responses -- one for each positively responsive electrode in that region of interest (see Table \ref{tab:responsive-counts}), and $9$ RDM matrices on the activations of the layers of DCNN. A pair of a frequency band and a time window, such as ``high gamma in 50-250 ms window'' is referred to as \emph{region of interest} in this work.

%
%
The second step was to compare the $\text{RDM}^{\text{probe}\ p}$ of each probe $p$ to RDMs of layers of DCNN. We used Spearman's rank correlation as measure of similarity between the matrices:
\begin{equation}
    \label{eq:rsa_score}
    \rho^{\text{probe}\ p}_{\text{layer}\ l}= \text{Spearman}(\text{RDM}^{\text{probe}\ p}, \text{RDM}^{\text{layer}\ l}).
\end{equation}
As a result of comparing $\text{RDM}^{\text{probe}\ p}$ with every $\text{RDM}^{\text{layer}\ l}$ we obtain a vector with 9 scores: $(\rho_\text{pixels},\rho_\text{conv1},\ldots,\rho_\texttt{fc8})$ that serves as a distributed mapping of probe $p$ to the layers of DCNN (see step 5 of the analysis pipeline on Figure \ref{fig:methods_pipeline}). The procedure is repeated independently for each probe in each region of interest.
To obtain an aggregate score of the correlation between an area and a layer the $\rho$ scores of all individual probes from that area are summed and divided by the number of $\rho$ values that have passed the significance criterion. The data for the Figures \ref{fig:all_areas_in_gammas} and \ref{fig:15_plates} are obtained in such manner.

Figure \ref{fig:rdm_corrs_within_dnn} presents the results of applying RSA within the DCNN to compare the similarity of representational geometry between the layers.

%
%
To assess the statistical significance of the correlations between the RDM matrices we ran a permutation test. In particular, we reshuffled the vector of brain responses to images $10000$ times, each time obtaining a dataset where the causal relation between the stimulus and the response is destroyed. On each of those datasets we ran the analysis and obtained Spearman's rank correlation scores. To determine score's significance we compared the score obtained on the original (unshuffled) data with the distribution of scores obtained with the surrogate data. If the score obtained on the original data was bigger than the score obtained on the surrogate sets with $p < 0.001$ significance, we considered the score to be significantly different. The threshold of $p = 0.001$ is estimated by selecting such a threshold that on the surrogate data none of the probes would pass it.

To size the effect caused by training artificial neural network on natural images we performed a control where the whole analysis pipeline depicted on Figure \ref{fig:methods_pipeline} is repeated using activations of a network that was not trained -- its weights are randomly sampled from a Gaussian distribution $\mathcal{N}(0, 0.01)$. 

For the relative comparison of alignments between the bands and the noise level estimation we took 1,000 random subsets of half of the size of the dataset. Each region of interest was analyzed separately. The alignment score was calculated for each subset, resulting in 1,000 alignment estimates per region of interest. This allowed us to run a statistical test between each pair of regions of interest to test the hypothesis that the DCNN alignment with the probe responses in one band is higher than the alignment with the responses in another band. We used Mann-Whitney~U~test~\citep{mann1947test} to test that hypothesis and accepted the difference as significant at p-value threshold of $0.005$ Bonferroni corrected~\citep{dunn1961multiple} to $2.22\mathrm{e}{-5}$.

%
%
\subsection{Quantifying properties of the mapping}
\label{sec:quantifying-mapping}

To evaluate the results quantitatively we devised a set of measures specific to our analysis. \emph{Volume} is the total sum of significant correlations (see Equation \ref{eq:rsa_score}) between the RDMs of the subset of layers $\mathbf{L}$ and the RDMs of the probes in the subset of brain areas $\mathbf{A}$:
\begin{equation}
    \label{eq:volume}
    V^{\text{areas}\ \mathbf{A}}_{\text{layers}\ \mathbf{L}} = \sum_{a \in \mathbf{A}}\sum_{l \in \mathbf{L}}\sum_{p \in \mathbf{S}^a_l} \rho^{\text{probe} \ p}_{\text{layer}\ l},
\end{equation}
where $\mathbf{A}$ is a subset of brain areas, $\mathbf{L}$ is a subset of layers, and $\mathbf{S}^a_l$ is the set of all probes in area $a$ that significantly correlate with layer $l$.

We express \emph{volume of visual activity} as
\begin{equation}
    \label{eq:visual-volume}
    V^{\mathbf{A} = \{17,18,19,37,20\}}_{\mathbf{L} = \text{all layers}},
\end{equation}
which shows the total sum of correlation scores between all layers of the network and the Brodmann areas that are located in the ventral stream: $17, 18, 19, 37$, and $20$.

\emph{Visual specificity} of activity is the ratio of volume in visual areas and volume in all areas together, for example visual specificity of all of the activity in the ventral stream that significantly correlates with any of layers of DCNN is
\begin{equation}
    \label{eq:visual-specificity}
    S^{\mathbf{A} = \{17,18,19,37,20\}}_{\mathbf{L} = \text{all layers}} = \displaystyle\frac{V^{\mathbf{A} = \{17,18,19,37,20\}}_{\mathbf{L} = \text{all layers}}}{V^{\mathbf{A} = \text{all areas}}_{\mathbf{L} = \text{all layers}}}
\end{equation}

The measures so far did not take into account hierarchy of the ventral stream nor the hierarchy of DCNN. The following two measures are the most important quantifiers we rely on in presenting our results and they do take hierarchical structure into account.

The \emph{ratio of complex visual features to all visual features} is defined as the total volume mapped to layers \texttt{conv5}, \texttt{fc6}, \texttt{fc7} divided by the total volume mapped to layers \texttt{conv1}, \texttt{conv2}, \texttt{conv3}, \texttt{conv5}, \texttt{fc6}, \texttt{fc7}:
\begin{equation}
    \label{eq:hhl}
    C^{\mathbf{A}} = \displaystyle\frac{V^{\mathbf{A}}_{\mathbf{L} = \{\texttt{conv5}, \texttt{fc6}, \texttt{fc7}\}}}{V^{\mathbf{A}}_{\mathbf{L} = \{\texttt{conv1}, \texttt{conv2}, \texttt{conv3}, \texttt{conv5}, \texttt{fc6}, \texttt{fc7}\}}}.
\end{equation}
Note that for this measure layers \texttt{conv4} and \texttt{fc8} are omitted: layer \texttt{conv4} is considered to be the transition between the layers with low and high complexity features, while layer \texttt{fc8} directly represents class probabilities and does not carry visual representations of the stimuli (if only on very abstract level).

Finally, the \emph{alignment} between the activity in the visual areas and activity in DCNN is estimated as Spearman's rank correlation between two vectors each of length equal to the number of probes with RDMs that significantly correlate with an RDM of any of DCNN layers. The first vector is a list of Brodmann areas $\mathbf{BA}^p$ to which a probe $p$ belong if its activity representation significantly correlates with activity representation of a layer $l$:
{\small
\begin{equation}
    \label{eq:alignment-areas}
    \textbf{A}_\text{align} = \Big\{\mathbf{BA}^p\ |\ \forall p\ \exists\ l : \rho(\text{RDM}^p, \text{RDM}^l)\ \parbox{5.5em}{\tiny{is significant according to the permutation test}}\ \Big\}.
\end{equation}
}$\mathbf{A}$ is ordered by the hierarchy of the ventral stream: BA17, BA18, BA19, BA37, BA20. Areas are coded by integer range from 0 to 4. The second vector lists DCNN layers $\mathbf{L}^p$ to which the very same probes $p$ were assigned:
{\small
\begin{equation}
    \label{eq:alignment-layers}
    \mathbf{L}_\text{align} = \Big\{\mathbf{L}^p\ |\ \forall p\ \exists\ l : \rho(\text{RDM}^p, \text{RDM}^l)\ \parbox{5.5em}{\tiny{is significant according to the permutation test}}\ \Big\}.
\end{equation}
}Layers of DCNN are coded by integer range from 0 to 8. We denote Spearman rank correlation of those two vectors as \emph{alignment}
\begin{equation}
    \label{eq:alignement}
    \rho_\text{align} = \text{Spearman}(\mathbf{A}_\text{align}, \mathbf{L}_\text{align}).
\end{equation}
We note that although the hierarchy of the ventral stream is usually not defined through the progression of Brodmann areas, such ordering nevertheless provides a reasonable approximation of the real hierarchy \citep{lerner2001hierarchical, grill2004human}. As both the ventral stream and the hierarchy of layers in DCNN have an increasing complexity of visual representations, the relative ranking within the biological system should coincide with the ranking within the artificial system. Based on the recent suggestion that significance levels should be shifted to 0.005~\citep{dienes2017redefine} and after Bonferroni-correcting for 15 time-frequency windows we accepted alignment as significant when it passed $p < 0.0003(3)$.

%
%
\section{Alignment between the layers of the DCNN and layers of human visual cortex}

This section present the results and observations that were achieved by comparing the two systems of vision. Here is a brief summary of our findings: activity in gamma band is aligned better than other frequencies to the hierarchical structure of a deep convolutional neural network, this alignment is mostly attributed to having two types of layer in DCNN: convolutional, that are representationally more similar to the activity of early visual areas, and fully connected layers, that are more similar to later visual and temporal areas of the ventral stream. The section describes the evidence in favor of those conclusion and presents more granular and deeper analysis focusing on specific areas of visual cortex and layers of the DCNN.

\subsection{Activity in gamma band is aligned with the DCNN}

\begin{figure}[h!]
    \centering
    \includegraphics[width=0.85\linewidth]{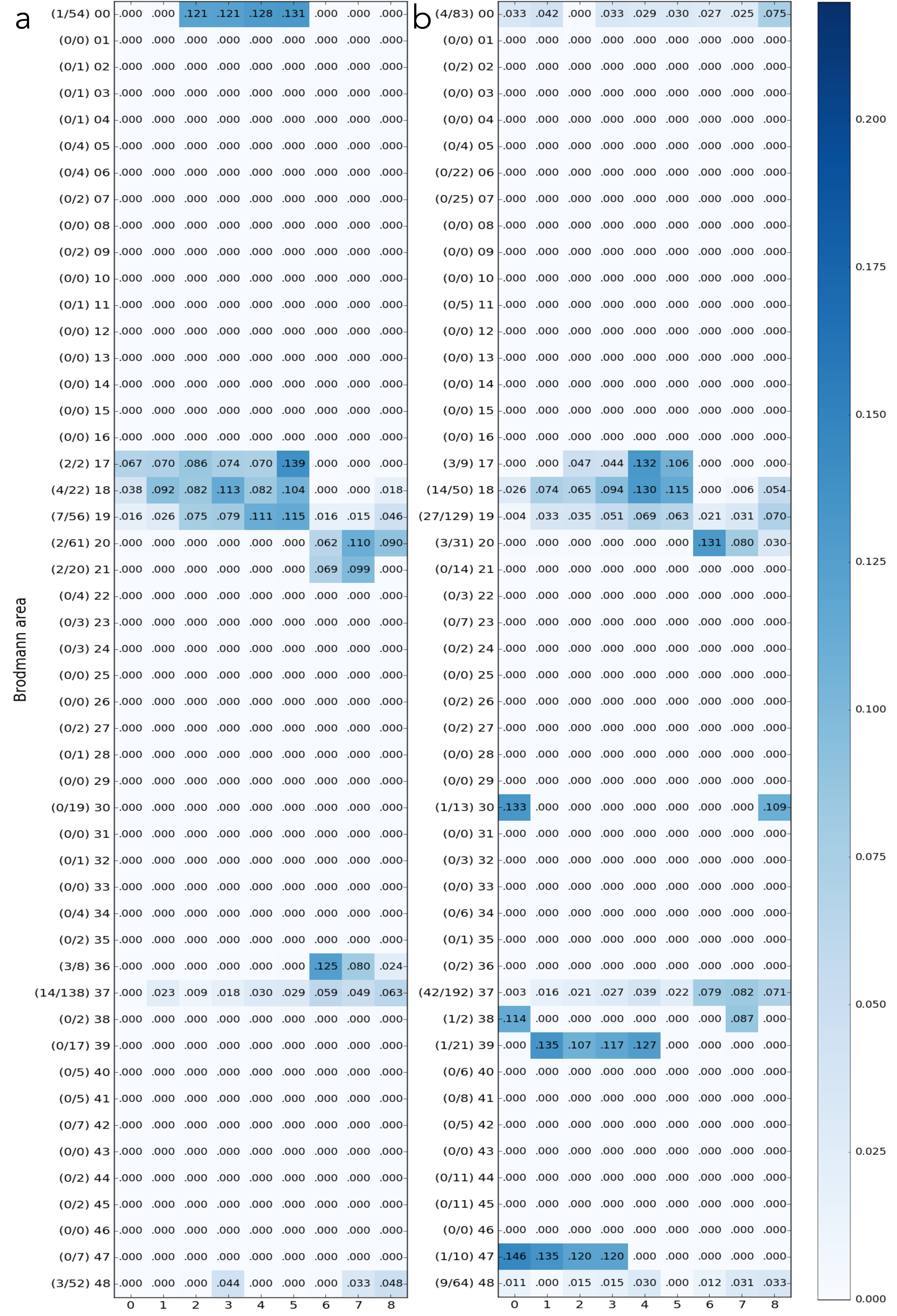}
    \caption{Mapping of the activity in Brodmann areas to DCNN layers. Underlying data comes from the activity in low gamma (31-70 Hz, panel a) and high gamma (71-150 Hz, panel b) bands in 150-350 ms time window. On the vertical axis there are Brodmann areas and the number of significantly correlating probes in each area out of the total number of responsive probes in that area. Horizontal axis represents succession of layers of DCNN. Number in each cell of the matrix is the total sum of correlations (between RDMs of probes in that particular area and the RDM of that layer) normalized by the number of significantly correlating probes in an area.}
    \label{fig:all_areas_in_gammas}
\end{figure}

We tested the hypothesis that gamma activity has a specific role in visual object recognition compared to other frequencies. To that end we assessed the alignment of neural activity in different frequency bands and time windows to the activity of layers of a deep convolutional neural network (DCNN) trained for object recognition. In particular, we used RSA to compare the representational geometry of different DCNN layers and the activity patterns of different frequency bands of single electrodes (see Figure \ref{fig:methods_pipeline}). We consistently found that signals in low gamma ($31 - 70$ Hz) frequencies across all time windows and high gamma ($71 - 150$ Hz) frequencies in $150-350$ ms window are aligned with the DCNN in a specific way: increase of the complexity of features along the layers of the DCNN was roughly matched by the transformation in the representational geometry of responses to the stimuli along the ventral stream. In other words, the lower and higher layers of the DCNN explained gamma band signals from earlier and later visual areas, respectively. Figure \ref{fig:all_areas_in_gammas}a illustrates assignment of neural activity in low gamma band and Figure \ref{fig:all_areas_in_gammas}b the high gamma band to Brodmann areas and layers of DCNN. Most of the activity was assigned to visual areas (areas 17, 18, 19, 37, 20). Focusing on visual areas revealed a diagonal trend that illustrates the alignment between ventral stream and layers of DCNN (see Figure \ref{fig:15_plates}).

\subsection{Activity in other frequency bands}

\begin{figure}[h!]
    \centering
    \includegraphics[width=0.7\linewidth]{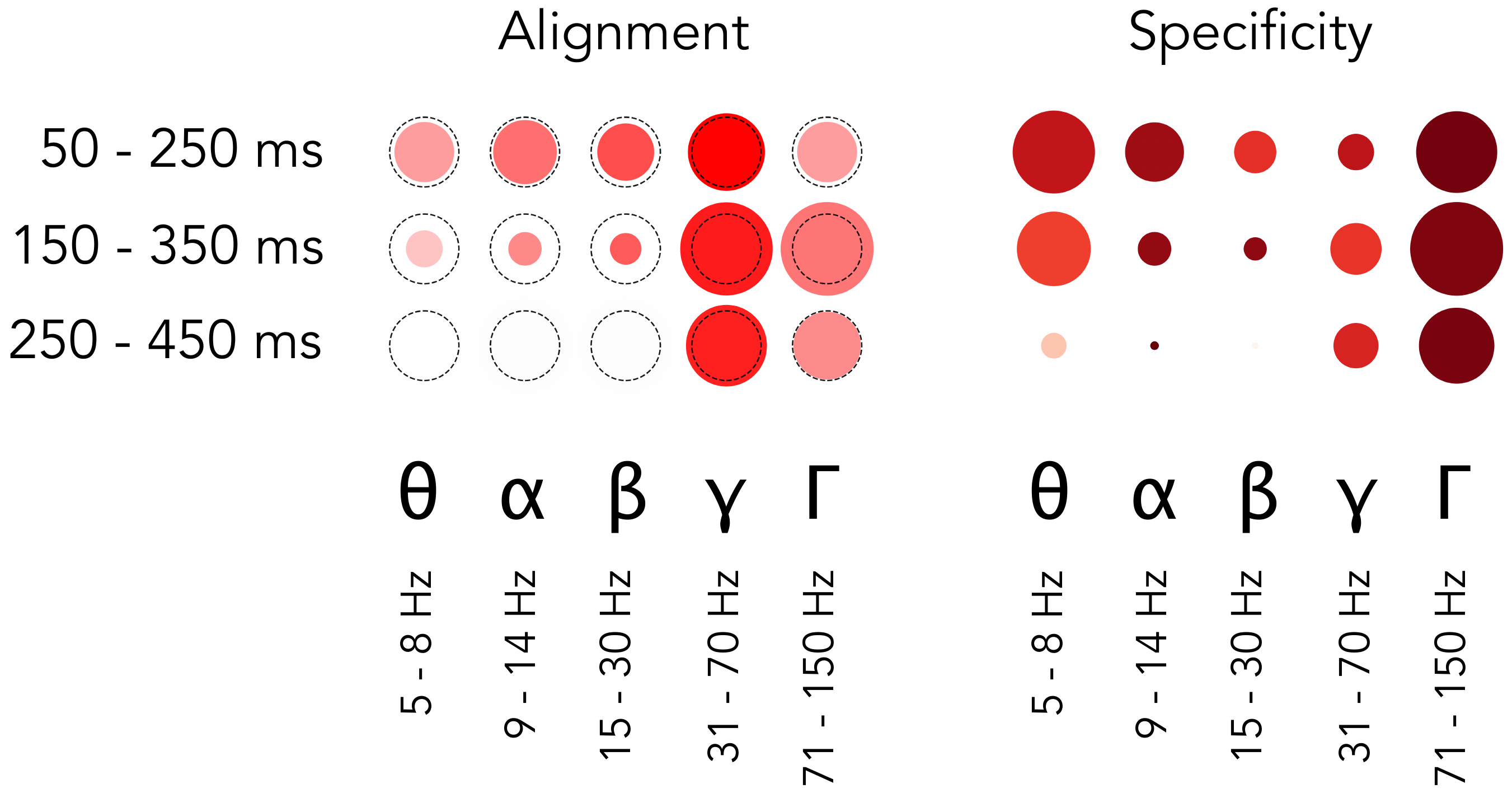}
    \caption{Overall relative statistics of brain responses across frequency bands and time windows. Panel \textbf{a} shows the alignment between visual brain areas and DCNN layers (see Equation \ref{eq:alignement}). The color indicates the correlation value ($\rho$) while the size of the marker shows the logarithm (so that not significant results are still visible on the plot) of inverse of the statistical significance of the correlation, dotted circle indicates $p = 0.0003(3)$ -- the Bonferroni-corrected significance threshold level of $0.005$. Panel \textbf{b} shows whether activity in a region of interest is specific to visual areas (see Equation \ref{eq:visual-specificity}): intense red means that most of the activity in that band and time window happened in visual areas, size of the marker indicates total volume (Equation \ref{eq:volume}) of activity in all areas. The maximal size of a marker is defined by the biggest marker on the figure.}
    \label{fig:diagonality_specificity_and_volume}
\end{figure}

To test the specificity of gamma frequency in visual object recognition, we assessed the alignment between the DCNN and other frequencies. Our findings across all subjects, time windows and frequency bands are summarized on Figure \ref{fig:diagonality_specificity_and_volume}a. We note that the alignment in the gamma bands is also present at the single-subject level (see supplementary Figure \ref{fig:single_plates} and supplementary materials \ref{sup:dcnn}). Apart from the alignment we looked at the total amount of correlation and its specificity to visual areas. Figure \ref{fig:diagonality_specificity_and_volume}b shows the volume of significantly correlating activity was highest in the high gamma range. Remarkably, 97\% of that activity was located in visual areas, which is confirmed by Figure \ref{fig:all_areas_in_gammas} where we see that in the gamma range only a few electrodes were assigned to Brodmann areas that are not part of the ventral stream. The detailed mapping results for all frequency bands and time windows are presented in layer-to-area fashion on Figure \ref{fig:15_plates}.

The results in the right column of Table \ref{tab:alignment-significance} show the alignment values and significance levels for a DCNN that is trained for object recognition on natural images. On the left part of Table \ref{tab:alignment-significance} the alignment between the brain areas and a DCNN that has not been trained on object recognition (i.e. has random weights) is given for comparison. One can see that training a network to classify natural images drastically increases the alignment score $\rho$ and its significance. One can see that weaker alignment (that does not survive the Bonferroni correction) is present in early time window in theta and alpha frequency range. No alignment is observed in the beta band. 

\begin{table}[h!]
    \vspace{0.5em}
    \centering
    {\small
    \begin{tabular}{cr|rr|rr|l}
        & & \multicolumn{2}{p{1.9cm}|}{\small{\parbox{2.5cm}{Alignment with layers of randomly initialized \texttt{AlexNet}\vspace{0.4em}}}} & \multicolumn{2}{p{1.9cm}|}{\small{\parbox{2.6cm}{Alignment with layers of \texttt{AlexNet} trained on \texttt{ImageNet}}}} &  \\ \hline
        \textbf{Band} & \textbf{Window} & \textbf{$\rho_\text{align}$} & \textbf{p-value} & \textbf{$\rho_\text{align}$} & \textbf{p-value} &              \\[1ex]
        $\theta$      &       50-250 ms &                   0.0632 &             0.71 &                   0.2257 &       0.00231575 & \textbf{*}   \\
        $\theta$      &      150-350 ms &                  -0.1013 &             0.59 &                   0.1396 &       0.08848501 &              \\
        $\theta$      &      250-450 ms &                   0.1396 &             0.59 &                   0.0695 &       0.78400416 &              \\[1.5ex]
        
        $\alpha$      &       50-250 ms &                  -0.2411 &             0.32 &                   0.3366 &       0.00103551 & \textbf{*}   \\
        $\alpha$      &      150-350 ms &                   0.0000 &             1.00 &                   0.2720 &       0.13199463 &              \\
        $\alpha$      &      250-450 ms &                       -- &               -- &                       -- &               -- &              \\[1.5ex]
        
        $\beta$       &       50-250 ms &                       -- &               -- &                   0.4166 &       0.00397929 &              \\
        $\beta$       &      150-350 ms &                       -- &               -- &                   0.3808 &       0.16141286 &              \\
        $\beta$       &      250-450 ms &                       -- &               -- &                       -- &               -- &              \\[1.5ex]
        
        $\gamma$      &       50-250 ms &                   0.1594 &             0.62 &                   0.5979 &       0.00004623 & \textbf{***} \\
        $\gamma$      &      150-350 ms &                  -0.1688 &             0.34 &                   0.5332 &       0.00000059 & \textbf{***} \\
        $\gamma$      &      250-450 ms &                  -0.1132 &             0.56 &                   0.5217 &       0.00001624 & \textbf{***} \\[1.5ex]
        
        $\Gamma$      &       50-250 ms &                   0.0869 &             0.42 &                   0.2259 &       0.00222940 & \textbf{*}   \\
        $\Gamma$      &      150-350 ms &                  -0.0053 &             0.96 &                   0.3200 &       0.00000051 & \textbf{***} \\
        $\Gamma$      &      250-450 ms &                  -0.1361 &             0.33 &                   0.2688 &       0.00047999 & \textbf{*}   \\[1ex]
    \end{tabular}}
    \caption{Alignment score $\rho_\text{align}$ and the significance levels for all 15 regions of interest. * indicates the alignments that pass p-value threshold of 0.05 Bonferroni-corrected to $< 0.003(3)$ and *** the ones that pass 0.005~\citep{dienes2017redefine} Bonferroni-corrected to $< 0.0003(3)$. Note how the values differ between random (control) network and a network trained on natural images. Visual representation of alignment and significance is given on Figure \ref{fig:diagonality_specificity_and_volume}a.}
    \label{tab:alignment-significance}
\end{table}

\setcounter{figure}{12}
\begin{sidewaysfigure}
    \centering
    \includegraphics[width=1.0\linewidth]{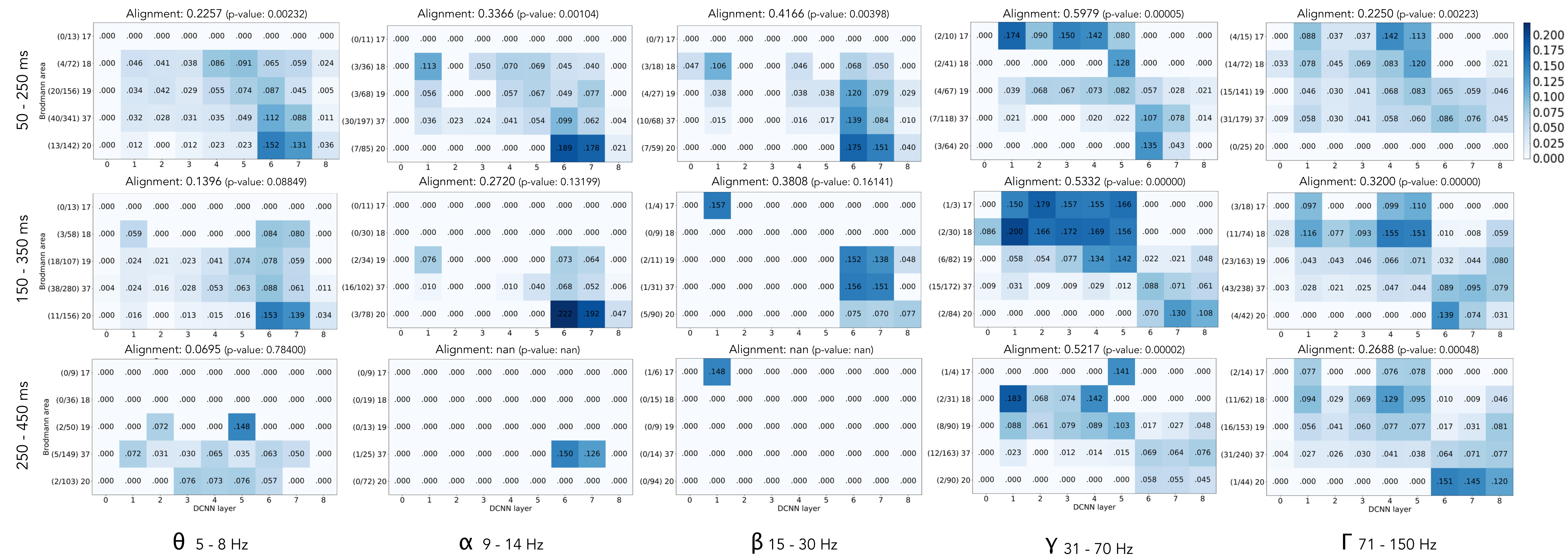}
    \caption{Mapping of activity in visual areas to activations of layers of DCNN across five frequency bands and three time windows. Vertical axis holds Brodmann areas in the order of the ventral stream (top to bottom), horizontal axis represents the succession of layers of DCNN. Number in each cell of a matrix is the total sum of correlations (between RDMs of probes in that particular area and the RDM of that layer) normalized by the number of significantly correlating probes in an area. The alignment score is computed as Spearman's rank correlation between electrode assignment to Brodmann areas and electrode assignment to DCNN layers (Equation \ref{eq:alignement}). The numbers on the left of each subplot show the number of significantly correlating probes in each area out of the total number of responsive probes in that area.}
    \label{fig:15_plates}
\end{sidewaysfigure}

In order to take into account the intrinsic variability when comparing alignments of different bands between each other, we performed a set of tests to see which bands have statistically significantly higher alignment with DCNN than other bands. See the section \ref{sec:mapping-brain-to-dcnn} for details. The results of those tests are presented in Table \ref{tab:pairwise-pvalues}. Based on these results we draw a set of statistically significant conclusions on how the alignment of neural responses with the activations of DCNN differs between frequency bands and time windows. In the low gamma range ($31-70$ Hz) we conclude that the alignment is larger than with any other band and that within the low gamma the activity in early time window $50-250$ ms is aligned more than in later windows. Alignment in the high gamma ($71 - 150$ Hz) is higher than the alignment of $\theta$, but not higher than alignment of $\alpha$. Within the high gamma band the activity in the middle time window $150-350$ ms has the highest alignment, followed by late $250-450$ ms window and then by the early activity in $50-250$ ms window. Outside the gamma range we conclude that theta band has the weakest alignment across all bands and that alignment of early alpha activity is higher than the alignment of early and late high gamma.

\begin{table}[h!]
    \vspace{0.3em}
    \centering
    \begin{tabular}{p{2.5cm}p{0.5cm}cp{6.8cm}}
        \normalsize$0.2079$ \tiny$\pm0.1381$\normalsize & $\theta^{50}$  & $>$ & - \vspace{0.2em}\\
        \normalsize$0.3352$ \tiny$\pm0.0989$\normalsize   & $\alpha^{50}$  & $>$ & $\theta^{50}$, $\Gamma^{50}$, $\Gamma^{250}$ \vspace{0.2em}\\
        \normalsize$0.5652$ \tiny$\pm0.1953$\normalsize   & $\gamma^{50}$  & $>$ & $\theta^{50}$, $\alpha^{50}$, $\gamma^{150}$, $\gamma^{250}$, $\Gamma^{50}$, $\Gamma^{150}$, $\Gamma^{250}$ \vspace{0.2em}\\
        \normalsize$0.4880$ \tiny$\pm0.1650$\normalsize   & $\gamma^{150}$ & $>$ & $\theta^{50}$, $\alpha^{50}$, $\Gamma^{50}$, $\Gamma^{150}$, $\Gamma^{250}$ \vspace{0.2em}\\
        \normalsize$0.4656$ \tiny$\pm0.2185$\normalsize   & $\gamma^{250}$ & $>$ & $\theta^{50}$, $\alpha^{50}$, $\Gamma^{50}$, $\Gamma^{150}$, $\Gamma^{250}$ \vspace{0.2em}\\
        \normalsize$0.2172$ \tiny$\pm0.1179$\normalsize   & $\Gamma^{50}$  & $>$ & -  \vspace{0.2em}\\
        \normalsize$0.3116$ \tiny$\pm0.1115$\normalsize   & $\Gamma^{150}$ & $>$ & $\theta^{50}$, $\Gamma^{50}$, $\Gamma^{250}$  \vspace{0.2em}\\
        \normalsize$0.2494$ \tiny$\pm0.1381$\normalsize   & $\Gamma^{250}$ & $>$ & $\theta^{50}$, $\Gamma^{50}$  \vspace{0.2em}\\
    \end{tabular}
    \caption{Comparison of the alignment across regions of interest. Alignment of the region of interest on the left is statistically significantly larger than the alignments of the regions of interest on the right. To obtain these results a pairwise comparison of the magnitude of alignment between the regions of interest was made. First column enlists significantly aligned regions, their average alignment $\rho$ score when estimated on 1000 random subsets of the data (each of the half of the size of the dataset), and standard deviation of the alignment. On the right side of the table we list the regions of interest of which the ROI on the left is larger. The hypothesis was tested using Mann-Whithney U test and only the results with the p-values that have passed the threshold of $2.2\mathrm{e}{-}5$ ($0.005$ Bonferroni corrected to take into account multiple comparisons) are presented in the table.}
    \label{tab:pairwise-pvalues}
\end{table}

\subsection{Alignment is dependent on having two types of layers in DCNN}

\setcounter{figure}{13}
\begin{sidewaysfigure}
    \centering
    \includegraphics[width=1.0\linewidth]{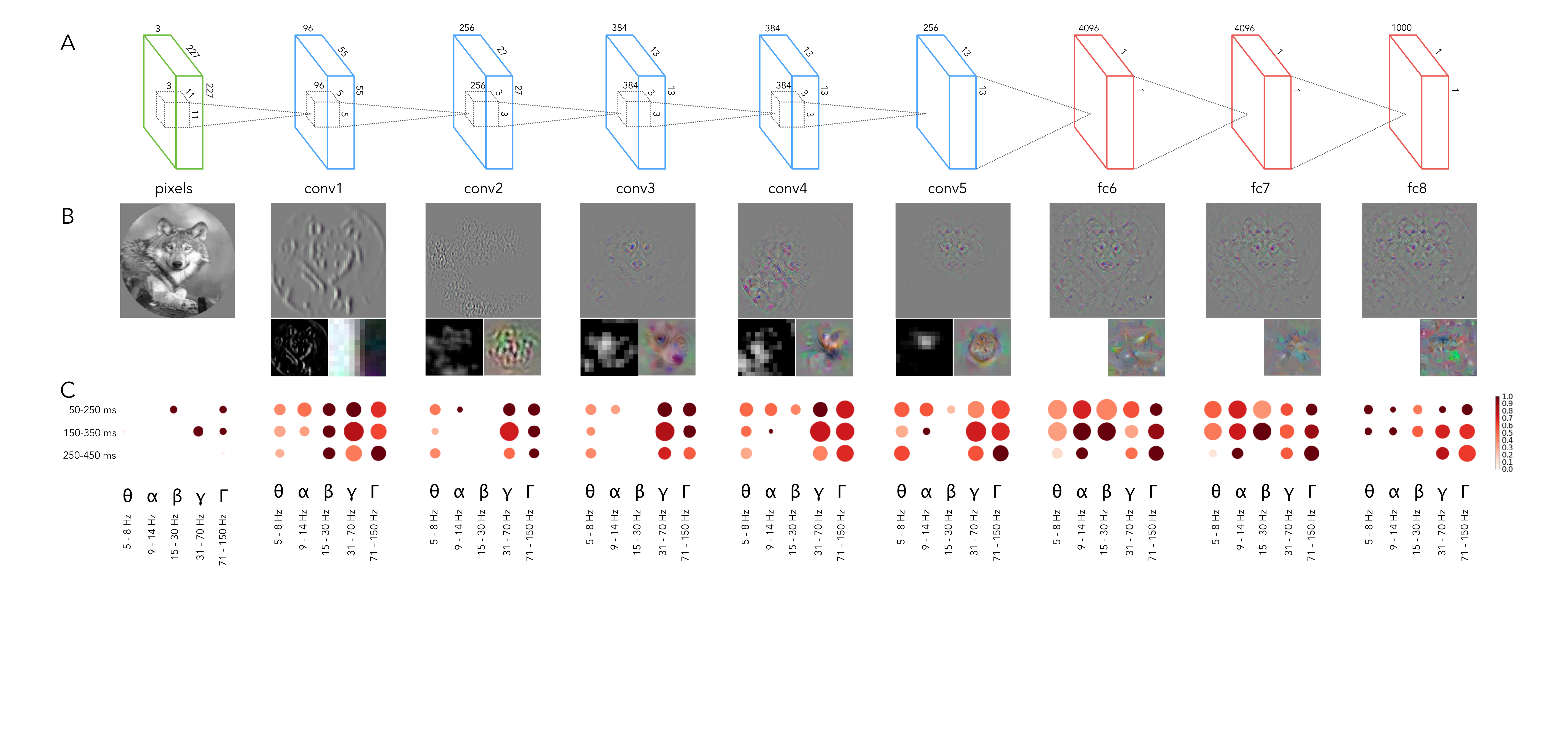}
    \caption{Specificity of neural responses to layers of DCNN across frequency bands and time windows. \textbf{a.} The architecture of the DCNN. Convolutional layer 1 consists of $96$ feature detectors of size $11 \times 11$, they take as input pixels of the image and their activations create 96 features maps of size $55 \times 55$, architecture of all consecutive convolutional layers is analogous. Five convolutional layers are followed by 3 fully-connected layers of sizes $4096, 4096$ and $1000$ respectively. \textbf{b.} The leftmost image is an example input image. For each layer we have selected one interesting filter that depicts what is happening inside of the neural network and plotted: a) a reconstruction of the original image from the activity of that neuron using the deconvolution \citep{zeiler2014visualizing} technique (upper larger image), (b) activations on the featuremap generated by that neuron (left sub-image) and (c) synthetic image that shows what input the neuron would be most responsive to (right sub-image). Visualizations were made with Deep Visualization Toolbox \citep{yosinski2015understanding}. All filters are canonical to AlexNet trained on ImageNet and can be explored using the above-mentioned visualization tool or visualized directly from the publicly available weights of the network. \textbf{c.} Specificity of neural responses across frequency bands and time windows for each layer of DCNN. Size of a marker is the total activity mapped to this layer and the intensity of the color is the specificity of the activity to the Brodmann areas constituting the ventral stream: BA17-18-19-37-20.}
    \label{fig:layer_specificity_and_volume}
\end{sidewaysfigure}

\begin{figure}[h!]
    \centering
    \includegraphics[width=0.7\linewidth]{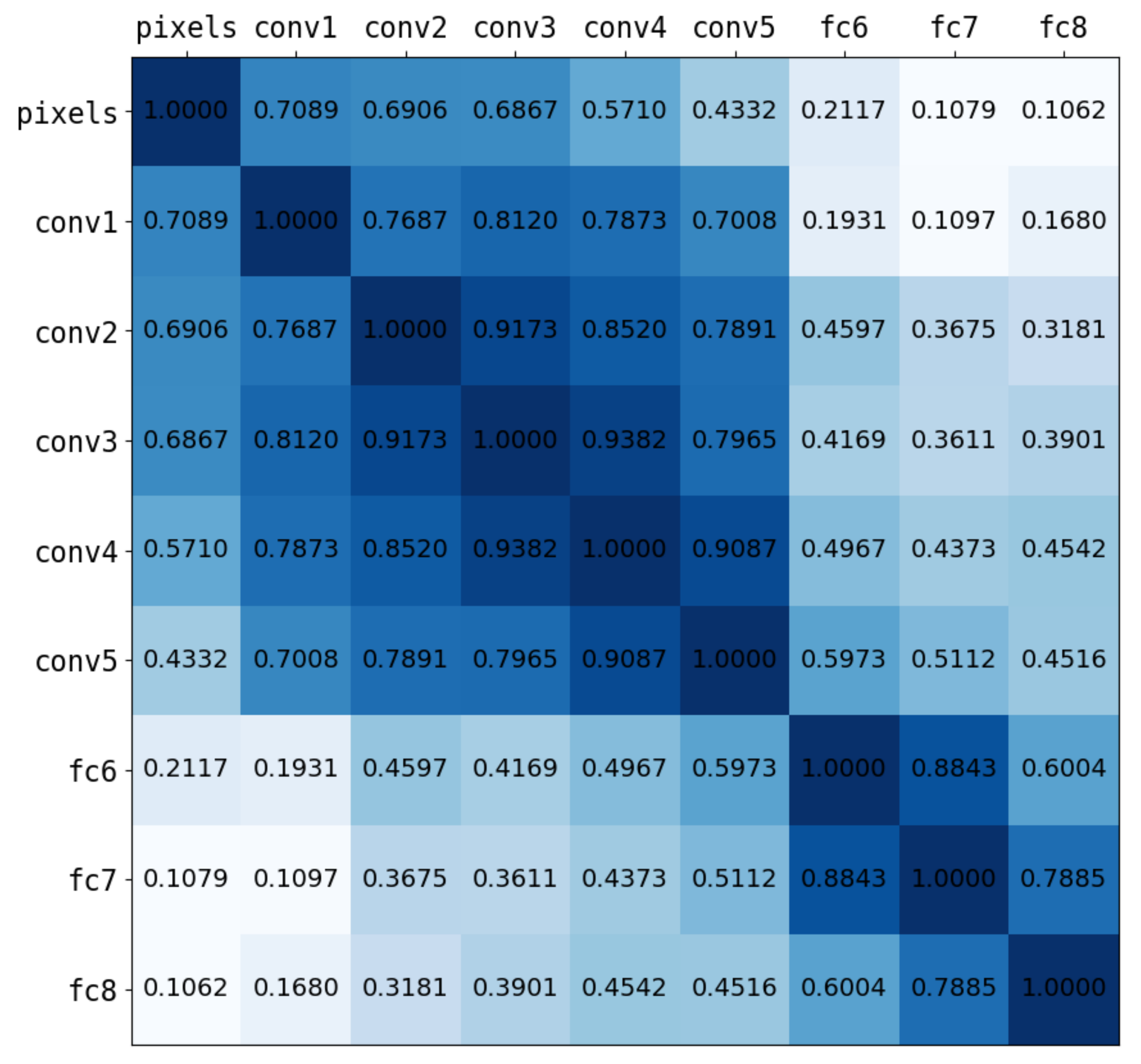}
    \caption{Correlations between the representation dissimilarity matrices of the layers of the deep convolutional neural network. All scores are significant.}
    \label{fig:rdm_corrs_within_dnn}
\end{figure}

On figures \ref{fig:all_areas_in_gammas} and \ref{fig:15_plates} one can observe that sites in lower visual areas (17, 18) are mapped to DCNN layers 1 to 5 without a clear trend but are not mapped to layers 6-8. Similarly areas 37 and 20 are mapped to layers 6-8 but not to 1-5. Hence we next asked whether the observed alignment is depending on having two different groups of visual areas related to two groups of DCNN layers. We tested this by computing alignment within the subgroups. We looked at alignment only between the lower visual areas (17, 18, 19) and the convolutional layers 1-5, and separately at the alignment between higher visual areas (37, 20) and fully connected layers of DCNN (6-8). We observed no significant alignment within any of the subgroups. So we conclude that the alignment mainly comes from having different groups of areas related more or less equally to two groups of layers. The underlying reason for having these two groups of layers comes from the structure of the DCNN -- it has two different types of layers, convolutional (layers 1-5) and fully connected (layers 6-8) (See Figures \ref{fig:layer_specificity_and_volume}a and \ref{fig:layer_specificity_and_volume}b for a visualization of the different layers and their learned features and a longer explanation of the differences between the layers in the \ref{sec:dcnn-discussion}). As can be evidenced on Figure \ref{fig:rdm_corrs_within_dnn} the layers 1-5 and 6-8 of the DCNN indeed cluster into two groups. Taken together, we observed that early visual areas are mapped to the convolutional layers of the DCNN whereas higher visual areas match the activity profiles of the fully connected layers of the DCNN.

\setcounter{joonis}{14}

\subsection{Visual complexity varies across areas and frequencies}
\setcounter{figure}{15}
\begin{sidewaysfigure}
    \centering
    \includegraphics[width=1.0\linewidth]{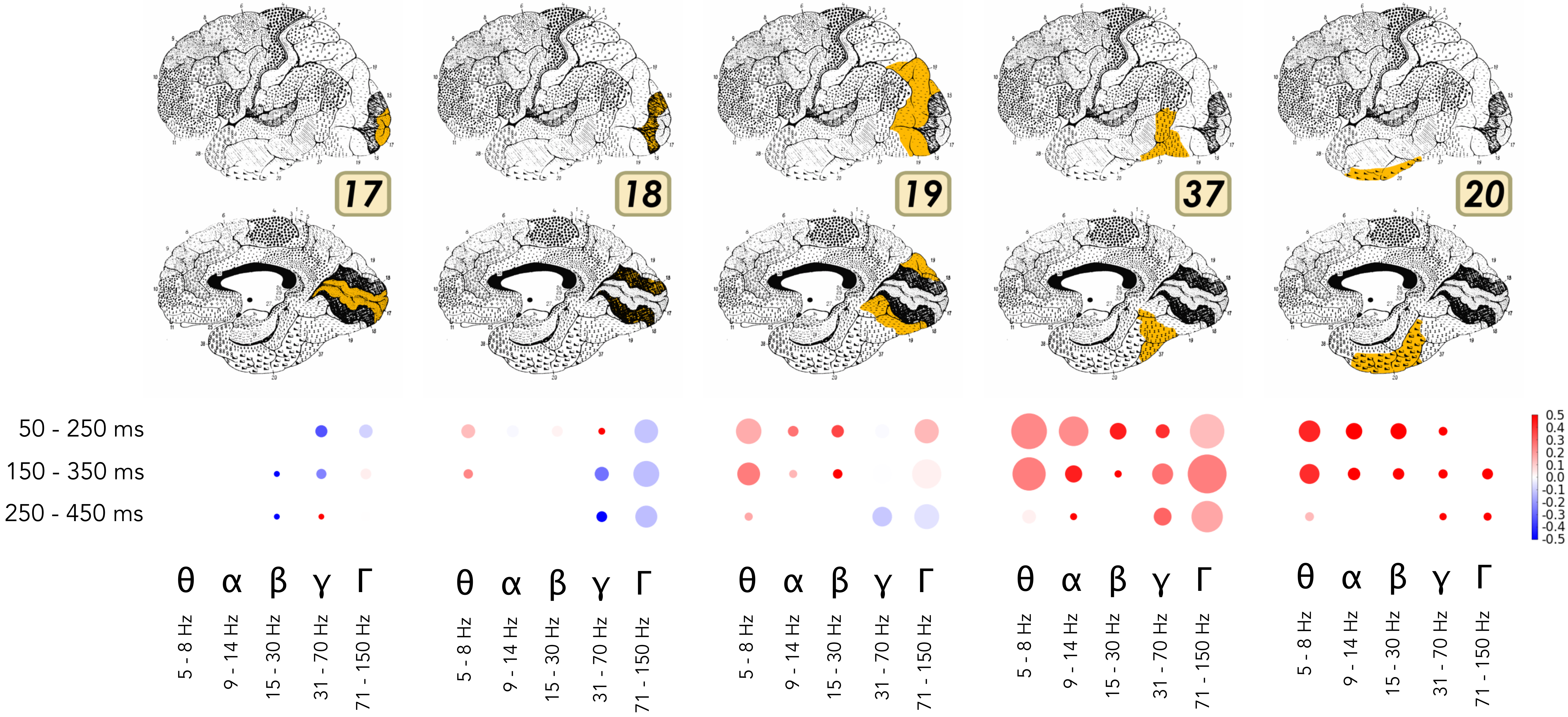}
    \caption{Area-specific analysis of volume of neural activity and complexity of visual features represented by that activity. Size of the marker shows the sum of correlation coefficients between the area and DCNN for each particular band and time window. Color codes the ratio of complex visual features to simple visual features, i.e. the comparison between the activity that correlates with the higher layers (\texttt{conv5, fc6, fc7}) of DCNN to the lower layers (\texttt{conv1}, \texttt{conv2}, \texttt{conv3}). Intense red means that the activity was correlating more with the activity of higher layers of DCNN, while the intense blue indicates the dominance of correlation with the lower areas. If the color is close to white then the activations of both lower and higher layers of DCNN were correlating with the brain responses in approximately equal proportion.}
    \label{fig:hhl_and_volume}
\end{sidewaysfigure}
\setcounter{joonis}{16}

To investigate the involvement of each frequency band more closely we analyzed each visual area separately. Figure \ref{fig:hhl_and_volume} shows the volume of activity in each area (size of the marker on the figure) and whether that activity was more correlated with the complex visual features (red color) or simple features (blue color). In our findings the role of the earliest area (17) was minimal, however that might be explained by a very low number of electrodes in that area in our dataset (less than 1\%). One can see on Figure \ref{fig:hhl_and_volume} that activity in theta frequency in time windows $50-250$ ms and $150-350$ ms had large volume and is correlated with the higher layers of DCNN in higher visual areas (19, 37, 20) of the ventral stream. This hints at the role of activity reflected by the theta band in visual object recognition. In general, in areas 37 and 20 all frequency bands reflected the information about high level features in the early time windows. This implies that already at early stages of processing the information about complex features was present in those areas.

\subsection{Gamma activity is more specific to convolutional layers}
We analysed volume and specificity of brain activity that correlates with each layer of DCNN separately to see if any bands or time windows are specific to particular level of hierarchy of visual processing in DCNN. Figure \ref{fig:layer_specificity_and_volume} presents a visual summary of this analysis. In section \ref{sec:quantifying-mapping} we have defined total volume of visual activity in layers $\mathbf{L}$ as $V_\mathbf{L}$. We used average of this measure over frequency band intervals to quantify the activity in low and high gamma bands. We noticed that while the fraction of gamma activity that is mapped to convolutional layers is high ($\frac{\bar{V}_{\mathbf{L} = \{\texttt{conv1} \ldots \texttt{conv5}\}}^{\gamma, \Gamma}}{\bar{V}_{\{\mathbf{L} = \texttt{conv1} \ldots \texttt{conv5}\}}^{\text{all bands}}} = 0.71$), this fraction diminished in fully connected layers \texttt{fc6} and \texttt{fc7} ($\frac{\bar{V}_{\mathbf{L} = \{\texttt{fc6}, \texttt{fc7}\}}^{\gamma, \Gamma}}{\bar{V}_{\mathbf{L} = \{\texttt{fc6}, \texttt{fc7}\}}^{\text{all bands}}} = 0.39$). Note that \texttt{fc8} was excluded as it represents class label probabilities and does not carry information about visual features of the objects. On the other hand the activity in lower frequency bands (theta, alpha, beta) showed the opposite trend -- fraction of volume in convolutional layers was $0.29$, while in fully connected it grew to $0.61$. This observation highlighted the fact that visual features extracted by convolutional filters of DCNN are more similar to gamma frequency activity, while the fully connected layers that do not directly correspond to intuitive visual features, carry information that has more in common with the activity in the lower frequency bands.

%
%
\section{Extending the methodology beyond the visual system}
\label{sec:dcnn-discussion}

The recent advances in artificial intelligence research have demonstrated a rapid increase in the ability of artificial systems to solve various tasks that are associated with higher cognitive functions of human brain. One of such tasks is visual object recognition. Not only do the deep neural networks match human performance in visual object recognition, they also provide the best model for how biological object recognition happens \citep{kriegeskorte2015deep, yamins2013hierarchical, yamins2014performance, yamins2016using}. Previous work has established a correspondence between hierarchy of the DCNN and the fMRI responses measured across the human visual areas \citep{gucclu2015deep, eickenberg2016seeing, seibert2016performance, cichy2016comparison}. Further research has shown that the activity of the DCNN matches the biological neural hierarchy in time as well \citep{cichy2016comparison, seeliger2017cnn}. Studying intracranial recordings allowed us to extend previous findings by assessing the alignment between the DCNN and cortical signals at different frequency bands. We observed that the lower layers of the DCNN explained gamma band signals from earlier visual areas, while higher layers of the DCNN, responsive for more complex features, matched with the gamma band signals from higher visual areas. This finding confirms previous work that has given a central role for gamma band activity in visual object recognition \citep{singer1995visual, singer1999neuronal, fisch2009neural} and feedforward communication \citep{van2014alpha, bastos2015visual, michalareas2016alpha}. Our work also demonstrates that the correlation between the DCNN and the biological counterpart is specific not only in space and time, but also in frequency.

The research into gamma oscillations started with the idea that gamma band activity signals the emergence of coherent object representations \citep{gray1989stimulus, singer1995visual, singer1999neuronal}. However, this view has evolved into the understanding that activity in the gamma frequencies reflects neural processes more generally. One particular view \citep{fries2005mechanism, fries2015rhythms} suggests that gamma oscillations provide time windows for communication between different brain regions. Further research has shown that especially feedforward activity from lower to higher visual areas is carried by the gamma frequencies \citep{van2014alpha, bastos2015visual, michalareas2016alpha}. As the DCNN is a feedforward network our current findings support the idea that gamma rhythms provide a channel for feedforward communication. However, our results by no means imply that gamma rhythms are only used for feedforward visual object recognition. There might be various other roles for gamma rhythms \citep{buzsaki2012mechanisms, fries2015rhythms}.

We observed significant alignment to the DCNN in both low and high gamma bands. However, when directly contrasted the alignment was stronger for low gamma signals. Furthermore, for high gamma this alignment was more restricted in time, surviving correction only in the middle time window. Previous studies have shown that low and high gamma frequencies are functionally different: while low gamma is more related to classic narrow-band gamma oscillations, high frequencies seem to reflect local spiking activity rather than oscillations \citep{manning2009broadband, ray2011different}, the distinction between low and high gamma activity has also implications from cognitive processing perspective \citep{vidal2006visual, wyart2008neural}. In the current work we approached the data analysis from the machine learning point of view and remained agnostic with respect to the oscillatory nature of underlying signals. Importantly, we found that numerically the alignment to the DCNN was stronger and persisted for longer in low gamma frequencies. However, high gamma was more prominent when considering volume and specificity to visual areas. These results match well with the idea that whereas high gamma signals reflect local spiking activity, low gamma signals are better suited for adjusting communication between brain areas \citep{fries2005mechanism, fries2015rhythms}.

In our work we observed that the significant alignment depended on the fact that there are two groups of layers in the DCNN: the convolutional and fully connected layers. We found that these two types of layers have similar activity patterns (i.e. representational geometry) within the group but the patterns are less correlated between the groups (Figure \ref{fig:rdm_corrs_within_dnn}). As evidenced in the data, in the lower visual areas (17, 18) the gamma band activity patterns resembled those of convolutional layers whereas in the higher areas (37 and 20) gamma band activity patterns matched the activity of fully connected layers. Area 19 showed similarities to both types of DCNN layers.

Convolutional layers impose a certain structure on the network’s connectivity -- each layer consists of a number of visual feature detectors, each dedicated to finding a certain pattern on the source image. Each neuron of the subsequent layer in the convolutional part of the network indicates whether the feature detector associated with that neuron was able to find its specific visual pattern (neuron is highly activated) on the image or not (neuron is not activated). Fully connected layers on the other hand, as the name suggests, connect every neuron of a layer to every neuron in the subsequent layer, allowing for more flexibility in terms of connectedness between the neurons. The training process determines which connections remain and which ones die off. In simplified terms, convolutional layers can be thought of as feature detectors, whereas fully connected layers are more flexible: they do whatever needs to be done to satisfy the learning objective. It is tempting to draw parallels to the roles of lower and higher visual areas in the brain: whereas neurons in lower visual areas (17 and 18) have smaller receptive fields and code for simpler features, neurons in higher visual areas (like 37 and parts of area 20) have larger receptive fields and their activity explicitly represents objects \citep{grill2004human, dicarlo2012does}. On the other hand, while in neuroscience one makes the broad differences between lower and higher visual cortex \citep{grill2004human} and sensory and association cortices \citep{zeki1993visual}, this distinction is not so sharply defined as the one between convolutional and fully connected layers. Our hope is that the present work contributes to understanding the functional differences between lower and higher visual areas. 
 
Visual object recognition in the brain involves both feedforward and feedback computations \citep{dicarlo2012does, kriegeskorte2015deep}. What do our results reveal about the nature of feedforward and feedback compoments in visual object recognition? We observed that the DCNN corresponds to the biological processing hierarchy even in the latest analysed time-window (Figure \ref{fig:diagonality_specificity_and_volume}). In a directly relevant previous work Cichy and colleagues compared DCNN representations to millisecond resolved magnetoencephalographic data from humans \citep{cichy2016comparison}. There was a positive correlation between the layer number of the DCNN and the peak latency of the correlation time course between the respective DCNN layer and magnetoencephalography signals. In other words, deeper layers of the DCNN predicted later brain signals. As evidenced on Figure 3 in \citep{cichy2016comparison}, the correlation between DCNN and magnetoencephalographic activity peaked between ca 100 and 160 ms for all layers, but significant correlation persisted well beyond that time-window. In our work too the alignment in low gamma was strong and significant even in the latest time-window 250-450 ms, but it was significantly smaller than in the earliest time-window 50-250 ms. In particular, the alignment was the strongest for low gamma signals in the earliest time-window compared to all other frequency-and-time combinations.

The present work relies on data pooled over the recordings from $100$ subjects. Hence, the correspondence we found between responses at different frequency bands and layers of DCNN is distributed over many subjects. While it is expected that single subjects show similar mappings (see also Supplementary Figure \ref{fig:single_plates}), the variability in number and location of recording electrodes in individual subjects makes it difficult a full single-subject analysis with this type of data. We also note that the mapping between electrode locations and Brodmann areas is approximate and the exact mapping would require individual anatomical reconstructions and more refined atlases. Also, it is known that some spectral components are affected by the visual evoked potentials (VEPs). In the present experiment we could not disentangle the effect of VEPs from the other spectral responses as we only had one repetition per image. However, we consider the effect of VEPs to be of little concern for the present results as it is known that VEPs have a bigger effect on low frequency components, whereas our main results were in the low gamma band. 

It must be also noted that the DCNN still explains only a part of the variability of the neural responses. Part of this unexplained variance could be noise \citep{gucclu2015deep,khaligh2014deep}. Previous works that have used RSA across brain regions have in general found the DCNNs to explain a similar proportion of variance as in our results \citep{cichy2016comparison,seibert2016performance}. It must be noted that the main contribution of DCNN has been that it can explain the gradually emerging complexity of visual responses along the ventral pathway, including the highest visual areas where the typical models (e.g. HMAX) were not so successful \citep{yamins2014performance, khaligh2014deep}. Recently it also has been demonstrated that the DCNN provides the best model for explaining responses to natural images also in the primate V1 \citep{cadena2017deep}. Nevertheless, the DCNNs cannot be seen as the ultimate model explaining all biological visual processing \citep{kriegeskorte2015deep, rajalingham2018large}. Most likely over the next years deep recurrent neural networks will surpass DCNNs in the ability to predict cortical responses \citep{kriegeskorte2015deep, shi2017deep}.

Intracranial recordings are both precisely localized in space and time, thus allowing us to explore phenomena not observable with fMRI. In this work we investigated the correlation of DCNN activity with five broad frequency bands and three time windows. Our next steps will include the analysis of the activity on a more granular temporal and spectral scale. Replacing representation similarity analysis with a predictive model (such as regularized linear regression) will allow us to explore which visual features elicited the highest responses in the visual cortex. In this study we have investigated the alignment of visual areas with one of the most widely used DCNN architectures -- AlexNet. The important step forward would be to compare the alignment with other networks trained on visual recognition task and investigate which architectures preserve the alignment and which do not. That would provide an insight into which functional properties of DCNN architecture are compatible with functional properties of human visual system.

To sum up, in the present work we studied which frequency components match the increasing complexity of representations of an artificial neural network. As expected by previous work in neuroscience, we observed that gamma frequencies, especially low gamma signals, are aligned with the layers of the DCNN. Previous research has shown that in terms of anatomical location the activity of DCNN maps best to the activity of visual cortex and this mapping follows the propagation of activity along the ventral stream in time. With this work we have confirmed these findings and have additionally established at which frequency ranges the activity of human visual cortex correlates the most with the activity of DCNN, providing the full picture of alignment between these two systems in spatial, temporal and spectral domains.

\chapter{State space visualization informs on representation of mental concepts in human brain}
\label{ch:mental-space-visualization-based}

Numerous studies in the area of BCI are focused on the search for a better experimental paradigm -- a set of mental actions that a user can evoke consistently and a machine can discriminate reliably. Examples of such mental activities are motor imagery, mental computations, etc. We propose a technique that instead allows the user to try different mental actions in the search for the ones that will work best. The system is based on a modification of the self-organizing map (SOM) algorithm and enables interactive communication between the user and the learning system through a visualization of user's mental state space. During the interaction with the system the user converges on the paradigm that is most efficient and intuitive for that particular user. Results of two experiments, one allowing muscular activity, another permitting mental activity only, demonstrate soundness of the proposed method and empirically validate the performance improvement over the traditional closed-loop feedback approach.

%
%
\section{The search for distinguishable mental patterns}

In many BCI experiments, participants are asked to perform certain mental actions. Consider an experiment, where a person is asked to control a point on a screen, and have it move to the left. In essense, the subject is requested to focus on a thought of ``moving the point leftwards''. This request is quite ambiguous -- should the user concentrate on the abstract notion of ``left'', engage in motor imagery or think about an unrelated concept? 

The problem of choosing the best kind of mental activity for BCI has been studied by \cite{curran2003learning, friedrich2012effect}. Most experiments first propose a particular paradigm and then evaluate its average effectiveness on a sample of users. Many paradigms have been evaluated this way~\citep{anderson1996classification, babiloni2000linear, alivisatos1997functional, allison2010toward, bacsar2007brain, cabrera2008auditory, chochon1999differential, curran2004cognitive}. As brain activity for a particular mental action differs across subjects \citep{miller2012individual, ganis2005understanding, tavor2016task}, any general paradigm will be suboptimal compared a user-specific one. In this work we propose a method that facilitates self-guided interactive search for a user-specific paradigm through communication between the user and the learning system. We demonstrate the feasibility of the approach on EEG recordings from two separate experiments on muscular and mental activity. The approach is general and does not depend on the neuroimaging method.

To achieve our goal we replace the traditional feedback~\citep{pfurtscheller2001motor} with a visualization of the feature space within which the underlying machine learning algorithm is operating. This visualization facilitates a `dialogue' between the learning algorithm and the user by visually explaining to the user why his current set of mental actions is suboptimal, which ones are being recognized well by the system and which ones should be changed. By exploring how their mental actions affect the visualization, a user can find a suitable mental action for each of the stimuli. The exploration of the mental state space can go for as long as needed to find mental actions that the user can produce consistently over time and that are distinguishable by the algorithm.

%
%
\section{BCI via topology-preserving visualization of the feature space}

\begin{figure}[h]
    \centering
    \includegraphics[width=1.0\linewidth]{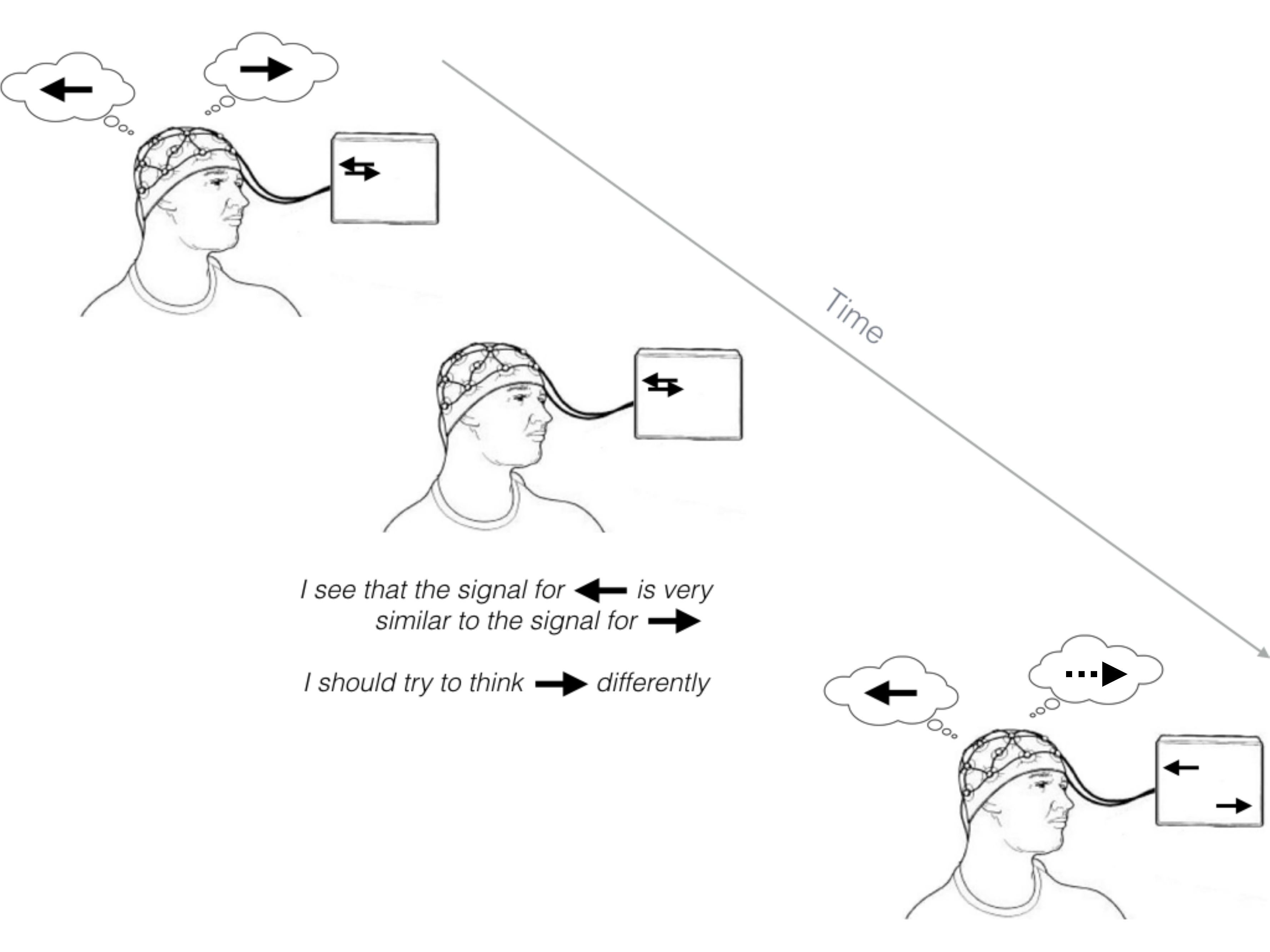} 
    \caption{Real-time interaction process between the system and the user, during which the user realizes that he must modify his mental activity for one of the actions to increase the system's performance.}
    \label{fig:interaction-process}
\end{figure}

At the core of almost any BCI system lies a machine learning algorithm that classifies user brain signal into desired actions~\citep{lotte2007review}. The algorithm sees the incoming data in a high-dimensional space and operates in that space. If an algorithm is unable to correctly discern the desired actions from the signal one can rely on visualization of the data and the space state to figure out why that is the case. Visualization allows to see particular data points in the context of other data, and allows to detect such issues as homogeneous data representation, failure to represent critical features of the data, biases in the data, insufficient flexibility of the feature space to present different data points differently, too high variance of the data points that should belong to the same group, and others. In the case of classification of mental actions we find that the two most important aspects a visualization could help evaluate are the cases where the data points from different classes look too much alike (one mental pattern is too similar to another) for the algorithm to differentiate between them, and the variance of the data within a class -- mental patterns that a user produces for the same action are not consistent and the algorithm is not able to group them together. With enough simplification we were able to present such a visualization to the user directly, allowing for a real-time evaluation of the above-mentioned issues during the training process. This allows the user to modify his mental activity in accordance with the feedback and try to produce more consistent and more distinguishable mental patterns. Figure~\ref{fig:interaction-process} depicts the interaction process between the user and the proposed feedback system.

Direct visualization of the space of mental signals provides more information to the user and allows to make more informed decisions than would be possible with the traditional approach \citep{pfurtscheller2001motor}. If in the case of usual system the subject has no information of why the system cannot distinguish the user's mental states, in the \emph{adaptive} paradigm, proposed in this work, the subject can see which mental actions are not being recognized or are colliding with others, previously attempted, mental states. The proposed framework naturally addresses a few limitations of the traditional approach, such as limited number of actions that can be trained simultaneously and makes a more efficient use of the training time by shifting training data class distribution towards more complicated cases. 

The concept described above poses several technological constraints on the choice of the underlying learning algorithm. To facilitate the real-time feedback loop the algorithm should work in an online setting and be fast enough to support real-time operation. In order to present the projection of the feature space to the user the algorithm must be compatible with topology-preserving dimensionality reduction techniques \citep{gisbrecht2015data}. In this section we describe a method that satisfies those requirements.

%
%
\subsection{Self-organizing map}
\label{sec:som}
Self-organizing map (SOM) \citep{kohonen1990self} is one of the topology-preserving dimensionality reduction techniques. These techniques try to preserve the relative distances through the transformation, such that the data points that were close in the original space remain close in the target space, and those that were apart, remain apart. SOM projects the data form the original space onto a \emph{map}, which is a collection of $m$ units organized into a multidimensional rectangular grid. Most commonly (and also in this work) a two-dimensional grid is used. Each SOM unit $u$ corresponds to a vector $\mathbf{w}_u \in \mathbb{R}^d$ in the space of input data points (signal samples from the EEG device, in our case). This way each unit effectively covers a region in the signal space. In this work the map has $625$ units ($25 \times 25$ square grid) with 630-dimensional vectors $\mathbf{w}$ initialized from uniform distribution $\mathcal{U}(0, 0.01)$.

The learning phase consists of updating vectors $\mathbf{w}$ with each new training sample $\mathbf{x}$. Once a new sample is obtained from a neuroimaging device the \emph{best matching unit} (BMU) for that sample is found according to Equation \ref{eq:bmu} with Euclidean distance used as the distance measure.
\begin{equation}
    \label{eq:bmu}
    BMU(\mathbf{x}) =  \underset{u\in\{1\ldots m\}}{\mathrm{argmin}}\ \mathrm{distance}(\mathbf{w}_u, \mathbf{x})
\end{equation}
Once BMU is found the weights $\mathbf{w}$ of unit $u$ and its neighbors are updated as shown in Equation \ref{eq:som-update}, where $s$ is the number of the current iteration.
\begin{equation}
    \label{eq:som-update}
    \mathbf{w}_u^{s + 1} = \mathbf{w}_u^s + \Theta(BMU, u, s) \alpha(s)(\mathbf{x} - \mathbf{w}_u^s)
\end{equation}
Default SOM is an offline learning algorithm that performs several passes over the training data. The update is thus repeated for each iteration $s \in \{1, \dots, S\}$, for each input data vector ($\mathbf{x}_1, \ldots, \mathbf{x}_n$) in the training set and for each unit in the map ($u_1, \ldots, u_m$). In total this procedure is being repeated up to $S \times n \times m$ times, where $S$ is the iteration limit, $n$ is the number of samples in the training data and $m$ is the size of the map. Not all units are updated with each new input vector, furthermore, not all units among the updated ones are updated equally. There are two functions in Equation (\ref{eq:som-update}), which are responsible for deciding which units will be updated and by how much. $\Theta(b, u, s)$ is called the \emph{neighborhood function}, it determines to what extent unit $u$ is neighbor of a unit $b$: for $b$ itself $\Theta(b, b, s) = 1$ and for some unit $u$, which is too far away to be considered to be a neighbor of $b$ $\Theta(b, u, s) = 0$. The parameter $s$ is used to decrease the number of neighbors on later iterations. The function $\alpha(s)$ outputs \emph{learning rate} that decreases with more iterations allowing the learning process to converge.

At the end of the learning process the units of the map represent centers of signal clusters in the training data. Each new data sample can be assigned to one of the clusters and this cluster will hold samples that are similar. The samples that are close in the original space will be assigned to map units that are close to each other on the map.

%
%
\subsection{Predictive online SOM}

We extend SOM to work in an online setting \citep{deng2000esom, somervuo2004online}, where the map is updated only once with each new data sample. We also assign a vector of probabilities $\mathbf{p}_u \in \mathbb{R}^C$ to each unit $u$ and use that vector to classify each new incoming data sample into one of $C$ classes. The class probability vector $\mathbf{p}$ of unit $u$ of Predictive Online SOM (POSOM) is initialized to a random vector of length $C$ with values sampled from uniform distribution $\mathcal{U}(0.0, 0.2)$. This vector holds action probability distribution for the unit $u$. It shows what is the probability that a signal $\mathbf{x}$, which was classified into unit $u$, has been produced in response to action $a$.

\begin{figure}[htb] 
    \centering
    \includegraphics[width=0.47\linewidth]{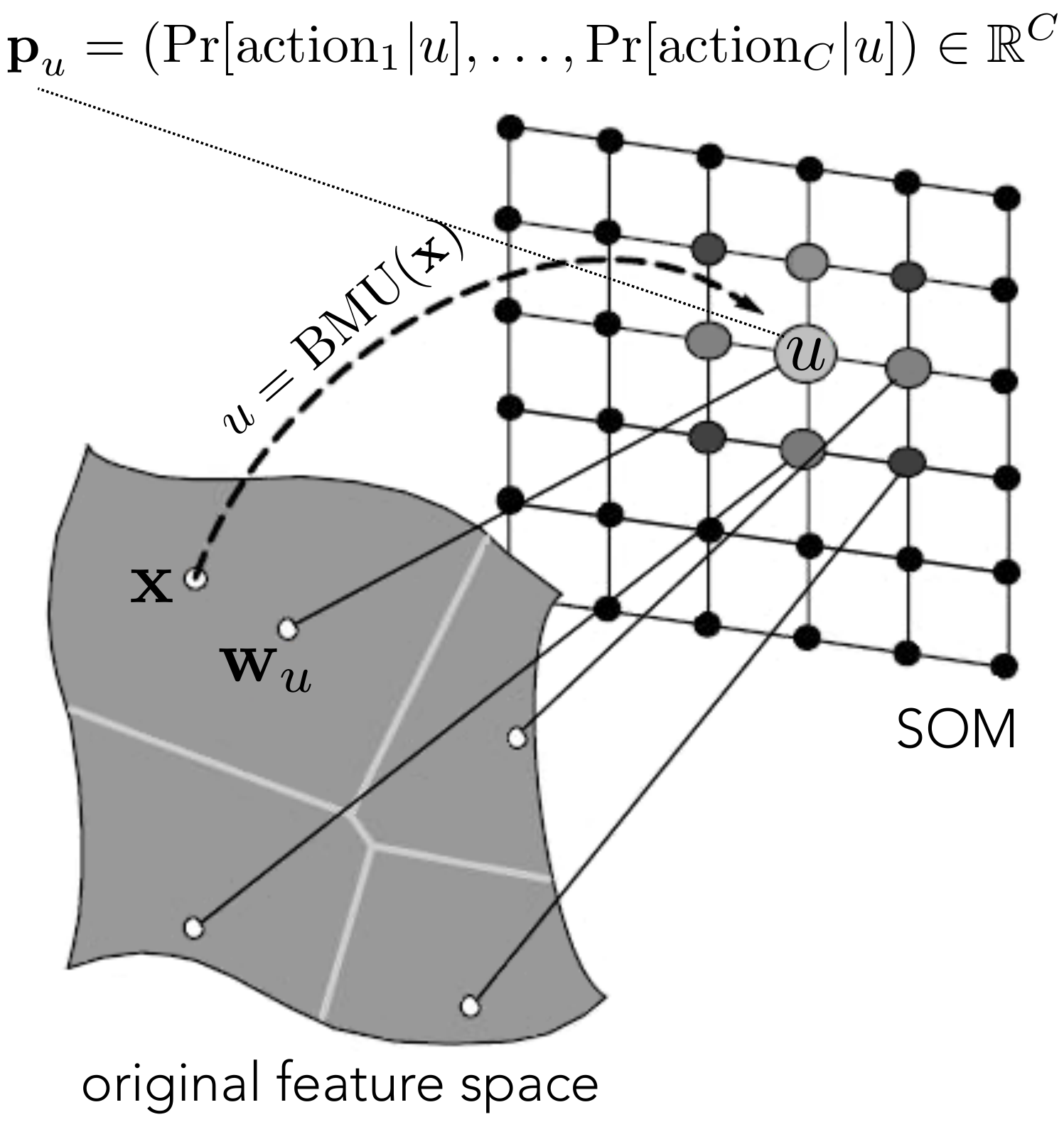} 
    \caption{SOM extended with class probability vectors. Signal representation $\mathbf{w}_u$ in the original feature space is mapped to a unit $u$ on a two dimensional map. This unit represents a cluster of signal samples similar to $\mathbf{w}_u$, such as sample $\mathbf{x}$. Unit $u$ holds a vector of class probabilities $\mathbf{p}_u$ that shows to which class a sample assigned to the cluster with centroid $u$ most probably belongs.}
    \label{fig:som-predictive}
\end{figure}

Class probability vectors are updated after each sample according to Equation \ref{eq:pred-som-update},
\begin{equation}
\label{eq:pred-som-update}
\mathbf{p}^{s+1}(u) = \mathbf{p}^s(u)(1 - \alpha) + \mathbf{c}\alpha
\end{equation}
where $s$ is iteration number, $\alpha\in(0,1)$ is a parameter, which specifies how fast the contribution of the older data samples deteriorates, and $\mathbf{c}$ is a bit vector, where for each class we have value 0 or 1. There can be only one non-zero value in the vector $\mathbf{c}$ and its position indicates the true class of a sample.

The probability vector $\mathbf{p}_u$ is used for classification as follows: for each new sample $\mathbf{x}$ we first identify POSOM's BMU $u$ for this sample, and predict the class of this sample by choosing the most probable class in the vector $\mathbf{p}_u$.

%
%
\subsection{POSOM-based BCI training system}
The learning method defined in the previous section satisfies all of the requirements of a system with an interactive feedback based on visualization on user's mental state space we have outlined in the beginning of the chapter.

\begin{figure}[htb]
    \centering
    \includegraphics[width=1.0\linewidth]{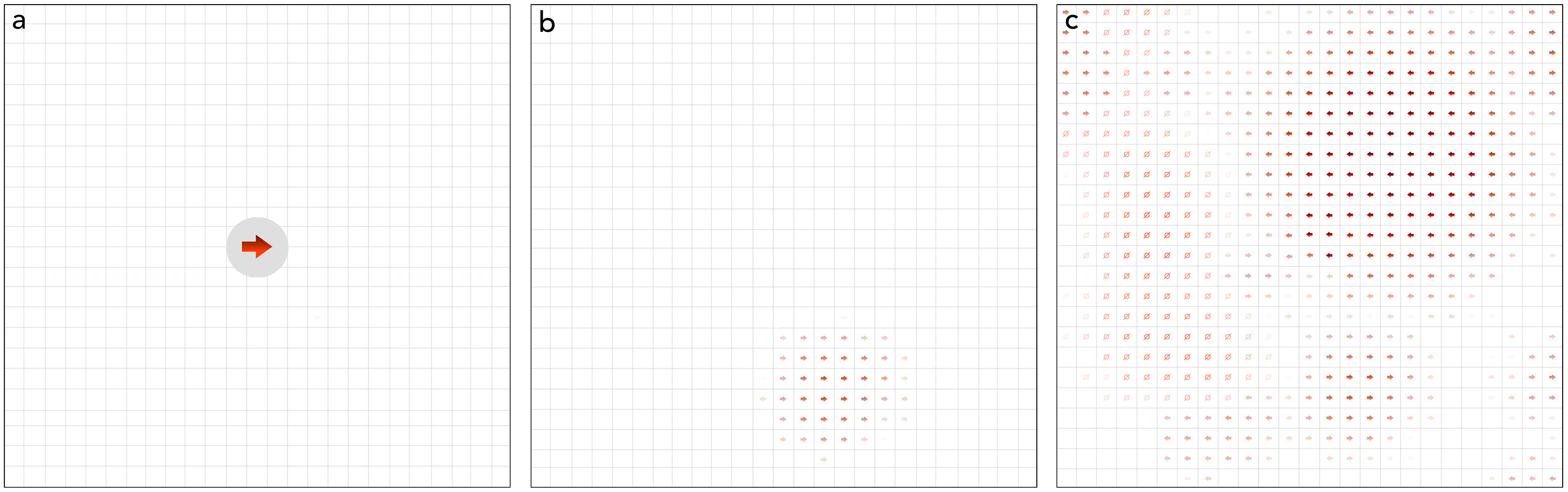}
    \caption{The process of forming the map. \textbf{a:} Visualization of an empty SOM and the very first stimulus cue. \textbf{b:} The first few samples are collected and assigned to the units on the map. \textbf{c:} Repeating steps (a) and (b) for all stimuli multiple times results in a map, where units are assigned mental state representations and corresponding actions.}
    \label{fig:posom-for-bci}
\end{figure}

The training process begins by presenting an empty SOM map to the user (Figure \ref{fig:posom-for-bci}a). A stimulus cue is displayed for a brief period of time and the system starts receiving samples from the neuroimaging device. It finds the best matching unit $u$ for each of the samples and updates the $\mathbf{w}_u$ and $\mathbf{p}_u$ vectors of the unit $u$ and its neighbours. Some of the randomly initialized SOM units now represent certain mental patterns and are mapped to corresponding actions, the action each unit is associated with is shown with a pictogram on the map (Figure \ref{fig:posom-for-bci}b).

The process continues until the user is satisfied with his ability to produce the required set of patterns consistently and system's ability to assign these patterns to correct units on the map (Figure \ref{fig:posom-for-bci}c). The user can see on the map which of the mental patterns are always assigned correctly and which ones are `jumping' across the map. This informs the user about the variance of a mental pattern, if the variance is too high it might be best to switch to another mental pattern instead. If a user can see that the patterns of two or more different actions tend to occupy the same region of the map he can conclude that the mental patterns he is producing for these actions are not different enough to be distinguishable and he should consider replacing one or all of them.

%
%
\section{Experimental validation on brain-computer interface for control}

Each of 5 test subjects completed a set of four experimental runs to compare maximal achievable classification accuracy of \emph{adaptive} (proposed) versus \emph{control} approach under two different conditions. In the first pair of experiments subjects were allowed to engage facial muscles to achieve the control of the system more easily. In the second pair only mental activity was allowed and the subjects were instructed to rely on mental imagery to control the system.

Subjects were seated in front of a computer screen in a quiet room with no distractions. All subjects had normal or corrected to normal vision. In all of experiments subjects were presented with 3 different stimuli (\texttt{left}, \texttt{right} and \texttt{none}) and were asked to engage different mental (or, in the case of the experiment where facial expressions were allowed, muscular) activity for each stimulus. A stimulus was shown for 7 seconds. Subjects were briefed on the usual mental paradigms including motor imagery~\citep{pfurtscheller2001motor, hwang2009neurofeedback}, vivid visual imagery~\citep{marks1973visual, neuper2005imagery}, mental computations~\citep{chochon1999differential} and auditory recollections~\citep{cabrera2010comparison}.

\begin{figure}[htb]
    \centering
    \includegraphics[width=1.0\linewidth]{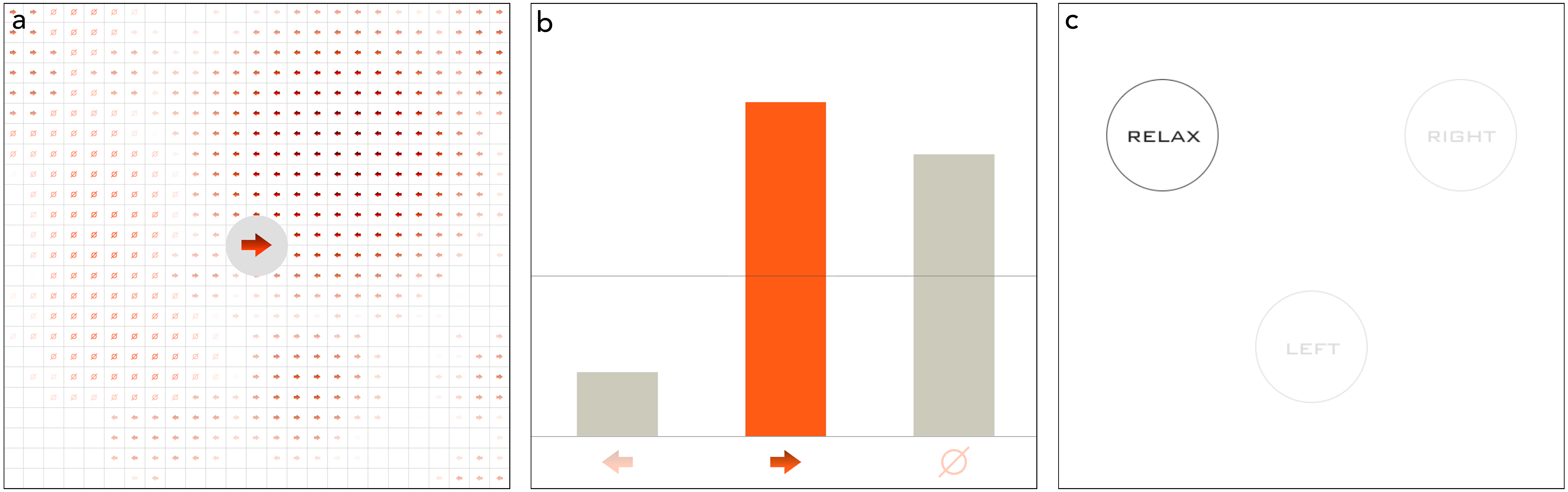}
    \caption{Interface of the experiment. \textbf{a:} In adaptive experiment the user is presented with a grid that visualizes 2D projection of decoder's feature space. The grid is updated with each new data sample received from the neuroimaging device, enabling the user to see how his mental actions affect the mental state space representation in real time. Cue to the next stimulus is shown in the center of the screen and disappears after 1 second, allowing the user to see the full grid. \textbf{b:} The control experiment provides the feedback by raising or lowering per-class performance bar, indicating which stimuli are being recognized well by the system. \textbf{c:} The decoding models resulting from both adaptive and control experiments are tested with the same interface, where a user is presented with a description of the mental activity he must engage in. We do not use the same cues as during training to measure the ability of the user to engage the metal activity associated with the action and avoid measuring involuntary reaction to the cue image.}
    \label{fig:interface}
\end{figure}
\ \\
The sequence of the experimental runs each test subject has completed was as follows:
\begin{enumerate}

    \item Training of the classification model in the traditional way. Stimuli were presented in a random order for 7 seconds each, for total time of 7 minutes, keeping the number of times each stimulus is shown balanced. The test subject received real-time feedback by observing the height of the performance indicator that was changing with each new data sample. Currently highlighted bar is the current action, height of the bar indicates the performance (Figure \ref{fig:interface}b). 
    
    \item Testing the traditional model. To avoid measuring the involuntary reaction to the cue image the user interface of the testing stage was different from the training stage and is shown on Figure \ref{fig:interface}c. Currently highlighted stimulus is the one the user should engage in. Stimuli were shown for 7 seconds in random order for a total length of 3 minutes.
    
    \item Training of the classification model in adaptive way. The user was presented with a visualization of the projection of the feature space onto 2D grid (Figure \ref{fig:interface}a). Each stimulus is shown for 7 seconds, the duration of the experiment was not fixed to allow the subject to test different mental activities for the same action until the one that works is found. The stimuli were presented in the order of their performance rate, the actions that have the lowest score are shown more frequently.
    
    \item Testing of the adaptively trained model. The procedure repeats the steps outlined in (2) exactly, making the testing runs comparable.

\end{enumerate}
\ \\
Upon finishing the trials the test subjects were asked of their subjective evaluation of the adaptive system in comparison with the traditional one and whether they were able to feel the interaction with the system and its efforts to adapt to test subject's mental efforts.

\subsection{Preprocessing of EEG data}
The data was recorded using the Emotiv EPOC \citep{stytsenko2011evaluation} consumer-grade EEG device. Signal from all 14 EEG channels was split into 1000 ms windows with a step size of 250 ms. Each 1000 ms recording was linearly detrended and converted to frequency domain using fast Fourier transform \citep{welch1967use}. Frequencies outside the 1 to 45 Hz range were excluded from further analysis.

A 1000 ms signal from one channel was represented by 45 frequency power measurements. By concatenating representations of all 14 channels we obtained feature representation of a signal with 630 dimensions. In machine learning terms a \emph{sample} $\mathbf{x}$ that represents 1000 ms of EEG recording has 630 \emph{features} and a categorical class label.

%
%
\section{Feedback based on mental state space visualization leads to higher decoding accuracy}
We have conducted two types of experiments to empirically validate the benefits of the adaptive search for mental BCI paradigm via visual exploration of a projection of subject's mental state space. During the first experiment the test subjects were allowed to engage in facial muscle activity in response to the stimuli \citep{huang2006application, heger2011online}. The second experiment was aimed at controlling the system via mental efforts only. In both experiments the proposed approach demonstrated statistically significant improvement in performance over the traditional method. Average performance of the model trained on facial expressions in traditional way was 23\% higher (Mann-Whitney U test $p = 0.006$) than that of the traditional approach. For the mental actions the adaptive approach resulted in a model that was significantly and consistently higher than the chance level (F1 score = $0.422$), while traditional approach failed to deliver a non-trivial model (F1 score = $0.354$). Comparatively, the adaptive approach yielded 19\% higher performance (Mann-Whitney U test $p = 0.018$).

\begin{figure}[h!]
    \centering
    \includegraphics[width=0.25\linewidth]{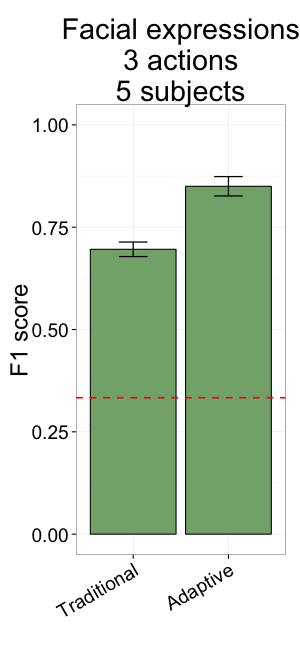}
    \includegraphics[width=0.25\linewidth]{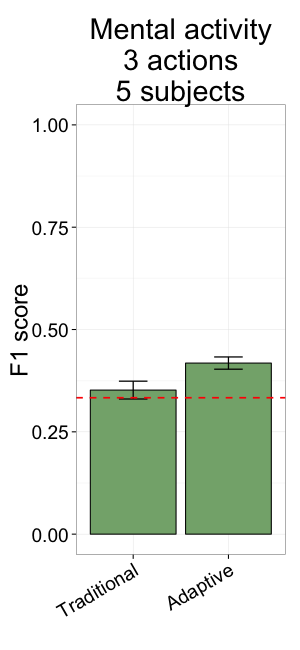}
    \caption{Average result of the experiments on facial expressions (left) and mental activity (right). In both cases adaptive approach demonstrates statistically significant improvement over the traditional approach.}
    \label{fig:results}
\end{figure}

\begin{figure}[h!]
   \centering
   \includegraphics[width=0.49\linewidth]{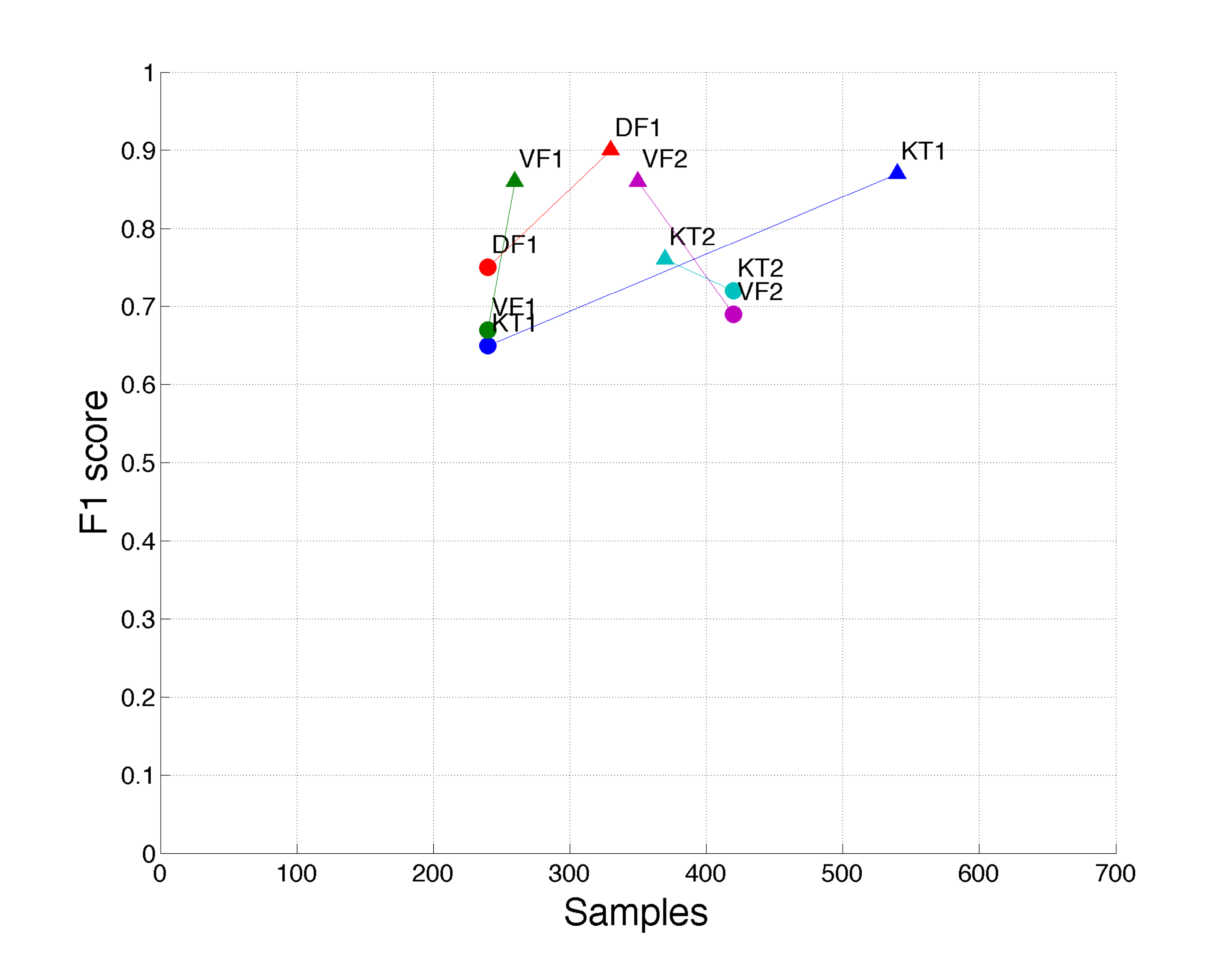}
   \includegraphics[width=0.49\linewidth]{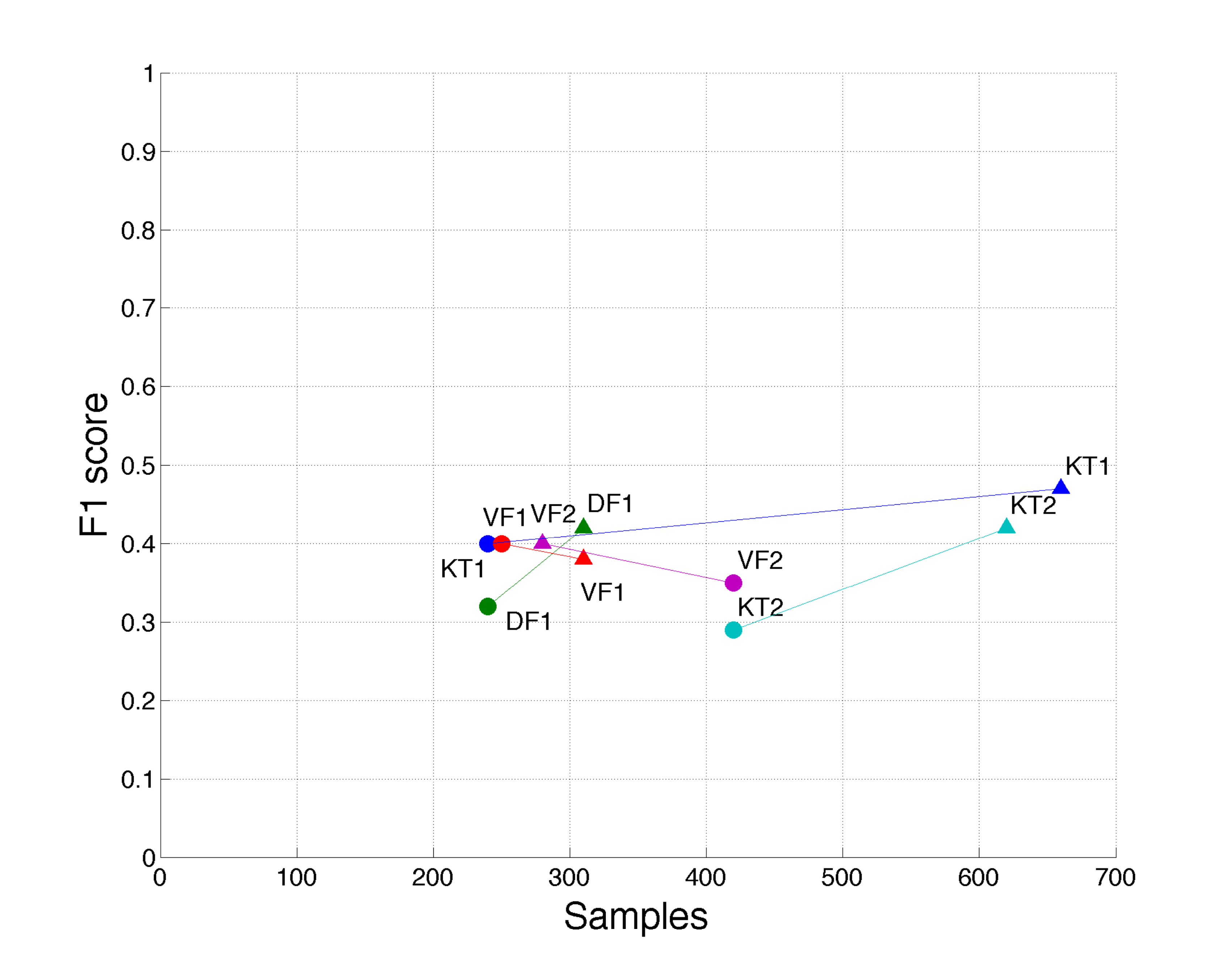}
   \caption{Details of performance on 3-class control problem. \textbf{Left:} 3-class training results using facial expressions. Circular markers denote the results achieved using the traditional approach and the triangular ones denote the adaptive approach. Each test subject is marked with a different color. On the $x$-axis we can see the number of samples the algorithm needed to reach the F1 score displayed on the $y$-axis. Traditional experiments were ran for 240 samples, or, if a subject felt that he would benefit from longer interaction with the system, the experiment was extended to 420 samples. \textbf{Right:} 3-class training results using power of thought via traditional (circle) and interactive (triangle) approach. Horizontal axis shows the number of sample is took to train the model, while the vertical one indicates the performance of the final model on the test run. The experiment continued for as long as test subject felt necessary.}
   \label{fig:real-results-facial-mental}
\end{figure}

Figure \ref{fig:real-results-facial-mental} (left) presents the detailed analysis of the results of the experiments involving facial expressions. Face muscle activity highly affects the EEG readings \citep{d1974contamination} and can be observed with the naked eye even on the raw signal. The primary goal of this series of experiments was to demonstrate the benefit of the adaptive approach in a high signal to noise ratio (SNR) scenario. Compared to the facial expressions experiment the task of distinguishing mental states was much harder \citep{haynes2006neuroimaging}. Since the effect of changing the activity was not immediately evident, it required more time for the test subjects to begin to understand how their efforts affect the learning system. Figure \ref{fig:real-results-facial-mental} (right)  shows the detailed results of the experiment.

%
%
\section{The general quest of navigating human mental state space}

In this work we have proposed, implemented and empirically tested an alternative approach to closed loop feedback BCI training that relies on the visualization of test subject's mental state space in real time facilitating the interaction with the learning system. In the traditional approach the feedback serves only one purpose -- to inform the test subject on his current performance. We expand the information available to the subject by allowing him to see not only that his current actions work poorly (or well), but also provide him with an insight into why a particular set of mental actions might not be a suitable one. We then provide him with an interactive way to experiment with other mental actions until he finds the ones that he can engage consistently and that are distinguishable by the learning system. By the sheer fact of sharing more information with the user we expect our system to achieve better performance. By facilitating the interactive training process we enable the test subject, given enough effort, to reach the desired level of performance.

In addition to the primary benefit described above we find that a few other properties of our approach are beneficial for training BCI systems, namely: the resulting paradigm is personalized to each particular test subject and thus can be tuned better than a one-fits-all paradigm such as motor imagery; system automatically takes care of failed trials and mistakes on the test subject's side -- a subject can rectify a mistake via further interaction with the system, the failed record does not taint the dataset forever, but is gradually phased out by further training; flexibility in training time allows to deviate from strict stimulation schedule and allows the test subject to focus on the most problematic actions, giving them more attention if more attention is needed. 

We would like to highlight the choice of the testing paradigm we employed in our work. We find that testing of a general-purpose BCI system must be decoupled from the training in terms of visual cues and protocol. This is necessary to avoid training the subject to simply react to the visual cues and not engage in the corresponding mental activity. By changing the cues we ensure that during the test time the test subject has to invoke the mental activity corresponding to each particular action. Such approach makes the resulting model more robust in the context of real-world applications.

We acknowledge the shortcoming of this study, such as low number of test subjects and a low-end EEG device. This work serves the purpose of initial validation of the concept that allows to plan a larger study, ideally involving intracranial neuroimaging techniques that would have sufficient SNR to make the approach applicable for real-world applications. Rectifying the above-mentioned issues and further exploring the possible topology-preserving dimensionality reduction techniques such as t-SNE \citep{maaten2008visualizing} and neural networks-based solutions are the primary directions for our future work.

The proposed methodology of visualizing test subject's mental state space in real time has wider applicability than BCIs. It gives a user the opportunity to roam the visualization space that is topologically consistent with the space of representation of user's mental signals. This means that if two mental state are close in terms of the brain activity they generate, they will also be close visually, which allows the user to see which thought, emotions, motor actions, and other activities that involve brain activity are close together and which are further apart. The basis for this approach lies in the ability of machine learning interpretability tools to explain what an artificial model is seeing to a human observer. Supplying the proposed system with a high quality neuroimaging device (such as intracortical electrode system) would allow a researcher to gain better understanding of the space of neural signals, or a general user to explore their own mind. Such use of this methodology could lead to new realizations about how human brain works and to new applications of neural technology.

\chapter*{Conclusion}\addcontentsline{toc}{chapter}{Conclusion}

Traditionally, neuroscience was and to a large extent is a hypothesis-driven science. The growing amount of data that is being generated by modern neuroimaging techniques merits, however, an increase in the role that exploratory, data-driven approach plays in modern neuroscience. In this thesis we make the case for the importance of adopting methods of interpretability of machine learning models in neuroscientific research. We discuss the benefit that machine learning brings by augmenting the ways of neuroscientific inquiry with an additional path of automatic hypothesis generation and validation.

The ultimate purpose of proposing new hypotheses and models of brain function is to discover the ones that describe the phenomenon well. In the hypothesis-driven approach most of the process relies on a human investigator, who first observes a certain phenomenon, then comes up with a model or a hypothesis to describe it, collects the data, validates the model based on the data, and, finally, rules the model to be true or insightful or discards it. In this scenario the role of automated data processing is confined to the process of obtaining and processing the data to provide measures on certain narrowly-defined experimental metrics that the investigator needs to reach a conclusion. This approach allows the human investigator to be in full control of the meaning of the model that is being created, but scales up only by increasing the number of investigators, naturally limiting the space of possible hypothesis that is humanly possible to test against the data. In the data-driven approach the process starts with the dataset that contains the observations of a phenomenon. Machine learning methods then automatically generate models (hypotheses) that attempt to explain the dynamics that was captured by the data. Model validation step automatically discards most of the models that merely capture shallow statistical dependencies in the particular data instance that was recorded and do not generalize to capture the underlying process. Some of the models, however, do, and, when validated, show to generalize well to correctly describe previously unseen data from the same phenomenon. When this happens we know that the process of automatic modeling has captured a description of the process that governs the phenomenon. All of the steps leading to this stage can be done with minimal human involvement and scale up with the amount of computational resources. This allows the space of models that are proposed and tested to be considerably larger than if we would only use humans to sift through the possible explanations. Now the human effort can be concentrated on the models and hypothesis that were identified by the automatic process as descriptive and general. Interpreting those models will show which ones are true and insightful and which ones are trivial.

The hypothesis-driven and data-driven approaches to hypothesis generation cover different parts of the conceptual space of unknown hypotheses and should both be exploited to advance our knowledge of the brain. The data-driven approach is designed to excel in the exploratory analysis, and, given the above-mentioned volumes of data, such exploration of this data has ever-growing chance of making a discovery. To properly facilitate this process, the interpretability techniques that have established their role in general machine learning community have to find their way into neuroscience research on a wider scale. We hope this thesis contributes to this process.

The exploration of the symbiosis between the fields of neuroscience and machine learning in Chapter \ref{ch:ml-ns-synergy} establishes the already existing and also the emerging track record of mutual benefit those two fields have provided each other. We find that one of the ways this benefit can prosper further is by adopting the view presented in Chapter \ref{ch:ml-for-modeling} of the machine learning approach playing the role of a builder of computational neurophysiological models. The need to interpret the knowledge the model has acquired and to articulate it in an intuitive manner leads us to the interpretability techniques. The need for a better understanding of the knowledge representation that artificial learning algorithms create require a structured approach to navigating the space of those representations. In Section \ref{sec:representation-taxonomy} we propose a taxonomy of machine learning methods based on knowledge representation and hope that this view angle proves useful when designing next neuroscientific study that involves machine learning methods.

The work that became the basis of this thesis serves as an example of adopting the proposed perspective and methodology and demonstrates its applicability on three different levels of organization. In Chapter \ref{ch:spectral-signatures-based} interpretable machine learning model is used to analyze neural dynamics at the level of localized activity across the human brain. This analysis allowed to characterize neural locations and their activity in during the task of visual perceptual categorization. The uncovered signatures of visual processing in human brain provided a multifaceted view on spectral, temporal and anatomical characteristics of this process. The comparison between biological and artificial systems of vision in Chapter \ref{ch:intracranial-dcnn-based} gives an example of the role machine learning models can play at a more abstract level, where the aim is to understand the functional organization of the human brain. In the last study described in Chapter \ref{ch:mental-space-visualization-based} the dimensionality reduction and visualization techniques provide an actionable insight into relative organization of mental concepts within a subject's mental state space. Visualizing the mental state space allows to analyze the behavior of our brain at the highest level of abstraction.

Taken together, the ideas and the results of this thesis highlight one of the roles machine learning could play in advancing our understanding of the human brain. The ability to uncover patterns and extract knowledge from data makes the method of machine learning a suitable tool for augmenting our capacity to create explanations of natural phenomena around us. Neuroscience is a particularly fitting area for application of this methodology due to its symbiosis with the area of artificial intelligence and machine learning. The shared goal of uncovering the mechanism of intelligence made the field of artificial intelligence follow and reapply the discoveries made in neuroscience. In some cases this has led to the realization that both systems, the biological and the artificial ones, if presented with the same functional goal, sometimes develop similar mechanisms of achieving it. The similarities between the mechanisms employed by biological and artificial systems that were discovered to this date, such as hierarchy of the visual system, the mechanism of periodic memory consolidation, grid-like representation of space for navigation, and others endorse the fact that an artificial learning system can emerge a mechanism that is similar to the one that is used by our brain. In this thesis we stress the importance of continued analysis of the ways how machine learning algorithms achieve their results as understanding of these mechanisms can shed light on the mechanisms employed by our brain.
\ \\

\emph{We hope you have found the perspective curious and the examples convincing enough to let the proposed approach occupy a part of your mental state space.}

\addcontentsline{toc}{chapter}{Bibliography}
\printbibliography

\appendix
\renewcommand\chaptername{Appendix}
\chapter{Code and data}

All the pre-processed spectrotemporal data that supports the findings reported in Chapter \ref{ch:spectral-signatures-based} are available for download under Academic Free License 3.0 from \url{https://web.gin.g-node.org/ilyakuzovkin/Spectral-Signatures-of-Perceptual-Categorization-in-Human-Cortex}. The code that was used to produce this data from the raw recordings and to perform all of the subsequent analysis steps is available at \url{https://github.com/kuz/Spectral-signatures-of-perceptual} \url{-categorization-in-human-cortex}.

The activations of biological and artificial systems of vision that are the bases for comparison and mapping reported in Section \ref{ch:intracranial-dcnn-based} are available for download under Academic Free License 3.0 from \url{https://web.gin.g-node.org/ilyakuzovkin/Human-Intracranial-Recordings-and-DCNN-to-Compare-Biological-and-Artificial-Mechanisms-of-Vision}. The full code of the analysis pipeline is publicly available at \url{https://github.com/kuz/Human-Intracranial-Recordings-and-DCNN-to-Compare-Biological-and-Artificial-Mechanisms-of-Vision} under the MIT license.

Raw human brain recordings that support the findings in chapters \ref{ch:spectral-signatures-based} and \ref{ch:intracranial-dcnn-based} are available from Lyon Neuroscience Research Center but restrictions apply to the availability of these data, which were used under license for the current study, and so are not publicly available. Raw data are however available from the author upon reasonable request and with permission of Lyon Neuroscience Research Center.

The data supporting the findings in Chapter \ref{ch:mental-space-visualization-based} along with the code of the analysis and experiments are publicly available at \url{https://bitbucket.org/ilyakuzovkin/adaptive-interactive-bci}. 

\chapter{Supplementary materials}

\section{Detailed visualizations of spectral signatures of visual processing filtered and feature importance maps}
\label{sup:signatures}
\noindent Full figures of averaged time-frequency importance maps, per-subject importance maps and per-area maps are available for download from \url{https://figshare.com/articles/Time-Frequency_Importance_Maps_average_per-subject_per-area/8223356}.
\ \\

\noindent Normalized per-probe time-frequency power activity plots with importance contour overlay are available at \url{https://figshare.com/articles/Normalized_TF_activity_with_importance_contour_overlay/8223389}.
\ \\

\noindent Full figures of comparison between polypredictive and monopredictive probes are available at \url{https://figshare.com/articles/Difference_between_polypredictive_and_monopredictive_neural_locations/8223398}.
\ \\

\noindent For each cluster we have visualized the cluster mean and also the TF activity of all individual probes that constitute that cluster. Full-resolution figures can be downloaded from \url{https://figshare.com/account/projects/64523/articles/8223383}.

\section{Mappings of Brodmann areas to layers of DCNN per area, layer and subject}
\label{sup:dcnn}
\noindent Visualizations of the alignment of all Brodmann areas to the layers of Deep Convolutional Neural Network based on representational similarity analysis are available at \url{https://figshare.com/articles/RSA_mapping_of_all_Brodmann_areas_to_DCNN_layers/8222579}. Mappings to ventral stream only are available at \url{https://figshare.com/articles/RSA_mapping_of_visual_Brodmann_areas_to_DCNN_layers/8222546}.
\ \\

\begin{figure}[h!]
    \centering
    \includegraphics[width=1.0\linewidth]{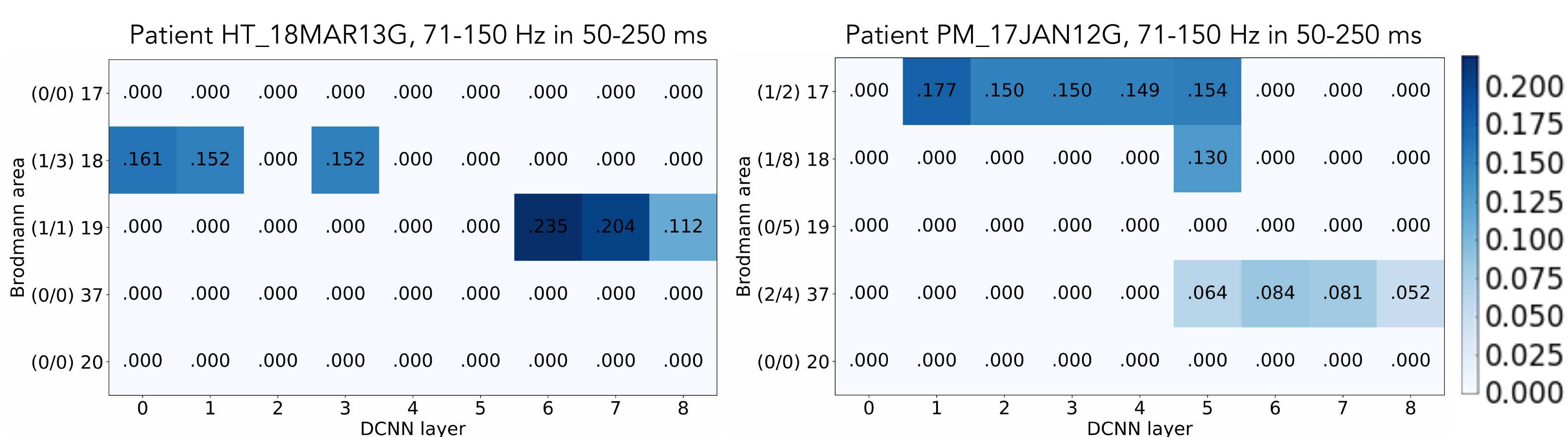}
    \caption{Single subject results from two different subjects. The numbers show the sum of correlations normalized by the number of probes in an area. On the left plot we see how a probe in Brodmann area 18 is mapped to the layers 0, 1, and 3 DCNN, while the activity in Brodmann area 19, which is located further along the ventral stream, is mapped to the higher layers of DCNN: 6, 7, 8. Similar trend is seen on the right plot. The numbers on the left of each subplot show the number of significantly correlating probes in each area out of the total number of responsive probes in that area.}
    \label{fig:single_plates}
\end{figure}

\noindent Visualizations of all single subject mappings of the Brodmann areas in the ventral stream to DCNN layers are available at \url{https://figshare.com/articles/Per-subject_RSA_mapping_of_visual_Brodmann_areas_to_DCNN_layers/8222939}.

\setcounter{secnumdepth}{-1}
\chapter{Acknowledgements}
I would like to thank the city of Tartu and its inhabitants for supporting the academic spirit of the city, University of Tartu, that is fueling that spirit and the Institute of Computer Science for getting me in touch with that spirit and for all of the knowledge and support it provided during my years here. I am very grateful to Raul Vicente for building the Computational Neuroscience Lab where I was able to apply my knowledge to fascinating challenges in neuroscience, and, of course, for his supervision during my PhD studies. I would like to thank Konstantin Tretyakov for introducing me to the field of machine learning and supervising my bachelor and master theses, and Sven Laur for his role in developing my understanding of machine learning further. This thesis would not be possible without the work that Juan R. Vidal and other co-authors from Lyon Neuroscience Research Center did to create the amazing dataset that facilitated our findings. I~am grateful to them for their work and for sharing that data with me. Anna Leontjeva, thank you for all the academic and philosophical discussions we had that helped to shape this thesis. It is hard to overestimate the role and contribution of the constant flow of ideas born at the seminars, lunch breaks, discussions I had with Tambet Matiisen, Ardi Tampuu, Kristjan Korjus and all the lab members and lab alumni, students, and co-authors. Thank you all for having that scientific passion and sharing it with others. \foreignlanguage{russian}{Спасибо} to my parents and family for instilling the value of knowledge and providing me with the opportunity to pursue it.
\ \\

This work has received financial support from Estonian Research Council through the research grants PUT438, PUT1476 and IUT20-40, Estonian Centre of Excellence in IT (EXCITE) funded by the European Regional Development Fund, and the Institute of Computer Science of University of Tartu.

\chapter*{List of abbreviations}
\addcontentsline{toc}{chapter}{List of abbreviations}
\begin{table}[h]
	\begin{tabular}{@{}ll@{}}
		AI		& Artificial Intelligence \\
		ANN 		& Artificial Neural Network \\
		BA		& Brodmann Area \\
		BCI		& Brain-Computer Interface \\
		BMU		& Best Matching Unit \\
		CNN 		& Convolutional Neural Network \\		
		CND		& Computational Neuroscience \\
		DCNN	\phantom{MMMI}	& Deep Convolutional Neural Network \\
		DL		& Deep Learning \\
		DMN		& Default Mode Network \\
		DNN		& Deep Neural Network \\
		DQN		& Deep Q-Network \\
		DRL		& Deep Reinforcement Learning \\
		EEG		& Electroencephalography \\
		FDR		& False Discovery Rate \\
		FFA		& Fusiform Face Area \\
		fMRI		& Functional Magnetic Resonance Imaging \\
		GMM		& Gaussian Mixture Models \\
		HMM		& Hidden Markov Models \\
		ICA		& Independent Component Analysis \\
		ITC		& Inferior Temporal Cortex \\
		iEEG		& Intracerebral electroencephalography \\
		kNN		& k-Nearest Neighbors \\
		LDA		& Linear Discriminant Analysis \\
		LFP		& Local Field Potential \\
		LSTM		& Long Short-Term Memory \\
		MDS		& Multidimensional Scaling \\
		MEG		& Magnetoencephalography \\
		ML		& Machine Learning \\
		MNI		& Montreal Neurological Institute \\
		MRI		& Magnetic Resonance Imaging \\
	\end{tabular}
	\label{tab:abbreviations-a}
\end{table}

\begin{table}[h]
	\begin{tabular}{@{}ll@{}}
		NS		& Neuroscience \\
		POSOM\phantom{MMM}	& Predictive Online Self-Organizing Map\\
		PCA		& Principle Component Analysis \\
		PPA		& Parahippocampal Place Area \\
		PPC		& Posterior Parietal Cortex \\
		RBM		& Restricted Boltzmann Machines \\
		RDM		& Representational Dissimilarity Matrix \\
		RF		& Random Forest \\
		RL		& Reinforcement Learning \\
		RNN		& Recurrent Neural Network \\
		ROI		& Region of Interest \\
		RSA		& Representational Similarity Analysis \\
		SNR		& Signal to Noise Ration \\
		SOM		& Self-Organizing Map\\
		SVM		& Support Vector Machines \\
		TF		& Time-Frequency \\
		t-SNE		& t-Distributed Stochastic Neighbor Embedding \\
		VEP		& Visual Evoked Potential \\
		VWFA		& Visual Word Form Area \\
	\end{tabular}
	\label{tab:abbreviations-b}
\end{table}
\phantom{x}
\ \\
\ \\
\ \\
\ \\
\ \\
\ \\
\ \\
\ \\
\ \\
\ \\

%
%
\begin{otherlanguage}{estonian}
\chapter*{Sisukokkuvõte}
\addcontentsline{toc}{chapter}{Sisukokkuvõte (Summary in Estonian)}

\section*{Inimaju arvutuslikke protsesside mõistmine\\masinõpe mudelite tõlgendamise kaudu\\ \Large\textmd{Andmepõhine lähenemine arvutuslikku neuroteadusesse}}

Käesolev doktoritöö uurib, millist rolli mängivad masinõppe meetodid selliste neuroteaduse mudelite loomisel, mis pakuvad modelleeritava nähtuse intuitiivset kirjeldust. Meie tahame näidata, et modelleerimisprotsessis võib asendada inimese poolt mudeli konstrueerimise masinõppe mudeli treenimisega. See lubab nihutada neuroteadlase töö neuroteaduslike mudelite loomiselt masinõppega treenitud mudelite tõlgendamisele. Valideeritud masinõppe mudeli puhul saame oletada, et see mudel peegeldab mehhanismi, mis toimus treeningandmed genereerinud ajus. Nüüd seisneb uurija roll selles, et tõlgendada masinõppe mudeli tööprintsiipi ja artikuleerida seda reaalsuse elegantse kirjeldusena.

Masinõppe meetodid suudavad töödelda palju suuremaid andmekoguseid ja uurida palju keerulisemaid seoseid kui inimene. Neuroandmestike hulk ja suurus kasvab väga kiiresti ja koos sellega kasvab ka andmepõhise analüüsi roll neuroteaduses. Selles töös me näitame, kuidas suurte andmemahtude peal treenitud masinõppe mudelit võib tõlgendada niimoodi, et see tõlgendus kirjeldab mitte ainult masinõppe mudeli mehhanismi ennast, vaid pakub ka seletust modelleeritava ajuprotsessi kohta. Teises peatükis töötame välja taksonoomia, mis grupeerib masinõppe meetodid selle järgi, kuidas teadmised, mida masinõppe algoritm järeldas andmetest, on väljendatud algoritmi sisemise esitusena. Erinevad esitusviisid lubavad erinevaid tõlgendusi, seega (neuro)teadlane, kelle eesmärk on mitte pelgalt treenida mudel, vaid ka aru saada, kuidas mudel saavutab oma tulemusi, peab arvestama algoritmi valides selle algoritmi sisemise esitusviisiga.

Me näitlikustame kirjeldatud lähenemist kolme uuringuga, mis kasutavad masinõppe tõlgendamismeetodeid kolmel erineval neuroloogilisel tasemel. Igas uuringus me näitame, kuidas masinõppe tõlgendamismeetodid olid rakendatud ja millist neuroteaduse teadmist see lubas meil avastada. Esimeses uuringus (peatükk 3) kasutame me tunnuste tähtsuse analüüsi juhumetsa (Random Forest) mudelitel, mis olid treenitud 100 inimese koljusiseste ajusignaalide peal. Analüüs lubas meil tuvastada, millised aegsagedus komponendid iseloomustavad inimese ajusignaali visuaalsete objektide tuvastamise ülesande puhul. Teine uuring (peatükk 4) kasutab esituste erinevuste analüüsi (Representation Dissimilarity Analysis, RDA), et võrrelda samade visuaalsete stiimulite puhul signaale inimese aju ventraalses piirkonnas ja konvolutsiooniliste tehisnärvivõrkude aktivatsioone erinevates kihtides. See metoodika võimaldas meil teha järeldusi inimese visuaalse ajukoore protsessidest ja kinnitada hüpoteesi, et mõlemad süsteemid, nii bioloogiline kui ka tehislik, kasutavad hierarhilist struktuuri visuaalsete objektide tuvastamise protsessis. Kolmas uuring (peatükk 5) pakub välja meetodi, mis lubab kasutaja arvutiekraanil reaalajas visualiseerida  tema ajuolekute ruumi projektsiooni. See funktsioonalsus on saavutatud kasutades topoloogiat säilitavat (topology-preserving) mõõtmelisuse vähendamise meetodit, mis teisendab mitmemõõtmelise ajusignaali kahemõõtmeliseks visualisatsiooniks. Visualiseeritud kujutis on inimese jaoks arusaadav ja lubab näha, millised ajusignaalid ja mõtteseisundid on üksteisele sarnased ja millised erinevad. Kõik kolm uuringut loovad ajuandmete peal teatud masinõppe mudeli ja siis kasutavad tõlgendamismeetodeid, et saada kätte teadmisi, mis on kasulikud neuroteaduse kontekstis. Samas, abstraktsiooni tase igas uuringus on erinev: esimeses me töötleme lokaalseid elektrofüsioloogilisi signaale, teises huvitab meid ajuprotsessi üldine struktuur ja signaali esitus, kolmandas me rakendame andmepõhist lähenemist mõtteseisundite visualiseerimiseks.

Kokkuvõttes, selles töös kirjeldatud ideed ja tulemused tõstavad esile rolli, mida masinõpe saaks mängida edendamaks meie arusaamist inimese ajust. Masinõppe meetodite oskus leida mustreid ja luua teadmisi on oluline täiendus meie pädevustele loodusprotsesside ja fenomenide seletamisel. Neuroteadus sobib eriti hästi nende meetodite rakendamiseks, sest sellel teadusharul on teatav sümbioos masinõppe ja tehisintellekti uuringutega. Nende teadusharude ühine eesmärk on avastada mehhanism, mis seletaks, kuidas töötab meie aju ja intellekt. Nagu on näidatud esimeses peatükis, kui bioloogilisel ja tehislikul süsteemil on sama eesmärk, siis tihti töötavad mõlemad süsteemid välja üllatavalt sarnased mehhanismid selle eesmärgi saavutamiseks. Mõned näited sellistest sarnasustest on: visuaalse objektituvastuse hierarhia visuaalses ajukoores ja konvolutsioonilistes tehisnärvivõrkudes; mälu konsolideerimine hipokampuses ja stiimulõppe algoritmides; heksagonaalne võre mida aju võrerakud ja stiimulõppe mudelid õpivad ruumilise navigatsiooni jaoks. Need näited tõendavad asjaolu, et mõned mehhanismid meie ajus on sarnased mehhanismidega, milleni jõuavad õppimise käigus masinõppe algoritmid. Mõned neist mehhanismidest on avastatud ja on ainult loogiline oletada, et on olemas veel mõned, mida me veel ei tea. Oma tööga me rõhutame masinõppe mudelite tõlgendamise tähtsust selliste mehhanismide avastamiseks.

\end{otherlanguage}

%
%
\chapter{Curriculum Vitae}

\section*{Personal data}
\begin{small}
\begin{tabular}{@{}l@{\hskip7mm}l}
Name:       	& Ilya Kuzovkin\\
Date of birth:	& 9 July 1988\\
Contacts:       & \texttt{ilya.kuzovkin@gmail.com} \\
		& \texttt{www.ikuz.eu}\\
Citizenship: 	& Estonia\\ 
\end{tabular}
\end{small}

\renewcommand{\arraystretch}{1.4}
\section*{Education}
\begin{small}
\begin{tabular}{@{}l@{\hskip7mm}p{100mm}}
2013 -- 2019	& PhD candidate at University of Tartu, Faculty of Science and Technology\\
2011 -- 2013    & MSc in Computer Science from University of Tartu, Faculty of Mathematics and Computer Science\\
2007 -- 2011 	& BSc in Computer Science from University of Tartu, Faculty of Mathematics and Computer Science
\end{tabular}
\end{small}

\section*{Employment}
\begin{small}
\begin{tabular}{@{}l@{\hskip7mm}p{100mm}}
2016 -- ...         & Machine Learning Architect at OffWorld Inc., USA \\
2014 -- 2019    & Junior Researcher at the Institute of Computer Science of University of Tartu\\
2012 -- 2016    & Software Engineer at Ideelabor OÜ\\
2012 -- 2015    & Teaching Assistant at the Institute of Computer Science of University of Tartu\\
2011 -- 2012    & Research Engineer at Department of Psychology of University of Tartu and Realeyes Inc.\\
2007 -- 2011    & Software Engineer at Surflink OÜ\\
\end{tabular}
\end{small}

%

%

\renewcommand{\arraystretch}{1.0}

%
%
\begin{otherlanguage}{estonian}
\chapter*{Elulookirjeldus}
\addcontentsline{toc}{chapter}{Elulookirjeldus (Curriculum Vitae in Estonian)}

\section*{Isikuandmed}
\begin{small}
\begin{tabular}{@{}l@{\hskip7mm}l}
Nimi:       		& Ilya Kuzovkin\\
Sünniaeg:		& 9 Juuli 1988\\
Kontaktandmed:       & \texttt{ilya.kuzovkin@gmail.com} \\
			& \texttt{www.ikuz.eu}\\
Kodakondsus: 	& Eesti\\ 
\end{tabular}
\end{small}

\renewcommand{\arraystretch}{1.4}
\section*{Haridus}
\begin{small}
\begin{tabular}{@{}l@{\hskip7mm}p{100mm}}
2013 -- 2019	& Doktorant, Tartu Ülikool, Loodus- ja täppisteaduste valdkond\\
2011 -- 2013  & Magister informaatikas, Tartu Ülikool, Matemaatika-informaatikateaduskond \\
2007 -- 2011 	& Bakalaureus informaatikas, Tartu Ülikool, Matemaatika-informaatikateaduskond \\
\end{tabular}
\end{small}

\section*{Teenistuskäik}
\begin{small}
\begin{tabular}{@{}l@{\hskip7mm}p{105mm}}
2016 -- ...         & Masinõpe arhitekt, OffWorld Inc., USA \\
2014 -- 2019    & Nooremteadur, Arvutiteaduse instituut, Tartu Ülikool \\
2012 -- 2016    & Tarkvarainsener, Ideelabor OÜ \\
2012 -- 2015    & Õppeassistent, Arvutiteaduse instituut, Tartu Ülikool \\
2011 -- 2012    & Teadusinsener, Psühholoogia instituut, Tartu Ülikool ja Realeyes Inc. \\
2007 -- 2011    & Tarkvarainsener, Surflink OÜ \\
\end{tabular}
\end{small}

%

%
\renewcommand{\arraystretch}{1.0}
\end{otherlanguage}

\chapter{List of original publications}
\vspace{-1.4em}
\noindent{\small\textsc{I. Kuzovkin, J. R. Vidal, M. Perrone-Bertlotti, P. Kahane, S. Rheims, J. Aru, J.-P. Lachaux, R. Vicente}}. \textbf{Identifying task-relevant spectral signatures of perceptual categorization in the human cortex}. \emph{In review}, 2020
\vspace{0.4em}\\
\emph{Lead author. Contributed to the idea, methodology, design of the experiments, development of the software, analysis, and writing of the manuscript.}
\vspace{0.6em}

\noindent{\small\textsc{I. Kuzovkin, R. Vicente, M. Petton, J.-P. Lachaux, M. Baciu, P. Kahane, S. Rheims, J. R. Vidal, J. Aru}}. \textbf{Activations of deep convolutional neural networks are aligned with gamma band activity of human visual cortex}. \emph{Communications Biology}, Vol. 1, No. 1, 2018
\vspace{0.4em}\\
\emph{Lead author. Contributed to the idea, methodology, design of the experiments, development of the software, analysis, and writing of the manuscript.}
\vspace{0.6em}

\noindent{\small\textsc{I. Kuzovkin, K. Tretyakov, A. Uusberg, R. Vicente}}. \textbf{Mental state space visualization for interactive modeling of personalized BCI control strategies}. \emph{Journal of Neural Engineering}, Vol. 17, No. 1, 2020
\vspace{0.4em}\\
\emph{Lead author. Contributed to the idea, methodology, design of the experiments, data collection, development of the software, analysis, and writing of the manuscript.}
\vspace{2.0em}\\

\noindent{\large Relevant publications not included in the thesis:}
\ \\


\noindent{\small\textsc{A. Leontjeva, I. Kuzovkin}}. \textbf{Combining static and dynamic features for multivariate sequence classification}. \emph{Proceedings of the 3rd IEEE International Conference on Data Science and Advanced Analytics (DSAA)}, pp. 21-30, 2016\\

\noindent{\small\textsc{A. Ingel, I. Kuzovkin, R. Vicente}}. \textbf{Direct information transfer rate optimisation for SSVEP-based BCI}. \emph{Journal of Neural Engineering}, Vol. 16, No. 1, 016016, 2018

\end{document}